\documentclass[12pt]{JHEP3}
\usepackage{epsfig}
\preprint{ {\tt hep-th/0510264} 
\\ {ITP-UU-05/48}\\
{SPIN-05/33} }
\newcommand{\be}[1]{ \begin{equation}\label{#1} }
\newcommand{\ee}{\end{equation}}
\newcommand{\bea}[1]{\begin{eqnarray}\label{#1} }
\newcommand{\eea}{\end{eqnarray}}
\newcommand{\eq}[1]{(\ref{#1})}

\title{  
Towards a string bit formulation of ${\cal N}=4$ super
Yang-Mills.}
\author{ Luis F. Alday$^{a}$,
Justin R. David$^{b,d}$, Edi Gava$^{b,c}$ , K. S. Narain$^b$ \\
$^a$Institute for Theoretical Physics and Spinoza Institute, \\
Utrecht University, 3508 TD Utrecht, \\
The Netherlands. \\
$^b$High Energy Section, \\
The Abdus Salam International Centre for Theoretical Physics,
\\Strada Costiera, 11-34014 Trieste, Italy.\\
$^c$Instituto Nazionale di Fisica Nucleare, sez. di Trieste, \\
and SISSA, Italy. \\
$^d$Harish-Chandra Research Institute, \\
Chhatnag Road., Jhunsi, Allahabad 211019, India.
\\
\email{L.F.Alday@phys.uu.nl, justin@hri.res.in, 
\\
gava, narain@ictp.trieste.it}
}
\abstract{
We show that planar ${\cal N}=4$ Yang-Mills theory at 
zero 't Hooft coupling can 
be efficiently described in terms of 8 bosonic and 8 fermionic
oscillators. We show that these oscillators can 
serve as world-sheet variables, the string bits,
of a discretized string.
There is a one to one correspondence between the on shell 
gauge invariant 
words  of the free Y-M theory and the states in the oscillators'
Hilbert space, obeying a local
gauge  and cyclicity constraints.
The planar two-point functions and the three-point functions of
all gauge invariant words are obtained 
by the simple delta-function overlap of the corresponding
discrete string world sheet. 
At first order in  the 't Hooft coupling, 
i.e. at one-loop in the Y-M theory,
the logarithmic corrections 
of the planar two-point and the three-point functions
can be incorporated by nearest neighbour interactions
among the discretized string bits.
In the $SU(2)$ sub-sector we show that the one-loop
corrections to the structure constants can be 
uniquely determined by the symmetries of the 
bit picture. 
For the $SU(2)$ sub-sector
we construct a gauged, linear, discrete world-sheet model
for the oscillators, 
with only nearest neighbour couplings, which reproduces 
the anomalous dimension Hamiltonian up to
two loops. This model also obeys BMN scaling to all loops.
}

\begin{document}
\baselineskip 3.5ex

\section{Introduction}

Among all the examples of field theories dual to a string theories,
the most precise and well studied is the case of the duality between
${\cal N}=4$ Yang-Mills theory in four dimensions with gauge group
$SU(N)$ and type IIB string theory on $AdS_5\times S^5$ 
\cite{Maldacena:1997re,Witten:1998qj,Gubser:1998bc,Aharony:1999ti}.
In units of the radius of  $AdS$, the string length is related to the 
't Hooft coupling of the gauge theory by
\be{ymmap}
\alpha' = \frac{1}{\sqrt{\lambda}}
= \frac{1}{\sqrt{g^2_{\rm YM}  N}}, 
\;\;\;\;\; G_N = \frac{1}{N^2},
\ee
where $g_{\rm YM}$ is the Yang-Mills coupling constant, $\alpha'$
the string length measured in units of the radius of $AdS$.
$G_N$ is the Newton's constant 
which is the effective string loop
parameter.

This duality has been tested in the supergravity regime on 
the string theory side, which, according to  \eq{ymmap} 
corresponds to strong 't Hooft coupling and large $N$
limit on the Y-M side. For applications of the duality to physically
more interesting cases, one would like to be able to control
the weak coupling regime in the gauge theory, which requires 
understanding string theory on  $AdS_5\times S^5$ at strong
sigma-model coupling.
The problem is that, at present at least, 
type IIB string on $AdS_5\times S^5$
has not been fully quantized. 
This puts strong limitations to the duality tests
and to the applications of the duality to physically relevant 
situations, confining them to the 
regime when type IIB string theory
can be approximated by the corresponding supergravity
or to sectors which involve large quantum numbers \cite{Gubser:2002tv}, 
for which semiclassical quantization is amenable. 
Early tests of the duality relied on 
protected quantities, 
like the conformal dimensions of chiral primaries of 
the Yang-Mills theory and the spectrum of Kaluza-Klein
supergravity states on 
$AdS_5\times S^5$ \cite{Witten:1998qj}. 
Their three-point functions were also shown
to agree with the three-point couplings evaluated from the supergravity 
action \cite{Lee:1998bx,Freedman:1998tz,Arutyunov:1999en}. 
At present there are no known means to extract information 
regarding the spectrum of string states or their correlation functions
from the sigma model on $AdS_5\times S^5$, other than in the plane wave limit
\cite{Berenstein:2002jq}\footnote{See 
\cite{Arutyunov:2003uj,Arutyunov:2004vx,Plefka:2005bk}
for developments in the use of spinning strings
to to study the quantization of  the sigma model on $AdS_5\times S^5$.} .
The field theory, on the other hand, is best understood at weak coupling,
in particular, as a perturbative expansion in the 't Hooft coupling. 
This has led to many efforts in trying to rewrite the spectrum of
free ${\cal N}=4$ Yang-Mills theory as a spectrum in a 
string theory \cite{Bianchi:2003wx,Beisert:2003te,Bianchi:2004xi}. 
There has also been an effort at reconstructing the string
theory world-sheet by rewriting the correlation function of gauge invariant
operators of the free theory as string-like amplitudes in 
$AdS$ \cite{Gopakumar:2003ns,Gopakumar:2004qb,Gopakumar:2005fx}. 

In this paper we formulate ${\cal N}=4$ Yang-Mills theory at weak coupling
in terms of a discrete world sheet made of bits, starting from the 
free point and then switching on the gauge coupling perturbatively.
There are two main motivations for 
a discrete world-sheet description of the Yang-Mills at weak coupling.
We detail these  below \footnote{Throughout this paper we will work
in the planar limit of the Y-M theory.}.

\vspace{.5cm}
\noindent
\emph{Motivations for a  string bit description}
\vspace{.5cm}

Let us again recall the map between the basic parameters of the 
string theory and ${\cal N}=4$ Yang-Mills.  
From \eq{ymmap} we see that
in the free limit,  the string 
length measured in units of the radius of $AdS$ 
becomes infinite and the string essentially becomes
tensionless,  there is no coupling between neighbouring
points on the  string. Therefore  the string breaks up into 
non interacting bits held together only by the $L_0 =\bar L_0$
constraint \cite{Dhar:2003fi}. 
This can be seen  more precisely  
using the Green-Schwarz type IIB superstring moving in 
$AdS_5\times S^5$ background. 
In the light cone gauge the Hamiltonian
which generates evolution in the light-cone time variable $x^+$ is
given by  \cite{Metsaev:1999gz,Metsaev:2000yu,Tseytlin:2002gz}
\be{lcham}
{\cal H}  =  \int d\sigma
\frac{1}{p_+} \left( P^2  + \frac{T^2}{r^4} (\partial_\sigma X)^2  \right).
\ee
Here we have written down only the bosonic co-ordinates, $T \sim
1/\alpha' = \sqrt{\lambda}$ the tension of the string,
$P$ stands for the momenta of the bosonic co-ordinates
and $X$ the position. $r$ refers to
the  radial distance in $AdS$, and we have used the Poincar\'{e}
co-ordinates of $AdS$ to write the sigma-model action.  The point to 
be emphasized is that the  terms with derivatives in the 
$\sigma$ co-ordinates of the world sheet are suppressed
by powers of the string tension. In the tensionless limit
we can neglect these terms and the string 
breaks up into bits of non-interacting particles held together only
through the level matching constraint $L_0=\bar L_0$. 

The second motivation to think of the string theory dual to 
free Yang-Mills in terms of bits arises from 
examining the AdS/CFT duality in the plane wave limit \cite{Berenstein:2002jq}.
BMN set up a dictionary between 
single trace Yang-Mills operators of large $R$-charge 
and states in the plane wave string theory. 
An example of such a correspondence is the following
\be{dictpp}
a_n^{\dagger i} a_{-n}^{\dagger j} 
| 0 , p_+\rangle_{\rm l.c} \leftrightarrow \frac{1}{
\sqrt{J N^{J +2} }} \sum_{l =1}^J  
{\rm Tr}  \left( \phi^i z^{l} \phi^j z^{J-l} \right)
e^{\frac{2\pi i n l}{J} }.
\ee
Here $i, j$ are directions which correspond to the $SO(4)
\subset SO(6)$ part of the
$R$-symmetry.
$z$ is the complex scalar which belongs to the Cartan
and has $R$-charge unity.  
$a_n^{\dagger i}$ refers to the oscillator modes of the 
plane wave string theory.  Though the above dictionary was set up 
for operators with large $R$-charge, $J\sim \sqrt{N}$, let us imagine 
extrapolating the dictionary for finite $J$ but with $g_{\rm YM} =0$.
As $g_{\rm YM}$ is set to  zero, one expects that there is no renormalization 
of the states. For finite $J$ from \eq{dictpp} it is clear that 
the Yang-Mills state is invariant for $n\rightarrow n+J$, thus we 
arrive at the conclusion that the modes of the string theory are
truncated at order $J$.  This implies that the world sheet 
is discrete, in fact the number of bits (or points ) of the world
sheet is of the order of the length of the Yang-Mills operator. 
Furthermore, it is easy to see from \eq{dictpp} that the cyclicity
of trace translates to the level matching condition on the string side.
In fact this discrete nature of the world sheet has already 
been noticed by 
\cite{Berenstein:2002jq,Constable:2002hw,
Kristjansen:2002bb,Verlinde:2002ig,Vaman:2002ka,Zhou:2002mi}

From the above discussion, it would seem that
one attempt to capture some features of the
string dual for free ${\cal N}=4$ Yang-Mills is just to discretize 
the plane wave string theory. Such an attempt runs into difficulties,
as the plane wave symmetry algebra is a contraction of the
symmetry algebra of $AdS_5\times S^5$.  We illustrate this by
considering the dual of operators which are in the traceless symmetric
representation of $SO(6)$, the chiral primaries. 
If the highest weight state of this representation has $R$-charge $J$, 
then one can suppose the string dual is 
the vacuum made of $J$ bits. 
According to the BMN dictionary the other
components then should be obtained by repeated action of the 
creation of zero
modes $a_0^{i\dagger}$. 
$a_0^i$'s just lower the $J$ charge  of the vacuum state.
From the correspondence with the Yang-Mills state we know that
number of $a_0^{i\dagger}$ cannot exceed $J$. This constraint
has to be imposed on the discrete plane wave theory.
Such constraints make it difficult to formulate the discrete 
theory in terms of plane wave oscillators.

In this paper we will introduce oscillators which naturally capture
the symmetries of $AdS_5\times S^5$ geometry.
We show that planar ${\cal N}=4$ Yang-Mills theory at zero 't Hooft coupling
can be efficiently described in terms of 8 bosonic and 8 fermionic oscillators.
These oscillators can serve as the discrete world-sheet variables 
of a string bit formulation of the Yang-Mills theory. 
We show that there exists a one to one correspondence 
between  the on-shell, gauge invariant words of the free Yang-Mills with 
the Hilbert space of these oscillators, together with a local 
$U(1)$ gauge symmetry
and the cyclicity constraint. 
An universal feature of any string theory is that the interactions are
described by the delta-function overlap of strings. In  fact
the structure constants of gauge invariant words, which in the planar
limit are proportional to $1/N$, should emerge from the joining or splitting 
of the strings. 
We will  show that the planar, two-point 
and three-point functions 
are obtained by the simple delta-function overlap 
of the corresponding discrete world-sheet.

We then turn on the 't Hooft coupling. From \eq{ymmap} we see that turning
on $\lambda$ renders the string tension finite, thereby introducing 
interactions between the bits. At one-loop in $\lambda$ and in the
planar limit only nearest neighbour bits would interact.
We will show that
nearest neighbour corrections to the global $SO(2,4)$ 
charges  in the string bit
formulation reproduces the logarithmic divergences of the 
one-loop corrected two-point functions and the three-point functions of the
Yang-Mills theory. For the $SU(2)$ sub-sector we shown that the symmetries
of the bit picture are sufficient to determine the one-loop
corrected structure constants evaluated in 
\cite{Okuyama:2004bd,Roiban:2004va,Alday:2005nd}, upto an overall coefficient.

We then focus on the anomalous 
dimension Hamiltonian for the $SU(2)$ sub-sector of the theory. 
Here the local $U(1)$
gauge symmetry of the bit picture is used to construct a 
gauged linear model of the corresponding oscillators. 
This model has only nearest neighbour couplings,
but it reproduces the anomalous dimensions in this sub-sector 
to two loops and obeys BMN scaling to all loops.

This paper is organized as follows. In section 2 we introduce
the oscillator variables and show the spectrum of the free
Yang-Mills theory is identical to the Hilbert space of these oscillators 
once the $U(1)$ gauge constraint and the cyclicity constraint are taken
into account.
In section 3 we construct the 2-string and 3-string overlap at $\lambda =0$
and show that they reproduce the two-point functions and the three-point
functions of all gauge invariant words of the Yang-Mills theory.
In section 4 we show that logarithmic corrections
in two-point functions and three-point functions can be 
reproduced due to nearest neighbour corrections to the global 
$SO(2,4)$ generators in the bit picture. We then show that the structure
constants in the $SU(2)$ sub-sector are entirely determined, upto an overall 
constant, from the symmetries of the bit picture.
In section 5 we introduce a gauged, linear, discrete world-sheet model
for the oscillators corresponding to the $SU(2)$ sub-sector, 
having only nearest neighbour interactions. This model
reproduces the anomalous dimensions up to two loops. It also obeys
BMN scaling to all loops.
Section 6 contains our conclusions, appendix A contains
the notations and conventions adopted in this paper. 
Appendices B and C contain details regarding the 
oscillator variables.  Appendix D works out a few examples of 
three-point functions at $\lambda =0$ and compares them to the
three string overlap. Appendix E introduces a
gauged oscillator model with non-nearest neighbour interaction
which reproduces the anomalous dimensions up to three loops.  
Appendix F contains details of the oscillator algebra necessary
for evaluating the anomalous dimension Hamiltonian from the
gauged oscillator model.

\section{${\cal N}=4$ Y-M spectrum in terms of oscillator variables}

The main objective of this section is to write the spectrum 
of free ${\cal N}=4$ Yang-Mills  entirely in terms of 
eight bosonic and eight fermionic oscillators. These 
oscillators will serve as the discretized  world sheet variables 
of a  string theory which we construct to describe the
Yang-Mills theory. We will refer to these oscillators as bits. 

The symmetries of $AdS_5\times S^5$ form the 
supergroup $PSU(2,2|4)$, 
the  bosonic component of this supergroup 
is given by $SO(2, 4)\times SO(6)$. 
The $SO(2,4)$ corresponds to  the conformal group in the 
${\cal N}=4$ super-Yang Mills in four dimensions, while $SO(6)$ 
to its $R$-symmetry.
We will first discuss the conformal algebra and
the strategy we will adopt in classifying the
representations of the conformal group. Then we use 
the oscillator variables  introduced  in 
\cite{Gunaydin:1984fk,Gunaydin:1998sw} \footnote{The latter reference 
contains a concise review of the oscillators variables.}
to write down  all the generators of $PSU(2,2|4)$.
We then  construct all the letters 
and the single trace gauge invariant words of ${\cal N}=4$ Yang-Mills 
using the oscillator variables.  
We evaluate the partition function of the single
letters and the single trace gauge invariant words using the 
oscillator  variables and show that they are in one to one
correspondence with  the gauge theory operators at zero coupling
modulo equations of motion and Bianchi identities.

\subsection{The conformal algebra}

The conformal group in four dimensions $SO(2,4)$ is generated by the
Lorentz generators $M_{\mu\nu}$, the four momentum $P_\mu$, the
generators of special conformal
transformations $K_\mu$, $(\mu, \nu = 0, 1, 2, 3)$ and the Dilatation 
generator $D$. The algebra of the generators is given by  
\bea{confal}
[M_{\mu\nu},M_{\rho\sigma}] & = &i (\eta_{\nu\rho}M_{\mu\sigma}-
\eta_{\mu\rho}M_{\nu\sigma} -\eta_{\nu\sigma}M_{\mu\rho}+
\eta_{\mu\sigma}M_{\nu\rho}),\cr [ P_{\mu}, M_{\rho\sigma} ] & = &
i (\eta_{\mu\rho}P_{\sigma}-\eta_{\mu\sigma} P_{\rho}),\cr
[K_{\mu},M_{\rho\sigma}]& = &i
(\eta_{\mu\rho}K_{\sigma}-\eta_{\mu\sigma} K_{\rho}),\cr
[D,M_{\mu\nu}]& = & [P_{\mu},P_{\nu}] = [K_{\mu},K_{\nu}]=0,\cr
[P_{\mu},D] & = &iP_{\mu}, \quad [K_{\mu},D]=-iK_{\mu},\cr
[P_{\mu},K_{\nu}]& = &2i(\eta_{\mu\nu}D-M_{\mu\nu}).
\eea
here $\eta_{\mu\nu} = {\rm diag }\; (-, +, +, +)$. The correspondence 
with the $SO(2,4)$ generators is made with the identifications
\be{soconi}
M_{\mu 5} = \frac{1}{2}( P_\mu - K_\mu), \;\;\;
M_{\mu 6} = \frac{1}{2}( P_\mu + K_\mu), \;\;\;
M_{56} = -D,
\ee
then these generators satisfy the $SO(2,4)$ algebra given by
\be{fullageb}
 [M_{AB}, M_{CD}] = i(\eta_{BC}M_{AD} - \eta_{AC}M_{BD}
-\eta_{BD}M_{AC} + \eta_{AD}M_{BC}).
\ee
Here $A, B, \ldots = 0, 1, 2, 3, 5, 6$ and 
$\eta_{AB} = {\rm diag} ( - , +, +, +, +, -)$. 
Note that the directions $0,6$ are time like and the directions $1, 2,
3, 5$ are space like.

A convenient way to classify representations of $SO(2,4)$ is to use
the maximal compact subgroup  $SO(2)\times SO(4) = U(1)_E \times
SU(2)_L \times SU(2)_R$. 
The compact generators correspond to 
\bea{compagen}
L_{m}&=&\frac{1}{2} ( \frac{1}{2} \varepsilon_{mnl}
M_{nl}+M_{m5}), \qquad : SU(2)_L,\cr 
R_{m}&=&\frac{1}{2} (
\frac{1}{2} \varepsilon_{mnl} M_{nl}-M_{m5}), \qquad
: SU(2)_R, \cr
E &=& M_{06} = \frac{1}{2}( P_0 + K_0), \qquad : U(1)_E,
\eea
where $m, n, l$ refer to the space like directions $1,2,3$. 
From these generators it is easy to see that the conformal algebra has
a decomposition as 
$SO(2,4) \rightarrow L^+\oplus L^0 \oplus L^-$ such that the following
commutation relations hold
\bea{gradcom}
[L^0, L^\pm] &\subset& L^{\pm}, \qquad [L^+, L^-] \subset L^0, \cr
[E, L^\pm] &=& \pm L^\pm, \qquad [E, L^0] =0.
\eea
We can now construct unitary representations from the state
$|\Omega\rangle$ with quantum numbers 
 $(j_L, j_R, E)$ and annihilated by the elements of $L^-$.
The representations are obtained by the action of the raising 
operators $L^+$ on the state $|\Omega\rangle$.
$E$ will have to be 
bounded from below for a physically relevant
 unitary representation. 
We denote the space of these representations by
${\cal H}_1$.

In the conformal invariant ${\cal N}=4$ Yang-Mills,
one usually represents the action of the conformal group on 
gauge invariant operators say at $x=0$. 
The stability group at $x=0$ is generated by the Lorentz generators
$M_{\mu\nu}$ which form $SL(2, C)$, 
the dilatations $D$ and the special conformal generators
$K_\mu$. The primary fields are those which are annihilated by 
$K_\mu$,  they carry the $SL(2,C)$ quantum numbers $(j_M, j_N)$ 
and the conformal dimensions $\Delta$.  
The generators of $SL(2,C)$  can be written down as two 
$SU(2)$'s given by
\bea{lorgen}
M_m &=& \frac{1}{2} ( \frac{1}{2} \epsilon_{mnl} M_{nl} + i M_{0m}
) , \cr 
N_m &=& \frac{1}{2} ( \frac{1}{2} \epsilon_{mnl} M_{nl} - i M_{0m}
) , \cr
D &=& -M_{56}.
\eea
It is sufficient to restrict
our attention to only primaries and generate all the 
secondaries by the action of momentum generator $P_\mu$.  
We denote the space of these representations by ${\cal H}_2$.

We now wish to find an isomorphism between ${\cal H}_1$ and ${\cal
H}_2$. Comparing \eq{compagen} and \eq{lorgen} it is easy to see if
there is a flip of the $5$-axis to  $i$ times the  $0$-axis the 
generators of $L_m$ and $R_m$ of
\eq{compagen} go over to $M_m$ and $N_m$ \eq{lorgen} 
and $E$ goes over to $-iD$.
Therefore to perform this rotation one needs to rotate by an angle
$\pi/2$ in the $0-5$ plane with a factor of $i$. This is performed by 
the following operation
\be{simtrans}
U = \exp{( \frac{\pi}{2} M_{05} ) } = 
\exp{\frac{\pi}{4} ( P_0 -K_0) }.
\ee
Note that in the above we do not have a factor of $i$ in the exponent,
this takes care of the fact that we need to rotate the $0$ axis to 
$i$ times the $5$ axis. A detailed proof will be given in the
appendix B.
The transformation $U$ has the following properties
\be{propu}
U^{-1}K_\mu U = L^-, \;\;\;\; U^{-1}P_\mu U = L^+, \;\;\; 
U^{-1}D U = iE.
\ee
Our strategy to obtain primary fields at the origin in ${\cal H}_2$
will be to start with the state $|\Omega\rangle$ in ${\cal H}_1$ which
is annihilated by all $L^-$ and then perform the transformation 
$U |\Omega\rangle$. It is now easy to see that using the first equation in
\eq{propu}, we have $K_\mu U|\Omega \rangle =0$. 
Therefore we can identify $U|\Omega\rangle$ with a gauge invariant operator
at $x=0$. Now to translate it to an arbitrary position $x$, we further
perform the transformation 
$\exp(ix P) U |\Omega \rangle$.

\subsection{The oscillator construction}

\noindent
\emph{ The conformal group} 
\vspace{.5cm}

We now discuss the method of constructing unitary infinite dimensional 
representation of $SO(2,4)$ using eight bosonic oscillators.
With this aim we organize the eight bosonic oscillators as 
into four complex oscillators transforming in the spinor 
representation of $SO(2,4)$, this given by. 
\be{defbspin}
\psi = 
\left(
\begin{array}{c}
a^{\dot 1} \\
a^{\dot 2} \\
-b_1^\dagger \\
-b_2^\dagger 
\end{array}
\right).
\ee
Here the oscillators $a_{\dot \gamma}$ and $b_{\gamma}$ 
obey the following commutation 
relations
\bea{bcomm}
[a^{\dot\gamma}, a_{\dot\delta\dagger}] = 
\delta^{\dot\gamma}_{\dot\delta}, \;\;&\;&\;\;
[b^{\gamma}, b_{\delta^\dagger}] = 
\delta^{\gamma}_{\delta}, \;\;\;\;\;
\cr
[a^{\dot\gamma}, b^{\gamma}] = 0 \;\;&\;&\;\;
[a^{\dot\gamma}, b_{\gamma}^\dagger] =0, \qquad \;\;\; 
\gamma, \delta , \dot\gamma, \dot\delta = 1,2.
\eea
\footnote{Note that we have defined the annihilation operators with 
raised indices.}
The action of  $SO(2,4)$ on the Fock space of these oscillators are
given by
\bea{genfob}
\hat{M}_{AB} = \bar\psi M_{AB} \psi, \qquad
{\rm where} \;\;\bar\psi = \psi^\dagger\gamma^0 = ( a_{\dot 1}^\dagger,
a_{\dot 2}^\dagger, b^1, b^2),
\eea
$\gamma^0$ and $M_{AB}$ are four dimensional irreducible but
non-unitary representation of $SO(2,4)$. They are  written down in 
terms of four dimensional gamma matrices, the details of these are
given in  appendix A. These generators 
satisfy \eq{fullageb} and they also have the property
\be{improp}
\gamma^0 M_{AB} = M_{AB}^\dagger \gamma^0, \quad\;\; \gamma^{0\dagger}
=\gamma^0. 
\ee
These properties ensure that 
the generators  
 $\hat{M}_{AB}$  of
\eq{genfob} which act on the Fock space of 
oscillators are Hermitian and they satisfy the $SO(2,4)$ algebra
\eq{fullageb}. This  is  due to the following
commutation  relations
\be{relhat}
[\hat{M}_{AB} , \hat{M}_{CD}] = \bar\psi[ M_{AB} , M_{CD} ] \psi.
\ee
which can be shown using the simple commutation rules of the
oscillators \eq{bcomm}. Since the generators $\hat{M}_{AB}$ 
are Hermitian, they generate an infinite dimensional 
but unitary representation
of $SO(2,4)$ in the Fock space of oscillators $a^{\dot\gamma}, 
b^{\gamma}$.

We will now discuss the properties of this Fock space. 
The vacuum is defined as $a^{\dot\gamma} |0\rangle = 
b^{\gamma}|0\rangle=0$. 
In this Fock space 
we find it convenient 
to work with the generators in the maximal compact sub-group
$U(1)_E\times SU(2)_L\times SU(2)_R $. As we mentioned in the 
previous subsection the transformation $U$ given in \eq{simtrans}
takes a representation in the maximal compact sub-group 
${\cal H}_1$ to the usual
classification of fields of the conformal group ${\cal H}_2$. 
The action of the transformation $U$ on the Fock space is given by
\bea{simtrfo}
\hat U = \exp(\frac{\pi}{2} \hat M_{05}  ) = \exp(-\frac{\pi}{4}
(a^\dagger b^\dagger + ba) )  \cr
{\rm {where} } \qquad
a^\dagger b^\dagger =  
a^\dagger_{\dot 1} b^\dagger_{1} 
+ a^\dagger_{\dot 2} b^\dagger_{2}, 
\qquad ba = 
b^{ 1} a^{\dot 1} 
+ b^{ 2} a^{\dot 2}. 
\eea
We now give the properties of $\hat U$ which corresponds to the ones 
given in \eq{propu}
\bea{fpropu}
\hat U ^{-1} \hat K_\mu \hat U = b \sigma_\mu a, &\qquad&
\hat U ^{-1} \hat P_\mu \hat U = -a^\dagger \bar\sigma_\mu b^\dagger, 
\cr
\hat U ^{-1}  \hat M_{mn} \hat U =  \hat M_{mn},  &\qquad&
\hat U ^{-1}  \hat M_{0m} \hat U =  i \hat M_{m5}. \
\\ \nonumber
\hat U ^{-} (-i\hat D)\hat U &=& 
\hat E,  \cr
&=& \frac{1}{2} ( a^\dagger a + bb^\dagger) =
\frac{1}{2} ( N_a + N_b) + 1.
\eea
Here $\sigma ^\mu = (-1, \sigma^m)$ and $\bar{\sigma}^\mu =(-1,
-\sigma^m)$,  
\be{defnumber}
N_a = a_{\dot 1}^\dagger a^{\dot 1} + a_{\dot 2}^\dagger a^{\dot 2}, 
\qquad
N_b= b_1^\dagger b^1 + b_2^\dagger b^2.
\ee
The properties given in \eq{propu} are most easily shown in 
the four dimensional representation of $SO(2,4)$ and then 
they are immediate in the Fock space representation because of 
\eq{relhat}, the details are provided in Appendix B. 
Other useful properties of $\hat U$ which will be used repeatedly
in the next sections is 
\be{usepropu}
U^\dagger U E = - U^\dagger U E, \qquad
U^2 E = - U^2 E. 
\ee
The above property is easily shown by repeatedly using the second
line of \eq{fpropu} and  $U^\dagger = U$.
Finally, the left and right $SU(2)$ generators are given 
\bea{lrsu2}
\hat{L}_m &=& \frac{1}{2} \left( \frac{1}{2} \epsilon_{mnl} \hat M_{nl}
+ \hat M_{m5} \right) = -a^\dagger \sigma_m a, 
\cr
\hat{R}_m &=& \frac{1}{2} \left( \frac{1}{2} \epsilon_{mnl} \hat M_{nl}
- \hat M_{m5} \right) = b \sigma_m b^\dagger. 
\eea
It is now clear that why  it is convenient to work in the 
maximal compact subgroup. From
\eq{fpropu} we see that the conjugate  of $\hat K_\mu$ are
annihilation operators while that of $\hat P_\mu$ are creation
operators. Therefore the vacuum in the Fock space satisfies
\be{vaccond}
\hat K_\mu \hat U | 0\rangle =0.  
\ee
The compact generator $\hat E$ is just the number operator in the 
Fock space, and  the $SU(2)_L$ corresponds to the $a$ oscillators
while the $SU(2)_R$ corresponds to the $b$ oscillators.
Thus we work with the Fock space of $a, b$ oscillators and 
the transform $\hat U$ maps to gauge invariant operators at $x=0$. 

It is important to note that the
Fock space is constrained, i.e. the 
the number of $a^\dagger$ 
oscillators is equal to the number of $b^\dagger$ 
oscillators. This can be seen as follows: first, note that the 
Fock vacuum corresponds to a conformal primary because of
\eq{vaccond}, and the descendents are obtained from the 
successive actions of $\hat P_\mu$, which, according to \eq{fpropu},
corresponds to the creation operator $a^\dagger \sigma_\mu b^\dagger$.
Therefore, the states have equal number of $a^\dagger$ 
and $b^\dagger$. From the generators listed in \eq{fpropu} and 
\eq{lrsu2}, we see that the bilinear operators of the form
$a^\dagger a^\dagger$, $b^\dagger b^\dagger$, $a^\dagger b$, 
and $b^\dagger a$  are missing. This constraint can be formalized by 
saying  that the $U(1)$ given by
\be{u1z}
Z_1 = N_a -N_b,
\ee
is gauged. Therefore in the Fock space we allow only $U(1)_{Z_1}$ 
neutral states. Let us 
make a count of the generators allowed: 
$a^\dagger_{\dot\gamma} b^\dagger_{\gamma}$,
$b^{\gamma} a^{\dot\gamma}$, 
$a_{\dot\gamma}^\dagger a^{\dot\delta}$, 
$b^{\gamma} b_{\delta}^\dagger$,
these form $15 +1$ generators, which form the 
generators of the conformal group and the central $U(1)_{Z_1}$ given by
\eq{u1z} respectively.

\vspace{.5cm}
\noindent
{\emph{The $SO(6)$  $R$ symmetry group} }
\vspace{.5cm}

We construct  representations of the $SO(6)$ group using eight fermionic
oscillators.
These are organized as 4 complex fermionic oscillators which transform in the
spinor representation of $SO(6)$ as
\be{ferspi}
\varphi = 
\left(
\begin{array}{c}
\alpha_{ 1} \\
\alpha_{ 2} \\
- \beta^{\dot 1 \dagger} \\
-\beta^{\dot 2 \dagger}
\end{array}
\right).
\ee
Here the fermionic oscillators obey the following anti-commutation
relations
\be{anticom}
\{\alpha^{\tau\dagger} , \alpha_{\upsilon} \} =\delta^\tau_{\upsilon}, 
\;\;\;\;
\{\beta^{\dot\tau\dagger} , \beta_{\dot\upsilon}\} =
\delta^{\dot\tau}_{\dot\upsilon}, \;\;\;\;
\{\alpha_{\tau} , \beta_{\dot\tau} \} =0 , \;\;\;\;
\{\alpha^{\tau\dagger} , \beta_{\dot\tau}\} = 0.
\ee
The Fock vacuum is defined as 
\be{ffvac}
\alpha_{\tau}|0 \rangle = \beta_{\dot\tau} |0\rangle = 0.
\ee
The action of the $SO(6)$ generators in the Fock space 
of these oscillators is given by 
\be{fgenf}
\hat M_{IJ}  = \varphi^\dagger M_{IJ} \varphi, 
\ee
here $M_{IJ}$ are four dimensional 
representation of generators of 
$SO(6)$ (more the details of these are given in appendix A). 
These generators satisfy the $SO(6)$ algebra given by
\be{compso}
 [M_{IJ}, M_{KL}] = i(\delta_{JK}M_{IL} - \delta_{IK}M_{JL}
-\delta_{JL}M_{IK} + \delta_{IL}M_{JK}).
\ee
$I, J, K, K = 1, \ldots 6$. 
The fact that these  generators are Hermitian
ensures  Hermiticity of $\hat M_{IJ}$. Furthermore $\hat M_{IJ}$ 
obeys the $SO(6)$ algebra since
\be{commso}
[\hat M_{IJ}, \hat M_{KL} ] = \varphi^\dagger [ M_{IJ}, M_{KL} ]
\varphi.
\ee
It is again convenient to work  in a graded decomposition of
$SO(6)$ as  $L^+ \oplus L^- \oplus L^0$. We now write down the
generators below.
\bea{sogenful}
& &L^+ = \{ \; 
\alpha^{\tau\dagger} \beta^{\dot\tau\dagger} \;  \}, \qquad
L^- = \{ \; \alpha_{\tau} \beta_{\dot\tau} \; \} \\ \nonumber
& &L^0 = SU(2)_{L'} \times SU(2)_{R'} \times U(1)_J, \\ \nonumber
& &  SU(2)_{L'} :  \;\;\; 
\alpha^\dagger \vec \sigma\alpha,  \qquad
SU(2)_{R'}  : \;\;\;   \beta^\dagger \vec \sigma \beta, \\ \nonumber
& & U(1)_J:\quad  J =  \varphi^\dagger M_{56} \varphi 
= \frac{1}{2} \left( 2 - N_\alpha - N_\beta \right).
\eea
Here $N_\alpha$ and $N_\beta$ are defined by
\be{defnumbfer}
N_\alpha = \alpha^{1\dagger} \alpha_1 +
\alpha^{2\dagger}\alpha_2, \qquad
N_\beta = \beta^{\dot1\dagger} \beta_{\dot 1} +
\beta^{\dot 2\dagger}\beta_{\dot 2}. 
\ee
Similarly to the  case of $SO(2,4)$,  
the Fock space of oscillators for $SO(6)$ is constrained.
It is clear from the generators listed in \eq{sogenful} that
the states must have zero charge under the following $U(1)$
\be{fu1gau}
B = N_\alpha -N_\beta.
\ee

Starting from the Fock vacuum we can obtain all the allowed states by
the repeated action of the generators in $L^+$. Since these generators
are fermionic, the action of $L^+$ truncates at the second level.
The table of states are given below.

\begin{center}
\begin{tabular}{c | c | c | c}
$|$ State $\rangle$
& $J$ &  $E$ & $\hat U|$ State $\rangle$
\\
\vspace{-.4cm}
$\;$ & $\;$ & $\;$ $\;$ \\
\hline
\vspace{-.4cm}
$\;$ & $\;$ & $\;$ $\;$ \\
$|0\rangle$  & $1$ & $1$& $z$ \\
\vspace{-.4cm}
$\;$ & $\;$ & $\;$ $\;$ \\
\hline
\vspace{-.4cm}
$\;$ & $\;$ & $\;$ $\;$ \\
$\alpha^{\tau\dagger}
\beta^{\dot\tau\dagger} |0\rangle$ & $0$ & $1$ & $\phi_i$; $i
=1,\ldots 4$ \\
\vspace{-.4cm}
$\;$ & $\;$ & $\;$ $\;$ \\
\hline
\vspace{-.4cm}
$\;$ & $\;$ & $\;$ $\;$ \\
$\alpha^{1\dagger}\alpha^{2\dagger}\beta^{\dot 1\dagger}
\beta^{\dot 2\dagger}|0\rangle$
& $-1$ & $1$ & $\bar{z}$ \\
\vspace{-.4cm}
$\;$ & $\;$ & $\;$ $\;$ \\
\hline
\end{tabular}
\\
\vspace{.5cm}
{\bf\small{Table 1}}: The $SO(6)$ scalar
\end{center}
Here we have written down the $J$-charge and the dimension $\Delta$ of
the corresponding states. 
Note that the fermionic creation operators  have $J$-charge $-1/2$.
In the last column of the above table
we have indicated the field at the origin corresponding to the state
which is created by the action of $\hat U$. 
Thus these states give the scalars of ${\cal N}=4$ Yang-Mills theory, which 
transform in the vector of $SO(6)$. 

\vspace{.5cm}
\noindent
{\emph{The supersymmetry generators} }
\vspace{.5cm}

${\cal N} =4$  Yang-Mills admits $16$ Poincar\'{e} supersymmetry
generators and $16$ superconformal supersymmetries. Their realization 
in terms of the oscillator construction is given by

\begin{center}
\begin{tabular}{c| c|c|c}
Supersymmetry & Operator  & $E$ & $J$ 
\\
\vspace{-.4cm}
 $\;$ & $\;$ &$\;$  & $\;$ 
\\ \hline
\vspace{-.4cm}
$\;$ & $\;$ &$\;$  &$\;$   \\
 $Q^+$ & $a_{\dot\gamma}^\dagger 
\alpha_{\tau}$ & $\frac{1}{2}$ & $\frac{1}{2}$ \\
\vspace{-.4cm}
$\;$ &$\;$ $\;$  &$\;$  &$\;$   \\
\cline{2-4}
\vspace{-.4cm}
$\;$ &$\;$  &$\;$  & $\;$  \\
 & $b_{\gamma}^\dagger \beta_{\dot\tau}$ & 
$\frac{1}{2}$ & $\frac{1}{2}$ \\
\vspace{-.4cm}
$\;$ &$\;$  &$\;$  &$\;$   \\
\hline
\vspace{-.4cm}
$\;$ &$\;$  &$\;$  &$\;$   \\
$Q^-$ & 
$a_{\dot\gamma}^\dagger \beta^{\dot\tau\dagger}$ & 
$\frac{1}{2}$ & -$\frac{1}{2}$ \\
\vspace{-.4cm}
$\;$ &$\;$  &$\;$  &$\;$   \\
\cline{2-4}
\vspace{-.4cm}
$\;$ & $\;$ & $\;$ &$\;$   \\
 & $b_{\gamma}^\dagger \alpha^{\tau\dagger}$ & 
$\frac{1}{2}$ & -$\frac{1}{2}$ \\
\vspace{-.4cm}
$\;$ &$\;$  &$\;$  &$\;$   \\
\hline
$S^-$ & $a^{\dot\gamma}\alpha^{\tau\dagger}$ &
 -$\frac{1}{2}$ & -$\frac{1}{2}$ \\
\vspace{-.4cm}
$\;$ &$\;$  &$\;$  &$\;$   \\
\cline{2-4}
\vspace{-.4cm}
$\;$ &$\;$  &$\;$  &$\;$   \\
 & $b^{\gamma} \beta^{\dot\tau\dagger}$ &
 - $\frac{1}{2}$ & -$\frac{1}{2}$ \\
\vspace{-.4cm}
$\;$ &$\;$  &$\;$  &$\;$   \\
\hline
\vspace{-.4cm}
$\;$ &$\;$  &$\;$  &$\;$   \\
$S^+$& 
$a^{\dot\gamma}\beta_{\dot\tau}$ &
 -$\frac{1}{2}$ & $\frac{1}{2}$ \\
\vspace{-.4cm}
$\;$ & $\;$ &$\;$  &$\;$   \\
\cline{2-4}
\vspace{-.4cm}
$\;$ &$\;$  &$\;$  &$\;$   \\
 & $b^{\gamma} \alpha_{\tau}$ & 
-$\frac{1}{2}$ & $\frac{1}{2}$ \\
\vspace{-.4cm}
$\;$ &$\;$  &$\;$  &$\;$   \\
\hline
\end{tabular}
\\
\vspace{.5cm}
{\bf\small{Table 2}}: Supersymmetry generators
\end{center}

The algebra of these generators realize the odd part of the ${\cal N}=4$
superconformal algebra. The (anti-)commutation relations of these generators 
are given in Appendix C. 
From the supersymmetry generators given in the above table
it is clear that the complete
$U(1)$ gauging is given by
\be{actgaug}
Z = Z_1 + B = N_a - N_b + N_\alpha - N_\beta.
\ee
As an example, consider the generators $Q^-$: the  number of
$a$'s  is equal to the number of $\beta$'s, thus respecting 
$U(1)_Z$ invariance.

\subsection{Partition function}

In this section we will first evaluate the  partition
function and show the Fock space of the oscillators 
together with the $U(1)_Z$ gauging is equivalent to all 
the  on shell single letter spectrum  of ${\cal N}=4$ Yang-Mills 
We then construct composite single trace operators 
of given length $l$ by considering $l$ copies of the Fock space of
oscillators together with the cyclicity constraint.

\vspace{.5cm}
\noindent
{\emph{Single bit partition function} }
\vspace{.5cm}

The partition function for the Fock space is given by
\be{defpart}
{\cal Z} =  \oint_{|x| =1}  \frac{dx}{x} {\rm Tr} (
q^{2E} x^Z ).
\ee
Here the trace is over all the states in the Fock space, 
the contour integral around the unit circle imposes the 
constraint $Z=0$ on the states.  
It is simple to evaluate the trace as these oscillators are free,
we obtain
\bea{partfn}
{\cal Z} &=& \oint_{|x| =1} \frac{dx}{x} q^2 \frac{(1+x)^2 (1+ x^{-1})^2
}{ ( 1-qx) ^2 ( 1-qx^{-1} )^2}, \\ \nonumber
&=& 2 q^2 \frac{(3-q)}{(1-q)^3} , \\ \nonumber
&=&  6q ^2 + 16q^3 + 30q^4 + 48q^5 + 70q^6 + \cdots.
\eea
Note that this single letter partition function agrees with 
the on shell single letter spectrum of ${\cal N}=4$ YM at 
$\lambda =0$ evaluated in 
\cite{Sundborg:1999ue,Polyakov:2001af,Aharony:2003sx,Bianchi:2003wx,Beisert:2003te}

We now give an account of the numbers occurring for the first few terms
in the expansion of the partition function in powers of $q$. 
From \eq{partfn} we note there are $6$ states at $E=1$, these
are due to the 6  scalars $\phi^I$ in the fundamental 
of $SO(6)$.  At $E=3/2$ we see that there are $16$ states, these 
are accounted for by  the $16$ fermions of ${\cal N}=4$ Yang-Mills,
which can be thought of as the $16$ component Weyl-Majorana fermion
in ten dimensions, $\psi$. 
For $E= 2$, we have the $24$ states from 
$\partial_\mu\phi^I$ \footnote{At $\lambda =0$, the
covariant derivative reduces to the ordinary derivative.}  
and $6$ states from $F_{\mu\nu}$. Together
they give rise to the term $30q^4$ in the partition function. 
At the next level $E= 5/2$, we have the states of 
$\partial_\mu \psi$, 
they give rise to $4\times 16 =64$ states, but on examining 
\eq{partfn} we see that there are only $48$ states. This is because
the states of the Fock space are in correspondence with the on shell 
single letters of the Yang-Mills, indeed on subtracting the 
$16$ components of the equations of motion 
$\gamma\cdot \partial \psi =0$
from $64$ we obtain the $48$ on shell states. 
Let us consider one more level to illustrate the fact  the Fock space
corresponds to on shell states. 
At $E=3$ there are the following states
$\partial_\mu \partial_\nu \phi^I $ and 
$\partial_\mu F_{\nu\rho}$, their 
number are $60 + 24$ respectively. 
$6$ states out of the $60$ 
 vanish due to the on shell condition $\partial^2 \phi^I=0$.
$8$ states 
out the $24$,  correspond to $\partial^\mu F_{\mu \nu} =0$ and
Bianchi identity $\partial_{[\mu} F_{\nu\sigma] } =0$. 
Subtracting these on shell degrees we get $ (60 + 24) - (6 + 4 +4) =
70$, which agrees with the partition function.
Below we list the 
basic Fock space states and the corresponding fields.  
\vspace{.5cm}

\begin{center}
\begin{tabular}{c | c}
State & $\hat U |$State$\rangle$
\\
\hline
\vspace{-.4cm}
$\;$  & $\;$ \\ 
$|0\rangle,\;  \;\;
\alpha^{\tau\dagger}\beta^{\dot\tau\dagger} |0\rangle, \;\;\;
\alpha^{1\dagger} \alpha^{2\dagger}\beta^{\dot 1\dagger} 
\beta^{\dot 2\dagger}|0\rangle$  &  $\phi^I$ \\ 
\vspace{-.4cm}
$\;$  & $\;$ \\ 
\hline
\vspace{-.4cm}
$\;$  & $\;$ \\ 
$a_{\dot\gamma}^\dagger \beta^{\dot\tau\dagger} |0\rangle, \; \;\;
b_{\gamma}^\dagger \alpha^{\tau\dagger} |0\rangle, \; \;\;
a_{\dot\gamma}^\dagger \beta^{\dot 1^\dagger} 
\beta^{\dot 2\dagger} \alpha^{\tau \dagger}
|0\rangle, \; \;\;
b_{\gamma}^\dagger \alpha^{1\dagger} \alpha^{2\dagger} 
\beta^{\dot\tau\dagger} |0\rangle$  & $\psi$ \\ 
\vspace{-.4cm}
$\;$  & $\;$ \\ 
\hline
\vspace{-.4cm}
$\;$  & $\;$ \\ 
$a_{\dot\gamma}^\dagger a_{\dot\delta}^\dagger 
\beta^{\dot 1\dagger} \beta^{\dot 2\dagger}|0\rangle,
\;\;\;
b_{\gamma}^\dagger b_{\tau}^\dagger 
\alpha^{1\dagger} \alpha^{2\dagger}|0\rangle$
& $F_{\mu\nu}$ \\ 
$\;$  & $\;$  
\vspace{-.4cm}
\\
\hline
\end{tabular}
\\
\vspace{.5cm}
{\bf\small{Table 3}}: Basic letters of ${\cal N}=4$ YM and 
oscillator states.
\end{center}

We  now examine the reason why the Fock space  includes only the 
on shell letters of ${\cal N}=4$ Y-M.
Consider the operator $\partial^\mu \partial_\mu$. From \eq{fpropu}
we see that on the Fock space the operator is represented by
$a^\dagger\bar \sigma^\mu b^\dagger a^\dagger 
\bar\sigma_\mu b^\dagger$.
One can easily see, using the identity \cite{Wess:1992cp}
\be{fierzid}
\bar\sigma^{\mu\dot\gamma\gamma} \bar\sigma_{\mu}^{\dot\tau\tau}  = 
-2 \epsilon^{\dot\gamma\dot\tau} \epsilon^{\gamma\tau}, 
\ee
and the fact that the $\epsilon$-tensor  is anti-symmetric in 
its indices, that
the representation of $\partial^2$ on the Fock space vanishes.
Similarly, consider the equation of motion of the 
fermion $a_{\dot\gamma}^\dagger \beta^{\dot\tau\dagger} |0\rangle$.
This is given by  
\be{eqomfer}
\bar\sigma^{\mu\dot\gamma \gamma} P_\mu
a_{\dot\gamma}^\dagger \beta^{\dot\tau\dagger} |0\rangle =
\bar\sigma^{\mu\dot\gamma \gamma} 
(a^\dagger \bar\sigma_\mu b^\dagger) 
a_{\dot\gamma}^\dagger \beta^{\dot\tau\dagger} |0\rangle = 0,
\ee
\footnote{From now on
we use the same symbol for
$P_\mu$ and its conjugate $(a^\dagger \bar\sigma b^\dagger)$}.
where we have again applied \eq{fierzid}.
Using the same method and \eq{fierzid}
one can show that the equation of motion of the other 
fermions and 
$\partial^\mu F_{\mu\nu} =0$ is identically satisfied in the
oscillator construction.  To show that the Bianchi identity
$\partial_{[\mu } F_{\nu\rho]} =0$ is satisfied, 
consider the anti-self dual component 
$a_{\dot\gamma}^\dagger
a_{\dot\delta}^\dagger \beta^{\dot 1\dagger} \beta^{\dot 2\dagger}
|0\rangle$. 
The Bianchi identity can be written as
\be{biancid}
\epsilon^{\lambda\mu\nu\rho}(a^\dagger \sigma_{\mu}b^\dagger)
(a^\dagger\bar \sigma_{\nu\rho}a^\dagger)
\beta^{\dot 1\dagger} \beta^{\dot 2\dagger}
|0\rangle = 2i 
(a^\dagger \sigma_{\mu}b^\dagger) 
(a^\dagger\bar \sigma^{\lambda\mu}a^\dagger)|0\rangle
=0.
\ee
To show the above term vanishes, one again uses \eq{fierzid} after
expressing $\bar\sigma^{\lambda\mu}$ in terms of $\bar\sigma$
and $\sigma$. Similarly, one can show that the Bianchi 
identity is satisfied by the anti-self dual component of the 
field strength.  Using the fact that the basic letters are 
on shell it is easy to see that derivatives of these letters 
will also be on shell.

\vspace{.5cm}
\noindent
\emph {Multi-bit partition function} 
\vspace{.5cm}

Once we have obtained 
all the letters of ${\cal N}=4$ Yang-Mills, it is easy to 
construct the gauge invariant words using the oscillators. 
Consider a single trace gauge invariant operator of length
$l$. To construct the 
oscillator representation of this operator one examines 
the Fock space 
$\otimes_{s=1}^l {\cal H}_s$, where ${\cal H}$ refers
to the Fock space of a single copy of the oscillators 
$a, b, \alpha, \beta$. The state at site $s$ of the Fock space
is constructed  such that it 
corresponds to the operator at the site $s$ of the 
gauge invariant word. Then one sums over all cyclic permutations
so that the oscillator state is invariant under the 
cyclic shift of the sites. Each term in the sum is weighted by
the sign of the permutation, the sign of the permutation is
minus if the total number of exchanges involving a fermion is odd.
This sum over cyclic permutation is performed since
a single trace operator is invariant under 
this weighted cyclic permutation.
Symbolically the Hilbert space of single trace operator is
denoted by
\be{singhib}
{\cal H}^{(l)} = \sum_{\pi} {\rm sign}(\pi) \otimes_{s=1}^l {\cal H},
\ee
where $\pi$ refers to a cyclic permutation of length $l$ and 
${\rm sign} (\pi) =-1$ if the number of exchanges involving a 
fermion is odd, else ${\rm sign}(\pi) =1$. 
To illustrate this construction, we provide the following 
simple example of the operator ${\rm Tr}(\phi^i Z^{l-1})$,
\be{dicexpm}
\sum_\pi \prod_{s=1}^l \left(O(x)^{\pi_s}|0\rangle^{(s)} \right)
\leftrightarrow  \frac{1}{\sqrt{N^l}} {\rm Tr}( \phi^i  Z^{l-1}).
\ee
Here 
\footnote{For the remaining part of the paper we will 
drop the hat, $\hat{\;}$ in the Fock space representation of 
the generators of $SO(2,6)$.} 
\bea{mutbidefop}
O(x)^{(1)} = \frac{1}{\sqrt{2}} \exp( ix P) U (\alpha^\dagger
\sigma^i\beta^\dagger)|0\rangle, \cr 
O(x)^{(2)} = O(x)^{(3)} \ldots = 
\exp( ix P) U | 0\rangle.
\eea
It is clear from the method of construction of the single
trace operators, that, if one evaluates the 
multi-bit partition function, it will agree with that 
of the single trace operators. To show this explicity 
we will evaluate the multi-bit partition function.
In doing this, in order  
to keep the discussion 
simple  we will not subtract the contributions 
of states that can be written as ${\rm Tr} ({\rm Fermion})^2$,
which vanish by Fermi statistics. 
Let us define $g$ as the generator of the
cyclic group, {\it i.e.}
\begin{equation}
g |s_1>|s_2>...|s_n>=|s_2>...|s_n>|s_1>.
\end{equation}
Inserting $g^t$ into the partition function of $l$ bits we obtain
\be{multipart}
\oint \prod_{s=1}^l \frac{dx_s}{x_s} {\rm Tr}( 
q^{2 \sum_{s=1}^l E(s) } \prod x_s^{Z(s)}g^t )
= 
\left( {\cal Z}(q^{\frac{l}{(t,l)}} \right).
\ee
here $E(s), Z(s) $ refer to the operators $E, Z$ at site $s$,
and the integrations over $x_s$ implement the 
local $U(1)$ constraint.
$(l,t)$  denotes the largest common divisor of $t$ and $l$.
To project on to 
the cyclically invariant states 
we insert the 
projector $P=\frac{1}{l}(1+g+g^2+...+g^{l-1})$.
into the multi-bit partition function. Using \eq{multipart}
we obtain
\bea{fulmubi}
{\cal Z}_{{l}} &=&
\frac{1}{l}\sum_{s=1}^l \left( {\cal Z}(q_2^{\frac{l}{(s,l)}})
\right)^{(s,l)}, \cr
& =&\sum_{d|l} \frac{\varphi(d)}{l} {\cal
Z}(q^d)^\frac{l}{d}.
\eea
In the second line of the above equation we have re-arranged the 
summation over $s$ to the sum  over the divisors of $n$. 
Then, $\varphi(d)$ denotes the
number of $s$, such that $(s,l)=\frac{l}{d}$. Since
$(s,l)=\frac{n}{d}$, this implies 
$s=\frac{l}{d}t$ with $(t,d)=1$. On the
other hand we also have  $s \leq l$ then $t < d$ (unless $d=1$) so
$\varphi(d)$ is given by the number of co-primes with $d$ and
smaller than $d$, with $\varphi(1)=1$. 
But this is the definition of Euler's totient function. 
To compare with the partition function 
of all the single trace operators of ${\cal N}=4$ Yang-Mills we
sum over all the lengths $l$ from $2$ to $\infty$, we neglect 
the case of a single bit. This gives
\begin{equation}
{\cal Z}_{\rm{Single trace}} = \sum_{l=2}^{\infty}
\sum_{d|l}\frac{\varphi(d)}{l} {\cal Z}_\Delta(q^d)^{\frac{l}{d}}
= 21 q^4 +96 q^5+392 q^6 +1344 q^7 +...,
\end{equation}
Note that we have also included the contributions
of states of the type ${\rm Tr}( {\rm Fermion})^2$, which 
can be subtracted easily.
It is clear that  using the oscillators it is also easy to 
evaluate partition functions which carry information 
of $SO(6)$ or $SO(2,4)$ quantum numbers as done in 
\cite{Bianchi:2003wx,Beisert:2003te}.

\section{String overlap at $\lambda =0$}

In the previous section we have seen that  gauge invariant
operators of, say, length $l$, can be represented
as a state in the Hilbert space 
${\cal H}^{(l)}$.
We refer to such a state in the Hilbert space as a string. 
The world sheet of such a string is discrete 
and composed of $l$ bits.
In this section we show that the 
planar two point and three point functions
of gauge invariant operators at $\lambda =0$ can be reproduced just 
by geometric overlap rules of their corresponding states in the
Hilbert space.
We also write the geometric overlaps in terms of 2-string vertex and 
a 3-string vertex for the two point and 3-point functions
respectively.

\subsection{Single bit overlap}

To show that the inner product of two string states in the Hilbert
space ${\cal H}^{(l)}$ reduces to the two point function of the
corresponding gauge invariant operator we first study the overlap of 
single bit states. 
This is done using three methods: the first method is a direct
evaluation of the inner product using just the oscillator algebra; 
in the second approach we show that the inner product satisfies
conformal ward identities which, this enables us to determine the 
overlap; finally we use the geometric meaning of the 
operators involved in the overlap to evaluate it. 

\vspace{.5cm}
\noindent
\emph{Direct evaluation of inner product} 
\vspace{.5cm}

Consider the vacuum state at position $x$, which is given by
$e^{i xP }U|0\rangle$. We have shown in the
previous section that this corresponds to the field $z(x)$. The inner
product of this state with the another vacuum state at position $y$
is given by
\be{sinin}
I(x, y)= \langle 0| U^\dagger e^{-i (x -y)P } U |0\rangle. 
\ee
Using the definition of $U$ in \eq{simtrfo} and the formulae in
\eq{fpropu}
the above expression can be written as
\bea{simsin}
I(x,y)&=& \langle 0| U^2 \exp\left[i 
(a^\dagger\bar\sigma b^\dagger)\cdot(y-x)
\right] |0\rangle,
\\ \nonumber
{\rm with }\;\;\;\;
U^2 &= &\exp\left[  -\frac{\pi}{2} ( a^\dagger b^\dagger +  ba )\right]. 
\eea
To evaluate the above inner product we use the identity 
\bea{defut}
U_t^2 &=& \exp {t( a^\dagger b^\dagger + ba ) }, \\ \nonumber
&=& \exp\left(  a^\dagger b^\dagger  \tan t \right)  
\exp \left[-  ( a^\dagger a +
bb^\dagger) \ln\cos t \right] \exp \left(  b a \tan t \right). 
\eea
This identity is shown as follows: 
first differentiate both sides of the above
equation with respect to $t$. Then 
move all the factors to the extreme left. 
This, together with the condition $U_0^2 =1$, results in the 
above identity.
Substituting \eq{defut} in \eq{simsin} we obtain
\bea{fin2pt}
I(x,y) &=&
\lim_{t\rightarrow -\pi/2} \frac{1}{\cos^2 t} \frac{1}{ 1 -2
(x^0-y^0)\tan t  +  (x-y)^2 \tan^2 t }, \\ \nonumber
&=& \frac{1}{(x-y)^2}.
\eea
Here we have also used the following property of  squeezed states,
\be{idnsq}
\langle 0| \exp( \frac{1}{2} a\cdot M\cdot a ) 
\exp( \frac{1}{2} a^\dagger\cdot N \cdot
a^\dagger ) |0\rangle = \left[ {\rm Det} ( 1- M\cdot N)
\right]^{-1/2}, 
\ee
where $a_i$'s refer to $n$ oscillators, $M$ and $N$ are $n\times n$
matrices.
From \eq{fin2pt} we see that the overlap of a single bit corresponding
to the operator $z(x)$  is identical to the two point function of 
$\langle \bar{z}(x) z(y) \rangle$. 
It is worth noting that the transformation $U$ is not a
unitary transformation, in fact the norm of the state at a given 
position $x$ is infinite, where as the norm of the state in 
the Hilbert space ${\cal H}_1$ is finite.

\vspace{.5cm}
\noindent
\emph{Method of conformal Ward identity}
\vspace{.5cm}

Another method to show that 
the overlap of the vacuum state reduces to the
two point function of the scalar $z(x_1)$,  is to show that the 
overlap satisfies the conformal Ward identities. 
This method does not rely on the direct manipulation of the 
oscillators but on the properties of the generators of the conformal
group. Using the properties of the operator $U$ given in \eq{fpropu}, 
the two point function $I(x, y)$ can also be written as 
\be{alter2pt}
I(x, y) = 
\langle 0| \exp(  i(b\bar\sigma^\dagger  a)\cdot y ) U^\dagger U
\exp ( -i (a^\dagger\bar\sigma  b^\dagger \cdot x) |0 \rangle.
\ee
We first consider the inner product,
\be{wardi}
\langle 0| \exp( i (b\bar\sigma^\dagger a)\cdot y ) U^\dagger U
E \exp ( -i (a^\dagger \bar\sigma b^\dagger )\cdot x) |0 \rangle
= (x\partial^x + 1) I(x, y), \\ \nonumber
\ee
where we have used the relation 
\be{ecomreldif}
[E, \exp( - i (a^\dagger \bar\sigma b^\dagger) \cdot x )]
= x\partial^x \exp(- i (a^\dagger \bar\sigma b^\dagger) \cdot x),
\ee 
to write the insertion of $E$ as a differential 
operator on the overlap.
The shift of 1  in the operator is due to the normal ordering constant in $E$.
We now can use the identity \eq{usepropu},
$U^\dagger U E = -E U^\dagger U$ to move the operator $E$ to
the left. Thus we obtain
\be{ward2}
\langle 0|  \exp( i (b\bar\sigma^\dagger  a )\cdot y)E U^\dagger U
\exp ( -i a^\dagger \bar\sigma b^\dagger )\cdot x ) |0 \rangle
= - ( y\partial^y + 1) I(x, y), \\ \nonumber
\ee
here again we have converted the action of $E$ as a differential
operator on the bra state. 
Now comparing the equations \eq{wardi} and \eq{ward2} we 
obtain
\be{wardi2}
( x\partial^x + y\partial^y + 2) I(x,y) =0.
\ee
The above equation is the conformal Ward identity for 
a primary field of weight $1$.
Using a similar procedure we can show that the overlap 
also satisfies
translational invariance. 
Inserting the operator 
$a^\dagger\bar\sigma_\mu b^\dagger$ 
which
is conjugate to the momentum operator 
(see \eq{fpropu})  in \eq{alter2pt} we
obtain
\be{momcons}
\langle 0| \exp( i (b\bar\sigma^\dagger a)\cdot y ) 
U^\dagger U
(a^\dagger\bar\sigma_\mu b^\dagger) 
 \exp ( -i (a^\dagger \bar\sigma b^\dagger )\cdot x ) |0 \rangle
= i \partial_\mu^x I(x, y).  \\ \nonumber
\ee
Here we have converted insertion of momentum operator to a derivative.
Using 
$U^\dagger U(a^\dagger\bar\sigma_\mu b^\dagger ) 
= (b\sigma_\mu^\dagger a) U^\dagger U$, which can be obtained
from \eq{fpropu}, 
we can move momentum 
operator to the left and then again convert it to
a derivative, this results in
\be{mcon1}
\langle 0| \exp( i (b\bar\sigma^\dagger a)\cdot y ) 
(b \bar\sigma_\mu^\dagger a) U^\dagger U
 \exp ( -i (a^\dagger \bar\sigma  b^\dagger)\cdot x  ) |0 \rangle
= - i \partial^y_\mu I(x, y).
\ee
Again comparing \eq{momcons} and \eq{mcon1} 
we obtain the momentum conservation equation.
\be{momcon1}
(\partial_\mu^x + \partial_\mu^y) I(x,y) =0
\ee
Using \eq{momcon1} we can replace the derivative with respect to $y$
in \eq{wardi2} to get the equation
\be{finward}
\left[ (x-y) \partial_x  +2 \right] I(x, y) = 0, 
\ee
the solution of this equation is the required two point function
$I(x, y) = 1/(x-y)^2$.

\vspace{.5cm}
\noindent
{\emph{The geometric method} }
\vspace{.5cm}

It is instructive to illustrate yet another method to show 
that the overlap of the vacuum state is the 
two point function of the scalar
$z(x)$. This method relies on expressing the 
action of the oscillators as
differential operators, this method is convenient when one wants to 
obtain the two point functions or three points functions at one-loop.
Consider the overlap again
\be{geover}
I(x, y) = \langle 0| \exp ( i(b\bar\sigma^\dagger a) \cdot y)
 U^\dagger U |x\rangle,
\ee
here $|x\rangle$ refers to the state 
$\exp(-i (a^\dagger \bar\sigma b^\dagger)\cdot x) |0\rangle$.
 Now we convert each of the 
operators  in \eq{geover} from oscillators
to differential operators 
acting on the state $|x\rangle$ from right to left.
The action of $U^\dagger U $ on the state 
$|x\rangle$ can be written in terms of the differential operator
\bea{udiffx}
U^\dagger U |x\rangle &=& 
\exp\left( \frac{\pi}{2} ( P_0 -K_0 )\right) |x \rangle , \\ \nonumber
 &=&  
 \exp\left[ - \frac{i\pi}{2} \left( 
 \partial_{0}^x +  2 x_0 + 2 x_0 x\cdot
\partial^x - x^2 \partial_{0}^x   \right) \right] 
| x\rangle.
\eea
Here we have used the fact that
\bea{diffopi}
P_\mu |x\rangle &=& -(a^\dagger  \bar\sigma_\mu  b^\dagger)
\exp(-i (a^\dagger \bar\sigma b^\dagger) \cdot x  ) |0\rangle,
\cr
&=& 
- i\partial^x_\mu |x\rangle = P_\mu^{(x)} |x\rangle,
\\ \nonumber
K_\mu |x\rangle &=& (b\sigma_\mu  a ) 
\exp ( -i (a^\dagger \bar\sigma b^\dagger)\cdot x ) |0\rangle, 
\cr
&=& i(2 x_\mu + 
2 x_\mu x\cdot \partial^x -x ^2 \partial_\mu^x ) |x\rangle,
\cr
&=& K_\mu^{(x)} |x\rangle.
\eea
Now one can convert the operator 
$\exp ( i (b\bar\sigma^\dagger a)\cdot y)$ as a
differential operator acting on the state $|x\rangle$, we then obtain
\bea{udiffxf}
I(x,y) = \langle 0| 
 \exp\left( \frac{\pi}{2} P_0^{(x)} - K_0^{(x)} 
) \right)  
\exp ( \tilde y \cdot K^{(x)}) |x\rangle.
\eea
Here $\tilde y = (y^0 , -y^m)$, this is due to the fact the
$K^{(x)}_\mu $ corresponds to the operator $(b\sigma_\mu a)$, 
while we have the operator $(b\bar\sigma^\dagger_\mu a)$.
Note that in the process of converting the oscillators to differential
operators there is a reversal in the order of action.
The next step is to explictly perform 
the action of the operators on the
state $|x\rangle$. This gives
\bea{finacdif}
I(x,y) &=& \langle 0 |
\exp\left( \frac{\pi}{2} ( K_0^{(x)}  -P_0^{(x)}) \right) 
\frac{1}{1+ 2 \tilde y\cdot x + \tilde y^2 x^2 }
| x'\rangle, \cr
{\rm where} \quad
x^{'\mu}  &=& 
\frac{x^\mu + \tilde y^\mu}{ 1 + 2 \tilde y \cdot x + y^2 x^2 },
\eea
here we have used the action of finite 
special conformal transformation $K_\mu^{(x)}$, by an amount
$\tilde y$ on a scalar primary of weight $1$.
To perform the action of the operation
$\exp( \frac{\pi}{2} ( P_0^{(x)} - K_0^{(x)}) ) $
one notes that
it is the rotation $\exp(\pi M_{05})$, 
and it acts on the coordinates as \footnote{ $\exp(\pi M_{05})$
acts linearly on the coordinates $\eta^A$ introduced in 
\cite{Ferrara},
then one can restrict its action on the light cone to 
realize its transformation  on the coordinate 
$x^\mu = \eta^\mu/(\eta^5 + \eta^6)$.
}
\be{fiuuactio}
x^0 \rightarrow -\frac{x^0}{x^2} 
\qquad x^m \rightarrow \frac{x^m}{x^2},
\ee
performing this operation  on \eq{finacdif}, we obtain
\bea{geofinac}
I(x,y)
&=& \frac{x^2}{(x-y)^2} \frac{1}{x^2} \langle 0  |x''\rangle , 
\qquad
x^{''0}  = \frac{-x^0 +y^0} {(x-y)^2},\; \;\;
x^{''m} = \frac{x^m -y^m}{(x-y)^2}, 
\\ \nonumber
&=& \frac{1}{(x-y)^2}.
\eea
In the above equation we have used $\langle 0|x''\rangle=1$ which is 
easy to see from the definition of $|x''\rangle$.
Thus again we have obtained the required two point
function of the scalar $\langle \bar z(x) z(y)\rangle$.

\vspace{.5cm}
\noindent
{\emph{ Single bit overlap for $SO(6)$ scalars}}
\vspace{.5cm}

It is now easy to see that the two point functions 
of the $SO(6)$ scalars also are given by
inner product of the corresponding
bit states. This is because the other scalars are obtained by the
action of creation operators $\alpha^{\tau \dagger}
\beta^{\dot\tau\dagger}$ on
the vacuum state, these
oscillators commute with $U^2$ and the translation operators 
$\exp(ix P)$. 
The inner product factorizes as the inner product of the bosonic 
oscillators and the inner product of fermionic oscillators.
Thus the position dependence of the overlap is entirely governed
by the bosonic oscillators $\{a, b\}$.  
The overlap of the fermionic oscillators ensure that the
Kr\"{o}necker delta in the two point function 
$\langle\phi^I(x) \phi^J(y)\rangle = \delta_{IJ} /(x-y)^2$
is reproduced. 

\vspace{.5cm}
\noindent
{\emph{Single bit overlap for fermions}}
\vspace{.5cm}

Let us  now examine the two point function of the fermions. 
Consider the state $\exp (ix  P ) U a_{\dot\gamma}^\dagger
\beta^{\dot\tau\dagger}|0\rangle$ which can also be written as
$ U a_{\dot\gamma}^\dagger \beta^{\dot\tau\dagger}
\exp( -i (a^\dagger \bar\sigma b^\dagger)\cdot x )|0\rangle$. 
The overlap of this state  
with the same state at position $y$ and with
indices $\dot\delta, \dot\upsilon$  is given by
\bea{ferminn}
F(x,y) &=& \delta^{\dot\tau}_{\dot\upsilon}
\langle 0|  \exp( i (b \bar\sigma^\dagger  a)\cdot y ) 
a^{\dot \delta} U^\dagger U 
a_{\dot\gamma} ^\dagger 
\exp ( -i ( a^\dagger \bar\sigma b^\dagger) \cdot x) |0\rangle, 
\\ \nonumber
&=& 
\delta^{\dot\tau}_{\dot\upsilon}
\langle 0|  \exp( i (b\bar\sigma^\dagger a)\cdot y) 
 U^\dagger U 
a_{\dot\gamma}^\dagger b_{\delta}^\dagger 
\exp ( -i  (a^\dagger \bar\sigma  b^\dagger)\cdot x) |0\rangle.
\eea
The Kr\"{o}necker delta is due to the inner product of the
fermions, in the second line we have moved $a^{\dot\delta}$
to the right of $U^\dagger U$ using the following equation
\be{anusid}
a^{\dot\delta} U^\dagger U = U^\dagger U b^\dagger_{\delta} \, .
\ee 
The above identity can  be shown using the 
from the basic identity in \eq{defut} and then commuting
$a^{\dot\delta}$ to the right.  
Rewriting the combination $a_{\dot\gamma}^\dagger b_\delta^\dagger$
as a derivative we obtain
\bea{fermoverd}
F(x,y)
&=&
-\frac{i}{2} 
\delta^{\dot\tau}_{\dot\upsilon}
\sigma_{\delta\dot\gamma} \cdot \partial^{x}
\langle 0|  \exp( i (b\bar\sigma^\dagger a)\cdot y)   U^\dagger U 
 \exp ( -i  (a^\dagger \bar\sigma  b^\dagger)\cdot x)  |0\rangle,
\\ \nonumber
&=&
-\frac{i}{2} 
\delta^{\dot\tau}_{\dot\upsilon}
\sigma_{\delta\dot\gamma} 
\cdot \partial^x \frac{1}{(x-y)^2}.
\eea
The last line has the required correlation  function of 
fermions. It is clear using the above manipulation 
the overlap of all the 
oscillators corresponding to fermions in Table 3, reduces 
to the required two point function.

\vspace{.5cm}
\noindent
{\emph{Single bit overlap for Field strength}}
\vspace{.5cm}

Finally consider the  two point function of bits corresponding to 
the field strength $F_{\mu\nu}$. From table 3. we see that this state
is given by
\be{deffiels}
F_{\mu\nu}(x) |0\rangle = 
\exp(i x P ) U \left(
(a^\dagger \bar\sigma_{\mu\nu}a^\dagger)
\beta^{\dot 1\dagger}\beta^{\dot 2\dagger} | 0\rangle  +
(b^\dagger \sigma_{\mu\nu}b^\dagger)
\alpha^{1\dagger}\alpha^{2\dagger} \right)| 0\rangle, 
\ee
here we have written the field strength as the sum of  the 
self-dual and anti-self-dual components.  
The  overlap of this state at position $x$ and $y$ is given by
\bea{infiest}
G_{\mu\nu;\rho\sigma} (x, y) &=& 
\langle 0| F_{\mu\nu}(y) F_{\rho\sigma}(x) |0\rangle,
\cr
&=& - \langle y| 
U^\dagger U
(a^\dagger \bar\sigma_{\mu\nu} a^\dagger) 
(b^\dagger\sigma_{\rho\sigma} b^\dagger)
|x\rangle
- \left( (\mu, \nu)  \leftrightarrow (\rho, \sigma) \right).
\eea
In obtaining the above equation we have repeatedly used the identity
\eq{anusid} and 
$\bar\sigma^\dagger_{\mu\nu}  = - \sigma_{\mu\nu}$.
Now  rewriting the pairs 
$a_{\dot\gamma}^\dagger b_\gamma^\dagger$
as derivatives on the state $|x\rangle$, we obtain
\bea{maninfis}
G_{\mu\nu;\rho\sigma} (x, y) &=& 
\frac{1}{16} \bar\sigma_{\mu\nu}^{\dot\gamma\dot\delta}
\sigma_{\rho\sigma} ^{\gamma\delta}
\left( 
\sigma^{\lambda}_{\gamma\dot\gamma} \sigma^{\varrho}_{\delta\dot\delta}
+\sigma^{\varrho}_{\gamma\dot\gamma} \sigma^{\lambda}_{\delta\dot\delta}
+\sigma^{\lambda}_{\delta\dot\gamma} \sigma^{\varrho}_{\gamma\dot\delta}
+\sigma^{\varrho}_{\delta\dot\gamma} \sigma^{\lambda}_{\gamma\dot\delta}
\right) \partial_\lambda\partial_\varrho I(x,y)
\cr
&+& ( (\mu,\nu) \leftrightarrow (\rho\sigma) ).
\eea
We now 
substitute the following identity \cite{Wess:1992cp} in the above equation 
\be{wbind1}
\sigma^\mu_{\gamma\dot\gamma} \sigma_{\delta\dot\delta}^\nu
+ \sigma^\nu_{\gamma\dot\gamma} \sigma_{\delta\dot\delta}^\mu
= - \eta^{\mu\nu}\epsilon_{\gamma\delta}
\epsilon_{\dot\gamma\dot\delta}
+ 4
 \sigma^{\rho\mu}_{\alpha\beta}
\bar\sigma^{\rho\nu}_{\dot\alpha\dot\beta},
\ee
to obtain 
\be{1stidgover}
G_{\mu\nu;\rho\sigma} (x,y) = \frac{1}{2}
\left[ \rm{Tr} (\sigma^{\rho\sigma} \sigma^{\kappa\varrho} ) 
\rm{Tr} ( \bar\sigma^{\mu\nu} \bar\sigma^{\kappa \lambda} ) 
+ ((\mu,\nu) \leftrightarrow (\rho,\sigma))
\right] \partial_\lambda
\partial_\varrho I(x, y).
\ee
To further simplify the above equation we need the following identities
\cite{Wess:1992cp}.
\bea{wbind}
\rm{Tr} ( \sigma^{\rho\sigma} \sigma^{\kappa\varrho} ) 
&=&  \frac{1}{2} \left(
\eta^{\rho\varrho}\eta^{\sigma\kappa} - \eta^{\rho\kappa}\eta^{\sigma
\varrho}
- i \epsilon^{\rho\sigma\kappa\varrho} \right),
\\ \nonumber
\rm{Tr} ( \bar\sigma^{\rho\sigma} \bar\sigma^{\kappa\varrho} ) 
&=&  \frac{1}{2} \left(
\eta^{\rho\varrho}\eta^{\sigma\kappa} - \eta^{\rho\kappa}\eta^{\sigma
\varrho}
+ i \epsilon^{\rho\sigma\kappa\varrho} \right).
\eea
Substituting the above identities in the last line of \eq{maninfis} we
obtain
\bea{fing}
G_{\mu\nu;\rho\sigma} (x, y) &=& 
\frac{1}{2} \left[ 
( \eta^{\rho\varrho}\eta^{\mu\lambda}\eta^{\sigma\nu}
- \eta^{\rho\varrho}\eta^{\sigma\mu} \eta^{\nu\lambda}
- \eta^{\sigma\varrho}\eta^{\mu\lambda}\eta^{\rho\nu}
+ \eta^{\rho\mu}\eta^{\sigma\varrho}\eta^{\nu\lambda}
\right. 
\\ \nonumber
&\;& \qquad - \left. \frac{1}{2}
( \eta^{\rho\mu}\eta^{\sigma\nu} \eta^{\varrho\lambda} 
 - \eta^{\rho\nu}\eta^{\sigma\mu}\eta^{\varrho\lambda}
)
\right]
\partial_{\lambda}\partial_{\varrho} I(x,y)
\\ \nonumber
&=& 
\frac{1}{2} 
\left( \eta^{\rho\varrho}\eta^{\mu\lambda}\eta^{\sigma\nu}
- \eta^{\rho\varrho}\eta^{\sigma\mu} \eta^{\nu\lambda}
- \eta^{\sigma\varrho}\eta^{\mu\lambda}\eta^{\rho\nu}
+ \eta^{\rho\mu}\eta^{\sigma\varrho}\eta^{\nu\lambda}
\right) 
\partial_{\lambda}\partial_{\varrho} I(x,y). \nonumber
\eea
Notice that
the second term in the square brackets just implements the
tracelessness condition in the indices $\lambda,\varrho$. 
The second line in 
the above equation is written with the understanding that 
the on shell condition, that is $\Box I(x, y)=0$ implements the 
tracelessness condition in $\lambda, \varrho$. 
\eq{fing} is 
 required two point function 
for the gauge field strength.

\subsection{Two-point functions}

In the previous subsection we have shown that the inner product of
all single bit states reduces to the two point function of the
corresponding fields. 
A gauge invariant operator of length $l$ corresponds  to a state in
the Hilbert space ${\cal H}^{(l)}$. 
Inner product in this space is simply the inner product inherited
from the inner product of the single bit states. 
To be more explicit, 
consider a state 
made of the product of single bit operators 
$\hat O^{(1)} \hat O^{(2)}\cdots \hat O^{(l)}$,
then the corresponding state in the Hilbert space ${\cal
H}^{(l)}$  is given by 
\be{stathl}
\hat {\cal O}(x)|0\rangle =  
\sum_{\pi} {\rm sign}(\pi) \prod_{i =1}^{l} 
\left( \hat O^{( \pi_i)  }(x) |0\rangle^{(i)}  \right),
\ee
here the summation runs over all the cyclic permutations $\pi$ of
the set of integers $\{1,2,  \ldots l\}$. The state $|0\rangle^{(i)}$,
refers to the $i$-th bit vacuum, while the vacuum $|0\rangle $ on the
left hand side of the above equation just refers to the product of
these single bit vacua.
The operator $O^{(\pi_i)}$ stands
for the  $O^{\pi(i)}$-th operator,  but written in terms of the
oscillators of the the $i$-th Fock space. 
The ${\rm sign}(\pi)$ just refers to the total sign one obtains after
the cyclic permutations of the operator 
$\hat O^{(1)} \hat O^{(2)} \cdots \hat O^{(l)}$. 
If in the cyclic permutation two fermionic single bit operators
are exchanged, there is negative sign. 
Bosonic single bit operators are
exchanged without any contribution to the  sign.
The sum of cyclic permutations with the sign as the weight essentially
implements the cyclic symmetry of the trace in the string bit
language.
The inner product of two
such states is defined by the following.
\be{defmul2p}
\langle 0| \hat{\cal O}^\dagger (x) \hat {\cal O}(y)|0\rangle
= \frac{1}{l} \sum_{\pi\sigma} 
{\rm sign}(\pi)
{\rm sign}(\sigma) \prod_{i=1}^l {}^{(i)} \langle 0|
O^{\pi_i\dagger}(x) O^{\sigma_i}(y) |0\rangle^{(i)}.
\ee
The above formula defines the inner product in the Hilbert space
${\cal H}^{(l)}$ in terms of the overlap of the single bit Hilbert
space, note that we have normalized the inner product by $1/l$ to avoid
over counting of the identical overlaps.
We will denote this overlaps as the 2-string overlap.
This  overlap can also be written as a 2-string vertex 
which is a state in the tensor product of the 2-string 
Hilbert space, this is given by
\bea{2vertexs}
|V_2\rangle &=& \frac{1}{l}\exp 
\sum_{s=1}^l \left(
a^{\dagger(1)}(s) a^{\dagger(2)}(s) +
b^{\dagger(1)}(s) b^{\dagger(2)}(s)   \right.\cr
&\;& \qquad
\left. +
\alpha^{\dagger(1)}(s) \alpha^{\dagger(2)}(s) 
+
\beta^{\dagger(1)}(s) \beta^{\dagger(2)}(s) 
\right) |0\rangle_{(1)} \otimes|0\rangle_{(2)}.
\eea
Here $(1)$ and $(2)$ refers to the Hilbert space of the
two strings,  $s$ stands for the sites on the bit strings
and
\bea{defcontaa}
a^{\dagger(1)} a^{\dagger(2)} = 
a^{\dagger(1)}_{\dot 1}
a^{\dagger(2)}_{\dot 1} +
a^{\dagger(1)}_{\dot 2}
a^{\dagger(2)}_{\dot 2}, 
\cr
b^{ \dagger(1)} b^{  \dagger(2)} = 
b^{\dagger(1) }_{ 1}
b^{\dagger(2)}_{1}
+
b^{\dot 2\dagger(1)}_2
b^{\dot 2\dagger(2)}_2,
\cr
\alpha^{\dagger (1)} \alpha^{\dagger(2)} =
\alpha^{1 \dagger (1) }\alpha^{1 \dagger (2)} +
\alpha^{2 \dagger (1) }\alpha^{2 \dagger (2)}, 
\cr
\beta^{\dagger (1)} \beta^{\dagger(2)} =
\beta^{\dot 1 \dagger (1) }\beta^{\dot 1 \dagger (2)} +
\beta^{2 \dagger (1) }\beta^{\dot 2 \dagger (2)}. 
\eea
It is clear from the structure of the 2-string vertex, it 
implements the delta function overlap $\delta(X^{(1)} -X^{(2)})$ 
where $X$ refers to the string bit world sheet.
Using the the 2-sting vertex, the 2-string overlap of 
operators \eq{defmul2p} can be written as
\be{commul2p}
\langle 0|\hat{\cal O}^\dagger(x) 
\otimes \langle 0| \hat{\cal O}^\dagger(y) |V_2\rangle.
\ee
It is easy to see from the definition of the 
2-string vertex \eq{2vertexs}, the above formula reduces to 
\eq{defmul2p}. 

We now show that this definition of the overlap is identical
to the rules of planar Wick contractions one performs in 
evaluating the two point function in the gauge theory.
The comparison of this overlap with planar Wick contractions  
performed in obtaining the two point function of two single trace
operators is show in the figure \eq{2pfiov}. 
\FIGURE{
\label{2pfiov}
\centerline{\epsfxsize=16.truecm \epsfbox{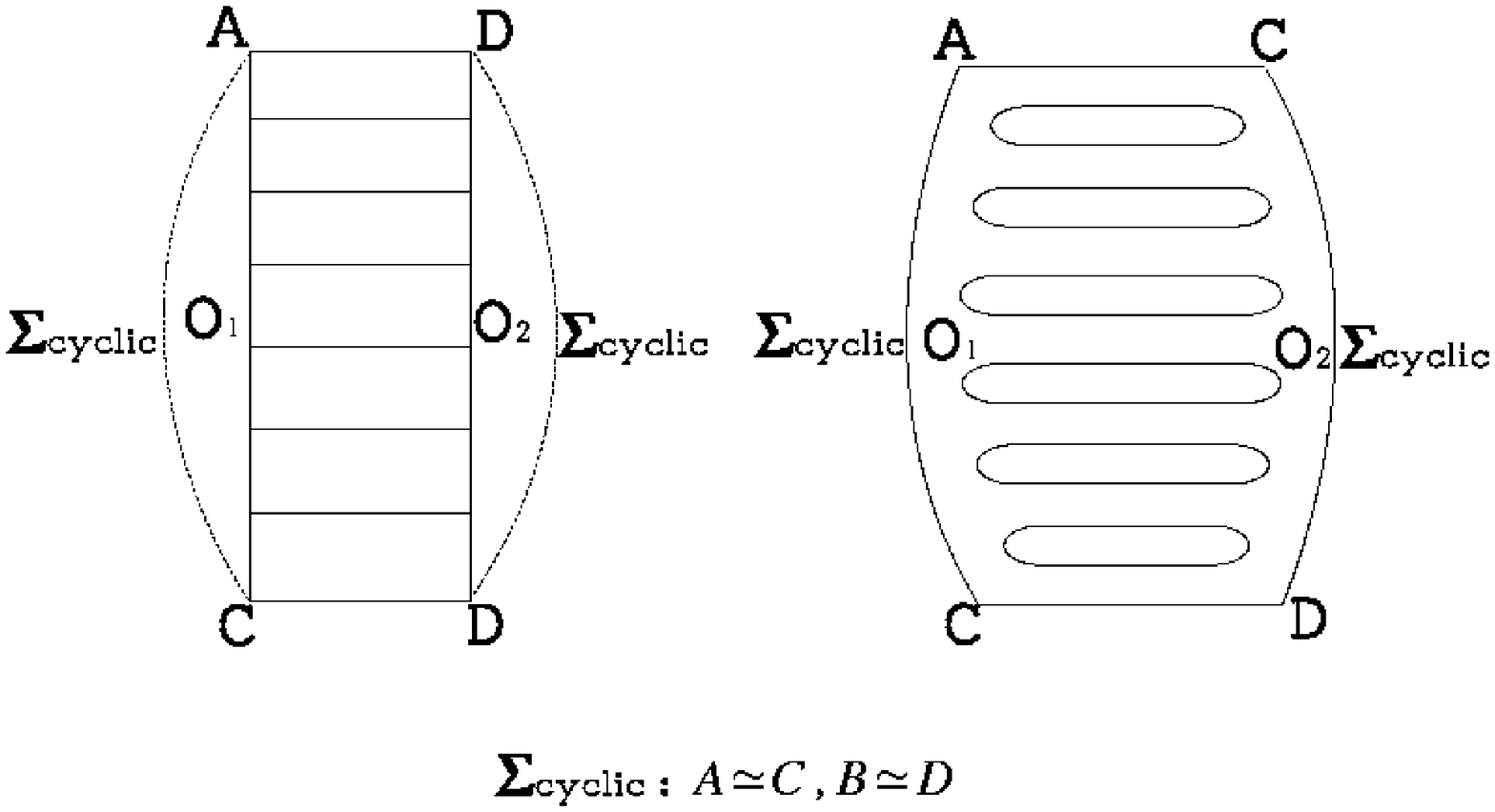}}
\caption{Bit overlap and two point function}
}
The horizontal lines in the bit string overlap denotes the single
bit inner product. In the previous subsection 
we have shown that single bit inner product
reduces to the two point function of the corresponding single letters
of the Yang-Mills. According to the definition in \eq{defmul2p} we
have to sum over the cyclic permutation of all the single bit states, 
the division by $l$ takes care of the over counting. In the evaluation
of the gauge theory correlator, one has to sum over the distinct
cyclic permutations in the trace. Thus the combinatorics involved in
the evaluation of the 2-string overlap is identical to the that of 
planar Wick contractions, the functional dependence is governed
by the product of the corresponding single bit overlaps. Therefore we
conclude that the 2-string overlap in \eq{defmul2p} reproduces 
the corresponding two point function of the gauge invariant operator.

\noindent
To illustrate the combinatorics involved we consider two simple
examples.

\vspace{.5cm}
\noindent
\emph{ Example (i)}
\vspace{.5cm}

Consider the most simple case with all the $\hat{O}^{(s)}$ being 
equal to the operator $  \exp(ix P)U$, that is  every one of the
single bit state is the vacuum state at position $x$. 
The gauge theory operator 
corresponding to this state is given by 
\be{mapgbi}
\sum_\pi \prod_{i=1}^l \left(O(x)^i|0\rangle^{(i)} \right)
\leftrightarrow  \frac{1}{\sqrt{N^l}} {\rm Tr}(Z^l)
\ee
The 2-string overlap of this state is given by
\be{2stovva}
\frac{1}{l} \sum_{\pi\sigma} \prod_{i=1}^l {}^{(i)}
\langle 0| U^\dagger \exp (-i P y) \exp (i P x) U |0\rangle^{(i)}
= \frac{l^2}{l}  \frac{1}{(x-y)^{2l} }.
\ee
Note that there is no sign involved in the permutation as all
operators are bosonic. Since all 
operators in each single bit Hilbert space is identical
the sum over $\pi$ and $\sigma$
permutations just give a factor of $l^2$, 
which cancels with the denominator to give the right power of $l$.
\eq{2stovva} is identical to the 
corresponding two point function of the operator in \eq{mapgbi}
A point to emphasize is that,  as in \eq{mapgbi},
we normalize all  gauge theory operators so that 
the planar two point functions do not have any $N$ dependence.

\vspace{.5cm}
\noindent
\emph{Example (ii)}
\vspace{.5cm}

Consider the  string state
corresponding to the operator 
$\frac{1}{\sqrt{N^l}} {\rm Tr}(\phi^i z^{l-1})$ given in 
\eq{dicexpm} and \eq{mutbidefop}.
Evaluating the 2-string overlap of this state we obtain
\bea{2ndopover}
\frac{1}{l}
 \sum_{\pi\sigma} 
 \prod_{i=1}^l 
 {}^{(i)} \langle 0|
O^{\pi_i\dagger}(x) O^{\sigma_i}(y) |0\rangle^{(i)}
&=& \frac{1}{l} \sum_{\pi} \prod_{i=1}^l 
 {}^{(i)} \langle 0|
O^{\sigma_i\dagger}(x) O^{\sigma_i}(y) |0\rangle^{(i)},
\\ \nonumber
&=& \frac{1}{(x-y)^{2l}}.
\eea
In the first line we have used the fact that the only when both
permutation $\sigma$ and $\pi$ are the same the overlap is non-zero.
Then the summation of the permutations gives a factor of $l$ which
cancels with factor of $l$ in the denominator. Finally, we have used
the single bit overlap for the scalars.
Thus the last line is the expected two point function for the
operator under consideration. 

\subsection{Three-point functions}

Here we construct the 3-string overlap which reproduces the
planar
3-point functions of the free theory.  Our discussion 
is divided into two parts, 3-point functions for which 
length of the operators is conserved and 3-point
functions for which the length is not conserved. 
The 3-vertex constructed for the
length conserving processes  has 
a natural interpretation of two strings joining into a third
string such that the length of the third string is 
the sum of the lengths of the first and the second. 
This is similar to the 3-vertex of light cone string field theory
of the critical string.
The length non-conserving process the 3-vertex 
can also be interpreted as string joining process, except the 
the first and the second string also overlap with each other.

\vspace{.5cm}
\noindent
{\emph{Length conserving process}}
\vspace{.5cm}

We first consider three point functions for which the length of one of
the operator say $\hat{\cal O}_3$ 
is the sum of the lengths of the remaining
two. We define the overlap of these states by the following,
\be{3ptin}
\frac{1}{N}\langle 0| 
\hat{\cal O}_1^\dagger(x_1) \otimes \langle 0| 
\hat{\cal O}_2^\dagger(x_2) 
\hat{\cal O}_3(x_3)|0\rangle,
\ee
where $\langle 0| \hat{\cal O}_1^\dagger(x_1)$ 
is a state in the Hilbert space
${\cal H}^{(l_1)}$, 
 $\langle 0| \hat{\cal O}_2^\dagger(x_2)$  
a state in the Hilbert space
${\cal H}^{(l_2)}$ 
and $  \hat{\cal O}^{3}(x_3)|0\rangle$ a state in the Hilbert space
${\cal H}^{(l_3)} $  and $l_1 + l_2 =l_3$.
$\otimes$ just refers to the tensor product of the states. The
inner product in \eq{3ptin} is the usual bit by bit inner product.
We have indicated this diagrammatically in figure \eq{f3ptov}. 
\FIGURE{
\label{f3ptov}
\centerline{\epsfxsize=16.truecm \epsfbox{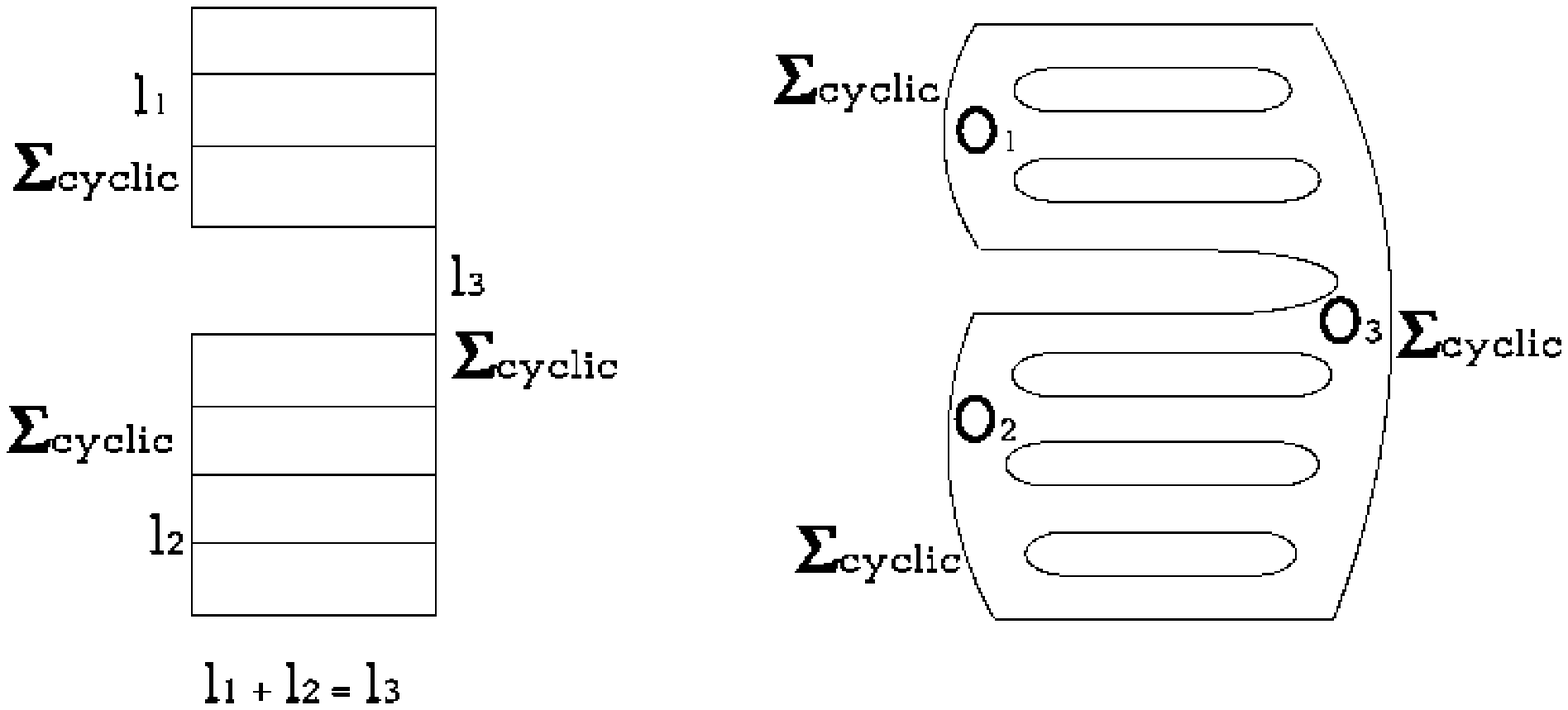}}
\caption{The three vertex and the 3-point function,
length conserving process.}
}
This overlap rule will be more familiar when written in terms of 
a state in the three string
Hilbert space. 
It is given by
\bea{3lcver}
|V_3 \rangle &=& \frac{1}{N}\exp \left[ \sum_{s=1}^{l^{(1)}}
(a^{\dagger (1)}(s) a^{\dagger(3)}(s) + 
b^{\dagger (1)}(s) b^{\dagger(3)}(s)  
\right. \cr
&\;& \qquad +
\alpha^{\dagger (1)}(s) \alpha^{\dagger(3)}(s) +
\beta^{\dagger(1)\dagger}(s) \beta^{\dagger(3)}(s))  
 \\
\nonumber & & 
\qquad +
 \sum_{s=l^{(1)}+1 }^{l^{(3)} }
(a^{\dagger(2) }(s) a^{\dagger(3)}(s) + 
b^{\dagger(2)}(s) b^{\dagger(3)}(s)  
\cr
&\;& \qquad +
\left.
\alpha^{\dagger(2)}_s \alpha^{\dagger(3)}_s +
\beta^{\dagger(2)}_s \beta^{\dagger(3)}_s  ) \right]
|0\rangle_{(1)}\otimes |0\rangle_{(2)}
\otimes \rangle_{(3)}. 
\eea 
Here the superscripts over the oscillators just indicate the 
Hilbert space they belong to and $s$
indicates the bit label. 
It is clear from the discussion of the 2-vertex, this vertex basically
implements the delta function overlap  $\delta(X^{(1)} +
X^{(2)} -X^{(3)} )$ where
the $X$'s indicate the world sheet coordinate of the bit string. 
In the above vertex, the bits from $1$ to $l^{(1)}$ of the 1st string 
overlaps the corresponding bits of the third string and the $l^{(2)}$ 
bits
of the second string labeled from $l^{(1)}+1 $ to $l^{(3)}$ 
overlaps the
corresponding bits on the third string. 
Note that as expected we have normalized the three vertex by
$1/N$ as it is the closed string coupling.
It is also easy to see from the discussion of the previous subsection
we have
\be{innovco}
\frac{1}{N}\langle 0|
\hat{\cal O}_1^\dagger(x_1) \otimes 
\langle 0| \hat{\cal O}_2^\dagger(x_2) 
\hat{\cal O}_3(x_3)^\dagger|O\rangle
= \langle 0| 
\hat {\cal O}_1^\dagger(x_1) 
\otimes \langle 0| \hat{\cal O}_2^\dagger(x_3) 
\langle 0| \hat{\cal O}_3^\dagger (x_3) |V_3\rangle.
\ee

From the figure \eq{f3ptov}, 
comparison of this overlap rules to
Wick contractions involved in obtaining the 
corresponding free and planar  Yang-Mills three point function
indicate that both the functional dependence of the correlation
function as well as the structure constants will be reproduced by the
3-string overlap. The overlaps rules just mimic planar Wick
contractions. 
The 
conformal Ward identities satisfied by the three-point functions 
are guaranteed to be satisfied, due to the fact that 
the single bit overlap satisfies the Ward identities, 
this will ensure the right functional dependence. 
In the appendix D we evaluate 
the 3-string overlap for a class of scalars and show that both the
functional dependence as well as the structure constants 
is in agreement with the gauge theory results.

\vspace{.5cm}
\noindent
{\emph{Length non-conserving process}}
\vspace{.5cm}

Consider  three operators $\hat{\cal O}_i(x_i)$ with lengths 
$l^{(3)}\geq l^{(2)}\geq
l^{(1)}$.  It is easy to see that for free Wick contractions such  operators 
will have nontrivial three point functions only if $2l = 
l^{(1)} +l^{(2)} - l^{(3)}$  and $l$ 
is an integer. The overlap rules for the states
corresponding to these operators are most clearly seen in the first
diagram of figure \eq{f3ptovn}. 
\FIGURE{
\label{f3ptovn}
\centerline{\epsfxsize=16.truecm \epsfbox{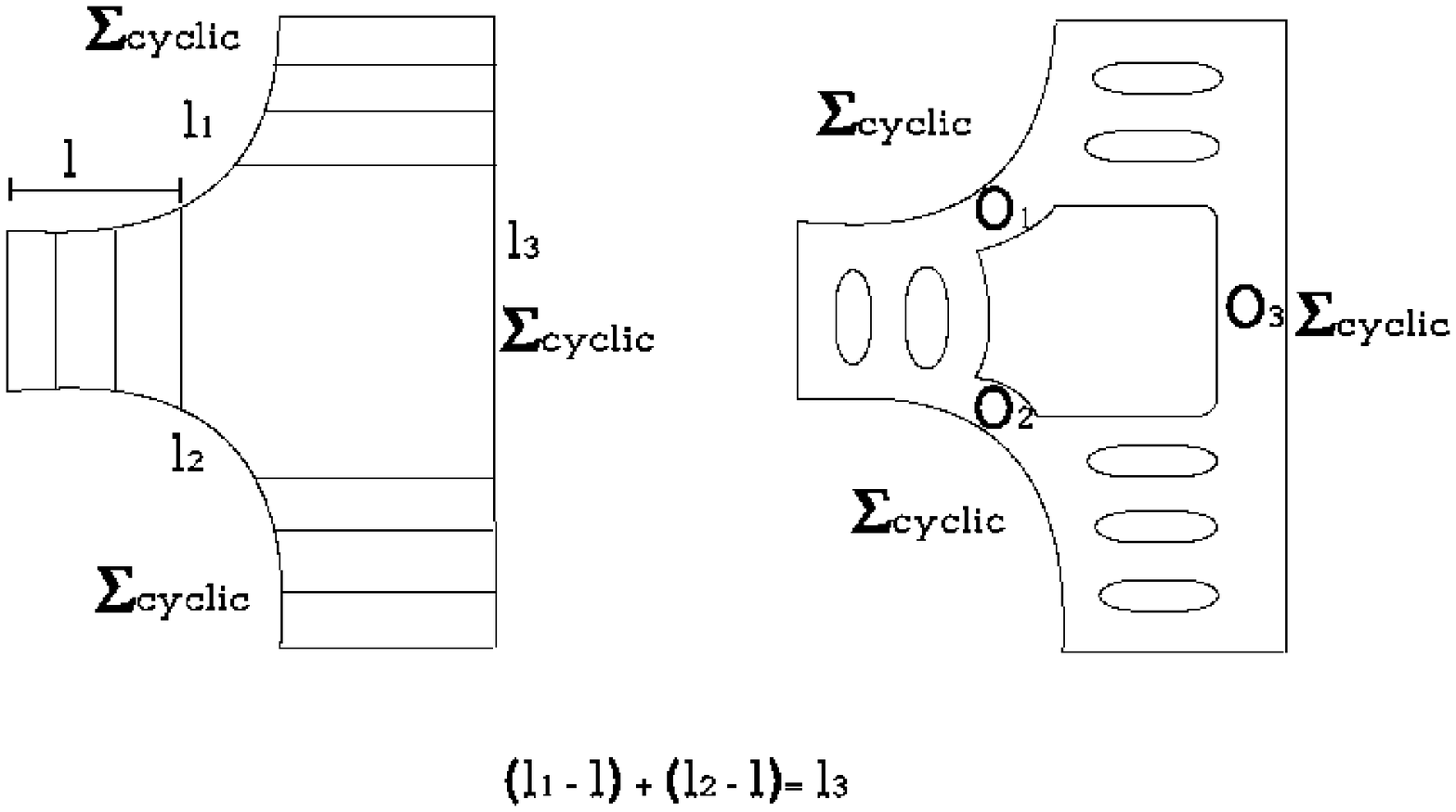}}
\caption{The three vertex and the 3-point function, length non-conserving
process.}
}
We overlap $l$ bits of the states $\hat{\cal O}(x_1)|0\rangle$ and
$\hat{\cal O} ( x_2)|0\rangle$ using the usual bit inner product. Then 
overlap the remaining $l^{(1)}-l$ bits of $\hat{\cal O}(x_1) |0\rangle$, and 
$l^{(2)}-l$ bits of $\hat{\cal O}(x_2)$ with that of the operator 
$\hat{\cal O}(x_3)$ as shown in
the figure. It is also clear from  figure \eq{f3ptovn}
that this rule mimics planar
Wick contractions of the corresponding operators in the gauge theory. 
Furthermore, the sum over the cyclic permutations of the bits 
corresponds to the sum over all the distinct planar Wick contractions
possible due to the cyclic property of the trace.
Again since the single bit overlap satisfies the conformal 
Ward identities 
the 3-string overlap constructed out of single bit overlaps will
satisfy the respective conformal Ward identities.
It is possible to write  the overlap rule for the length
non-conserving process as the following state in the three Hilbert
space
\bea{ncv3} 
|V_3\rangle &=& \exp\left[ N^{(13)} + N^{(23)}
+ N^{(12)} \right]|0\rangle_1\otimes
|0\rangle_2\otimes|0\rangle_3\frac{1}{N}, 
\\ \nonumber N^{(13)} &=& \sum_{s= 1}^{l^{(1)}
 -l } 
a^{(1)\dagger}(s) a^{(3)\dagger}(s) + 
b^{(1)\dagger}(s) b^{(3)\dagger}(s) +
\alpha^{(1)\dagger}(s) \alpha^{(3)\dagger}(s) + 
\beta^{(1)\dagger}(s) \beta^{(3)\dagger}(s) , 
\\ \nonumber N^{(23)} &=& 
\sum_{s=l^{(1)} -l+1}^{l^{(3)}   } \left(
a^{(2)\dagger}(s+l) a^{(3)\dagger}(s) + b^{(2)\dagger}(s+l)
b^{(3)\dagger}(s) \right.
\\ \nonumber 
& &  \quad  +\left. 
\alpha^{(2)\dagger}(s+l) \alpha^{(3)\dagger}(s) + 
\beta^{(2)\dagger}(s+l) \beta^{(3)\dagger}(s) \right),  
\\ \nonumber 
N^{(12)} &=& \sum_{s=l^{(1)}
-l+1}^{l^{(1)}   } \left( 
a^{(1)\dagger}(2l^{(1)} +1 -s-l) a^{(2)\dagger}(s) + 
b^{(1)\dagger}(2l^{(1)} +1 -s-l)
b^{(2)\dagger}(s)  \right. 
\\ \nonumber
& & \quad +\left.
\alpha^{(1)\dagger}(2l^{(1)} +1 -s-l)
\alpha^{(2)\dagger}(s) + \beta^{(1)\dagger} (2l^{(1)} +1 -s-l)
\beta^{(2)\dagger}(s)
\right). 
\eea 
This 3-vertex implements the overlap rule whereby $l$ bits of  the 
states $\hat{\cal O}_1(x_1) |0\rangle$ and 
$\hat{\cal O}(x_2)|0\rangle$ overlaps with each
other and the remaining $l^{(1)}-l$ and $l^{(2)}-l$ bits of these states
overlaps with the state $O_3(x_3)|0\rangle$. 
Note that in the 3-vertex \eq{ncv3}, the bits of the second string
are labeled from $l^{(1)} -l +1$ to $l^{(3)} +l$, thus there 
are $l^{(2)} = l^{(3)} - l^{(1)} + 2l$ bits. 
As in the discussion of the length conserving process, the three point
function of the states $\hat{\cal O}_1(x_1)|0\rangle ,
 \hat{\cal O}_2(x_2) |0\rangle,
\hat{\cal O}_3(x_3)|0\rangle$ is given by
\be{fulform3v}
\langle 0|\hat{\cal O}\dagger_1(x_1)
\otimes \langle 0|\hat{\cal O}^\dagger_2(x_2)
\langle 0|\hat{\cal O}^\dagger_3(x_3) |V_3\rangle.
\ee
In the appendix D we evaluate the three-point function for a length
non-conserving process and show that both the functional dependence as
well as the structure constants is in agreement with the gauge theory
results.

\section{String overlap at one-loop in $\lambda$ }

From the previous section 
we see that the two-point functions and the 
three-point functions at $\lambda =0$ were essentially determined by
the following: 
i) The position dependence of the correlators was governed by the 
equation $U^\dagger U E = -EU^\dagger U$. 
This equation guaranteed conformal Ward identities which are
required to be satisfied for the correlation functions.
ii) The combinatorics involved in the planar Wick contractions was
reproduced in the bit picture by the delta function overlap rule. This 
guaranteed that the structure constants in the bit picture was
identical to that of the gauge theory.
In this section, we derive the modifications to the above picture when
on rendering $\lambda$ finite. From the AdS/CFT dictionary \eq{ymmap},
we see that rendering $\alpha'$ finite would introduce 
interactions between bits. At first order in $\lambda$ and in the 
planar limit, only the nearest neighbour bits would interact. 
Therefore, turning on $\lambda$ modifies the free propagation 
of bits in the bit string theory.  The nearest neighbour interactions
should be such that it reproduces the logarithmic corrections 
as well as the finite corrections to structure constants of the 
three-point functions.
We now outline the strategy by which we will introduce these
nearest neighbour interactions between the bits so as to reproduce 
the logarithmic divergences and the structure constants at one loop
in $\lambda$.

\vspace{.5cm}
\noindent
{\emph{Logarithmic corrections}}
\vspace{.5cm}

From the discussion in section 3.1 of the single bit overlap
and, in particular, from  the method of conformal Ward identities, we see
that the position dependence  essentially arose from to the identity
$U^\dagger U E = -EU^\dagger U$. Examining the quantities involved in
the identity in some detail will provide the clue of how to reproduce
the logarithms. $E$ is the operator conjugate to the dilatation
operator \eq{fpropu}. At $\lambda =0$, 
for a bit string of length $l$ we see that  
$E$ is the sum of the conformal dimensions of all the letters
composing the string, therefore from 
the discrete world sheet point of view  it is a global charge on the 
world sheet. As we have argued from \eq{ymmap}, 
we see that turning on $\lambda$ renders $\alpha'$
finite, which introduces interaction between the
bits at one loop. 
At the planar level, 
this interaction just involves  only nearest neighbours at
one-loop in $\lambda$.
Therefore,  the global charge $E$ gets corrected to 
$E + \lambda \sum_{s =1}^l  H_{s, s+1}$, where $H_{s, s+1}$ is the 
anomalous dimension Hamiltonian which contains the information of the 
interactions. Similarly, 
$U^\dagger U = \exp ( \frac{\pi}{2} ( P_0-K_0))$, where $P_0$ and
$K_0$ are global charges on the discrete world sheet. Again at one
loop in $\lambda$ the charges $P_0-K_0$ gets corrected to 
$P_0 - K_0 + \lambda \sum_{s=1}^l \delta K_{0 s, s+1 }$. Here we have
chosen a scheme in which the global Lorentz generators are not 
corrected and the interactions are present only in the corrections of
the special conformal generators. 
We will  show  that it is possible to choose  such a scheme. 
The corrections 
$H$ and $\delta K_\mu$ are such that they have to respect 
global $SO(2,4)$ invariance to one-loop in $\lambda$. 
This  will  ensure that the identity 
$U^\dagger U E = -EU^\dagger U$ is  satisfied  to one-loop in
$\lambda$, which in turn guarantees conformal Ward identities 
at one-loop and thus reproduces the logarithmic divergences. We will 
explicity implement this strategy in this section to obtain the 
logarithmic corrections in both the two-point function and the 
three-point functions in the bit string picture.

\vspace{.5cm}
\noindent
{\emph{Structure constants}}
\vspace{.5cm}

To indicate how to incorporate the corrections to 
structure constants, let us recall how they arise in the gauge theory.
Consider a set of conformal 
primary operators $O_i$ \footnote{We will work with scalars and 
choose a basis such that 
their anomalous dimension matrix is diagonal for simplicity.}.
By conformal invariance, the general form for the two-point
functions of these operators at one-loop in $\lambda$ and 
at large $N$ is given by
\be{def2ptgt}
\langle O_i (x_1) O_j(x_2) \rangle = 
\frac{1}{(x_1-x_2)^{2\Delta_i}} \left(
\delta_{ij} + \lambda g_{ij} - \lambda\gamma_i \delta_{ij}
\ln((x_1-x_2)^2 \Lambda^2 )\right).
\ee
Note that at one-loop one also obtains a finite constant 
mixing matrix $g_{ij}$ proportional to $\lambda$.  
Though this matrix in scheme dependent, it contributes  
through a scheme independent combination to the 
one-loop corrections to the structure constants.
We now detail the precise combination by which it contributes 
to the one loop corrections to the structure constants.
The three point function of these operators at one loop is given by
\bea{3ptscop}
\langle O_i(x_1) O_j(x_2) O_k(x_3) \rangle
&=& \frac{1}{
|x_{12}|^{\Delta_i + \Delta_j - \Delta_k}
|x_{12}|^{\Delta_i + \Delta_k - \Delta_j}
|x_{23}|^{\Delta_j + \Delta_k - \Delta_i} }
\times  \\ \nonumber
&\; &  \left( C^{(0)}_{ijk} \left[
1 
- \lambda\gamma_i \ln | \frac{x_{12}x_{13} \Lambda}{x_{23}} |
- \lambda\gamma_j \ln | \frac{x_{12}x_{23} \Lambda}{x_{13}} | \right.
\right.
\cr
&\;& \left. \left.
- \lambda\gamma_i \ln | \frac{x_{13}x_{23} \Lambda}{x_{12}} | \right]
+ \lambda \tilde C^{(1)}_{ijk}  \right).
\eea
Here again the finite constant $\tilde C^{(1)}_{ijk}$ is not
renormalization scheme independent, but the following combination is the
scheme independent correction to the structure constant
\be{schindep}
C_{ijk}^{(1)} = \tilde C_{ijk}^{(1)} - \frac{1}{2} g_{ii'}
C_{ijk}^{(0)} - \frac{1}{2} g_{jj'} C^{(0)}_{ij'k} 
- \frac{1}{2} g_{kk'} C^{(0)}_{ijk'}.
\ee
A more detailed discussion of the one loop corrections
to structure constants is given in \cite{Alday:2005nd}.

We will now argue the that the metric corrections $g_{ij}$ 
can be understood as the changes in the inner product of
the states built with oscillators $a, b, \alpha, \beta$ .
A convenient way to think of the metric correction $g_{ij}$ in the
two-point function is that it is a correction to the scalar product
of the corresponding states. It can be
isolated from the logarithmic 
corrections  by evaluating the two-point function 
 with the operator $O_j$ at $x_2=0$ and the operator 
$f \circ O_i(0)$, where $f$ refers to the 
inversion $x^\mu \rightarrow -x^\mu/x^2$. 
Evaluating the two-point function we then obtain
\bea{new2pf}
\langle f\circ O_i (0) O_j(0)  \rangle
&=& \lim_{x\rightarrow 0} \frac{1}{x^{2\Delta_i + \lambda\gamma_i }}
x^{2\Delta_i + \lambda\gamma_i} \left( \delta_{ij} + 
\lambda g_{ij} \right), \\
\nonumber
&=& \delta_{ij} + \lambda g_{ij}.
\eea
Note that here we have performed the inversion also at one-loop in
$\lambda$ to cancel the logarithmic corrections. We will refer to this
two-point functions as to the norm.
In the bit language performing the inversion on one of the states is 
equivalent 
to acting on it with the operator $U^2= \exp (\pi M_{05})$, which
corresponds to the inversion \eq{fiuuactio}. 
Then the norm of the operators $O_i$ and $O_j$ is given by 
\bea{normbi}
\langle 0|O^\dagger_i (0) U^2 O_j (0) |0\rangle  &=&
\langle 0|s_i ^\dagger U^\dagger U^2 U s_j|0\rangle, \\ \nonumber
&=& \langle 0|s_i^\dagger s_j  |0\rangle,
\eea
where $O_j(0)$ refers to the state in the string bit Hilbert space. 
To obtain the first line of the above equation we use the definition that 
a state in the string bit Hilbert space is a obtained by the action of 
$U$ on the fock space of oscillators. Here we have denoted this state 
by $s_i, s_j$. For instance the $s_i$'s for the basic
letters of ${\cal N}=4$ YM is given in the first column
of table 3.
To obtain the last line we have used the fact that
$U^\dagger = U $ and $U^4=1$.  Thus the norm of two operators is just
the inner product in the Fock space of the oscillators. This fact is 
more obvious when one considers only the scalar $SO(6)$ sector, since
the in this sector the states are created only by the fermionic 
$\alpha, \beta$ oscillators and 
the norm is entirely governed by the inner product of these fermionic
oscillators.

We need a strategy to incorporate the change in the norm at one loop
in $\lambda$. 
At first sight it would be hard to imagine one can 
consistently deform the inner product of the oscillators 
to incorporate the nearest neighbour interactions. This deformation 
also should respect the global $PSU(2,2|4)$ symmetry of the theory.
In  critical string theory compactified on a circle 
in the direction $9$ say, the norm of the
oscillators in the direction of the circle is given by
$\langle \alpha_1^9 \alpha_{-1}^9 \rangle =G^{99}$. 
Here the change in the norm of the oscillator $\alpha_9$ is due
the change in the corresponding kinetic term.
This 
provides us the clue to incorporate the change in the norm.
We must first write down a world-sheet Lagrangian 
for the oscillators $a, b, \alpha,\beta$ 
with nearest neighbour interactions 
proportional to $\lambda$ such that the quantization of the 
Lagrangian gives the usual inner product along with a correction 
proportional to $\lambda$. This will provide us a consistent
deformation of the norm. 
A natural choice for the world sheet Hamiltonian is  
the operator $E$ \footnote{One can possibly assume that there exists a
a gauge in which the world sheet Hamiltonian is the conformal
dimension of operators just as in the case of
the plane wave limit where $\Delta -J$ is the light 
cone Hamiltonian.}, 
the world sheet Lagrangian for the 
$\lambda =0$ case is given by 
\be{wsacti}
{\cal S} = \sum_s^l \int dt \frac{i}{2} \left( 
\bar\psi_s\partial_t \psi_s -  \partial_t \bar\psi_s \psi_s
\right) +  \frac{1}{2}\psi^\dagger_s \psi_s + 
\frac{i}{2} \left( \varphi^\dagger_ s\partial_t \varphi_s  
- \partial_t \varphi^\dagger_s \varphi_s \right).
\ee
Here $\psi$ and $\varphi$ refer to the $SO(2,4)$ spinor and $SO(6)$
spinor defined in \eq{defbspin} and \eq{ferspi} respectively.
Note that the potential term $\psi^\dagger\psi$ breaks 
global $SO(2,4)$ invariance \footnote{$\bar\psi\psi$ is $SO(2,4)$
invariant},  
it is clear that this is necessary 
as the Hamiltonian is one of the generators of $SO(2,4)$. Performing
canonical quantization on this action we obtain the world sheet
Hamiltonian 
\be{wsham}
{\cal H} =  \frac{1}{2}\sum_s^l \psi^\dagger_s \psi_s = 
\frac{1}{2}\sum_s ( a_s^\dagger a_s + b_s b_s^\dagger) = E
\ee
The commutation relations of the oscillators are the ones given in
\eq{bcomm} and \eq{anticom}.  
As we have discussed earlier for the case of 
critical string theory compactified on the
circle, a world-sheet deformation  which modifies the commutation
relations 
of the oscillators $\{a, b, \alpha, \beta\}$ 
must be proportional to the kinetic energy term in the
Lagrangian \eq{wsacti}. 
In  section 4.2,  for the $SU(2)$ subsector, we will introduce 
such a deformation in the Lagrangian which modifies the 
commutation relations of the oscillators to include interactions
with nearest neighbour oscillators. We show that such a deformation 
is unique upto a proportionality constant and it preserves the
global $SU(2)$ symmetry.
We then use the 
approach developed by  \cite{Kugo:1992md} to construct the deformed 
two-string vertex and the three-string vertex.  Using these new
vertices we show that the structure constants 
obtained from the bit picture in the $SU(2)$ sector
agree with those of the gauge theory 
evaluated in \cite{Alday:2005nd}, up to a proportionality constant.

\subsection{Logarithmic corrections at one loop}

\vspace{.5cm}
\noindent
{\emph{Two point functions}}
\vspace{.5cm}

As we discussed earlier, 
at first order in $\lambda$ we have nearest neighbour
interactions between the bits. We have indicated this correction to
the string bit overlap for the two-point functions schematically in
figure \eq{2plogc}.
\FIGURE{
\label{2plogc}
\centerline{\epsfxsize=10.truecm \epsfbox{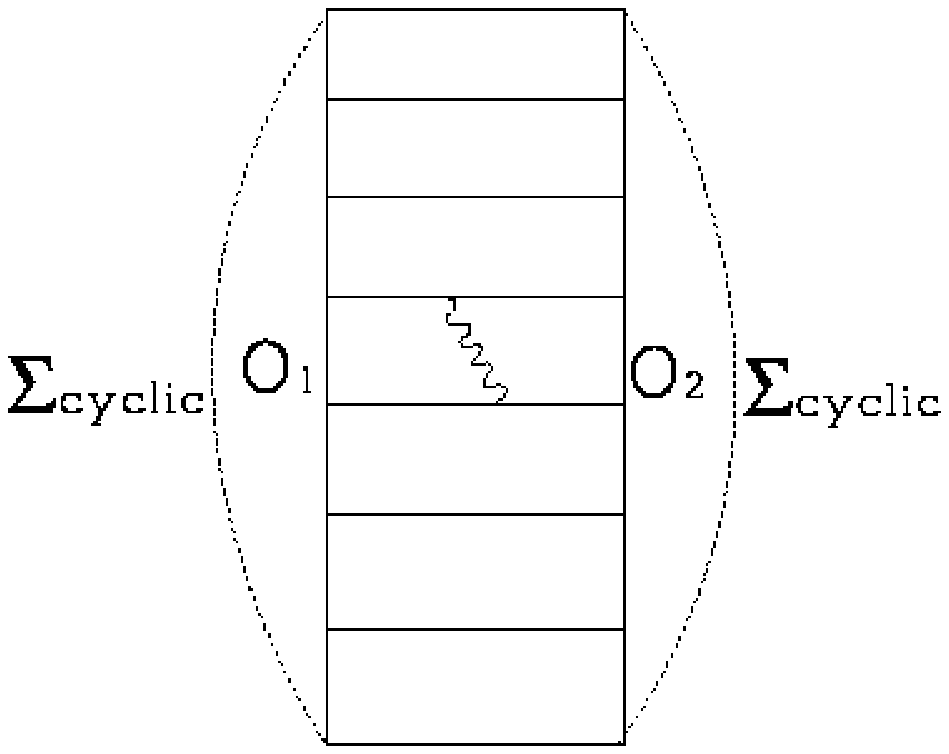}}
\caption{Nearest neighbour interactions in two-string overlap}
}
From 
this figure it is clear that it suffices to focus on the
corrections to the two bit overlap, just as for the case of $\lambda =0$
it was sufficient to focus on the single bit overlap.
The first step in our strategy is to evaluate the corrections 
to the global charges $P_\mu$ and $K_\mu$. It is most convenient to 
write the corrected charges in terms of differential operators
acting on the oscillators states.
For simplicity we will work in the $SO(6)$ subsector,
though the discussion can be easily generalized to 
all sectors. 
The charges at one-loop in 
$\lambda$ for  a two bit state in the $SO(6)$ sector is given by
\bea{corchr}
P_{\mu}  
\exp(-ix  P) )|s_1s_2\rangle &=& 
-i\partial_\mu \exp(-ix P)  |s_1s_2\rangle ,\\ \nonumber
 K_\mu  
\exp(-ix P) |s_1s_2\rangle &=& 
i( 2( \Delta^{(0)} + \lambda H_{12} )x_\mu 
\cr
&\;& + 2 x_\mu x\cdot\partial 
- x^2 \partial_\mu ) 
\exp(-ix  P) |s_1s_2\rangle,   \\ \nonumber
&=& (K_\mu^{(0)} + 2 \lambda H_{12}x_\mu ) 
\exp(-ix  P ) |s_1s_2\rangle,   \\ \nonumber
E  
\exp(- ix P  ) |s_1s_2\rangle &=& 
(( \Delta^{(0)} + \lambda H_{12}) + x\cdot \partial ) 
\exp(-ix  P  ) |s_1s_2\rangle, \\ \nonumber 
&=& (E^{(0)} + \lambda H_{12}) 
\exp(-ix P) |s_1s_2\rangle . 
\eea
Here $\Delta^{(0)} =2$ for the two bit  state 
$|s_1s_2\rangle$, and $s_1s_2$ refer to the two
$SO(6)$ labels of the states. This state can be written   
in terms of the fermionic oscillators $\alpha, \beta$ acting
on the two bit vacuum. $P = P_1 +P_2$ is 
the global momentum of the two bits and
similarly $K$ stands for the  global special conformal 
generator of the two bits.
$H_{12}$
refers to the anomalous dimension Hamiltonian which has the
information of the interactions between the two bits, 
it can be written in terms of the fermionic oscillators.
It is easily shown 
that these corrected generators satisfy the conformal algebra
\eq{confal} to first order in $\lambda$. 
In doing this one uses the 
fact that $H_{12}$ is a Casimir of the $\lambda$=0 algebra 
\cite{Beisert:2003jj},
therefore in the manipulations 
we can treat $H_{12}$  as a c-number.
Let us now show that  the identity $U^\dagger UE = -EU^\dagger U$ is
also true to one loop when $U$ is constructed out of the corrected 
generators given in \eq{corchr}.
We first evaluate $U^\dagger U$ to first order in $\lambda$
\bea{uuexp}
U^\dagger U &=& \exp\left( \frac{\pi}{2}( P_0 - K_0^{(0)} - 2i
\lambda x_0 H_{12} )\right), \\ \nonumber
&=& \exp\frac{\pi}{2} \left( P_0 - K_0^{(0)} \right)
\\ \nonumber
&\;& - i\lambda H_{12} 
\int_0^\pi  dt 
\exp( \frac{t}{2} ( P_0^x -K_0^{(0)} ))  x_0 
\exp \left(  \frac{1-t}{2} ( P_0^x   -K_0^{(0)} ) \right). 
\eea
Here we have used the following expansion, which 
is valid to first order in $\lambda$,  for any two operators $A$ and
$B$.
\be{idopexp}
\exp(A + \lambda B) = 
\exp( A) + \int_0^1  dt \exp( tA)  \lambda B \exp((1-t)A) 
+ O(\lambda^2).
\ee
In \eq{uuexp} we have also used the fact that $H_{12}$ is the Casimir
of the algebra at $\lambda =0$ to move  it to  the extreme left.
Now, 
to evaluate the integral on the last line in \eq{uuexp} we use the
following relation 
\bea{geomac}
&\;&\int_0^\pi  dt 
\exp( \frac{t}{2} ( P_0 -K_0^{(0)} ))  x_0 
\exp ( - \frac{t}{2} ( P_0   -K_0^{(0)} ) ) 
\cr
&=&
\int_0^\pi dt \frac{ 2 x_0 \cos t  +  i(1-x^2) \sin t
}{2 ix_0 \sin t  +   ( 1- x^2)\cos t  +  (1+x^2) }, \\ \nonumber
&=& -i\log (x^2). 
\eea
The simplest way to obtain the first line of the above equation is to
realize that $ P_0 -K_0  = 2 M_{05}$ and use the geometric action of
the rotation $M_{05}$ on $x_0$. This action is not linear, 
but $M_{05}$ acts linearly on the light cone coordinates introduced in
\cite{Ferrara} as a rotation. 
Using the relation of the light cone
coordinates to $x^\mu$ one obtains the relation in \eq{geomac}.
Note that the occurrence of $i$ in the 
above equation is due to the fact that transformation 
$\exp(t M_{05})$ 
acts as a boost but with imaginary angle. 
Furthermore, for convenience,  we have worked in units so that 
the coordinates are dimensionless. 
Reinstating the dimensions of the coordinates would give a scale
in the logarithm of the above equation.
Substituting \eq{geomac} in \eq{uuexp} we obtain  the formula for
$U^\dagger U$ to linear order in $\lambda$
\bea{uuexp1}
U^\dagger U &=& ( 1 -\lambda H_{12} \log (x^2) ) 
\exp \frac{\pi}{2} ( P_0 - K_0^{(0)} ), \\ \nonumber
&=& 
( 1- \lambda H_{12} \log (x^2) ) (U^\dagger U)^{(0)}.
\eea
It is now possible to verify the relation $EU^\dagger U = -U^\dagger U
E$ to first order in $\lambda$. 
The steps involved are indicated below
\bea{stepsuu}
E U^\dagger U &= &
(\Delta^{(0)} + x\cdot \partial + \lambda H_{12} ) U^\dagger U, \\ \nonumber
&=& 
(\Delta^{(0)} + x\cdot \partial + \lambda H_{12} ) 
( 1- \lambda H_{12} \log (x^2) ) (U^\dagger U)^{(0)},
\\ \nonumber
&=&  
U^\dagger U 
( - \Delta^{(0)} - x\cdot\partial - 2 \lambda H_{12} + \lambda H_{12}),
\\ \nonumber
&=& -U^\dagger U  E.
\eea
The ingredients we have used in the manipulations are the fact that
the zeroth order operator $E^{(0)}$  anti-commutes with 
$(U^\dagger U)^{(0)}$, $H_{12}$ is a Casimir and finally 
$\;$ 
$[x\cdot \partial, \log (x^2)] = 2$. In fact it is this last fact which 
gives the extra $-2\lambda H_{12}$ in the third line of the above
equation. We have of course retained only terms to order $\lambda$. 

It is clear from the method of conformal Ward identities of the
previous section, that,  once the  $U^\dagger U E = -EU^\dagger U$
is true to  order $\lambda$, the equation \eq{wardi2} becomes
\be{nwwardi}
(x\partial_x + y\partial_y + 2 \Delta^{(0)} + 2 \lambda H_{12} )I(x,y)
=0.
\ee
This together with translation invariance results in the following
solution for the two-point function for the 
two bit states $|s_1s_2\rangle$
at $x$ and $|s_1's_2'\rangle$ at $y$ 
\be{sol2pfir}
I(x,y) = \frac{1}{(x-y)^{2\Delta^{(0)}}} 
( 1 - 2\lambda \Delta \log (x-y)^2). 
\ee
In the above equation $\Delta$ refers to the c-number 
$\langle s_1s_2| H_{12}  |s_1's_2' \rangle$.
This is the required two-bit overlap of the two-bit states, which
on summation over all the nearest neighbour bits 
reproduces the required logarithmic corrections in the
two-point functions.
Thus the overlap with the corrected $U$ as given in \eq{uuexp1}
reproduces the logarithmic corrections in the two-point functions.

\vspace{.5cm}
\noindent
{\emph{Three-point functions}}
\vspace{.5cm}

It is clear that once the logarithmic divergences of 
all primaries in the two-point
functions are fixed,  then from conformal invariance one can also 
determine all the logarithmic divergences in the three-point functions.
Here we want to see how  these arise in the three-string overlap. 
As before, to keep the discussions simple,
we will restrict our attention
to the $SO(6)$ subsector, but it can be easily generalized
to all primaries.

The figures \eq{3plog2} and \eq{3plogc}, schematically 
indicate the nearest neighbour interactions 
that need to be included in the three string overlaps for both the
length conserving process and the length non-conserving process.
The interactions can be divided into two basic types:
(i) Two body interactions, which involve interactions between 
nearest neighbour bits belonging to any two of the strings
participating in the three string overlap. These are shown in 
figure \eq{3plog2}.
(ii) Genuine three body interactions as shown in figure \eq{3plogc}
\FIGURE{
\label{3plog2}
\centerline{\epsfxsize=16.truecm \epsfbox{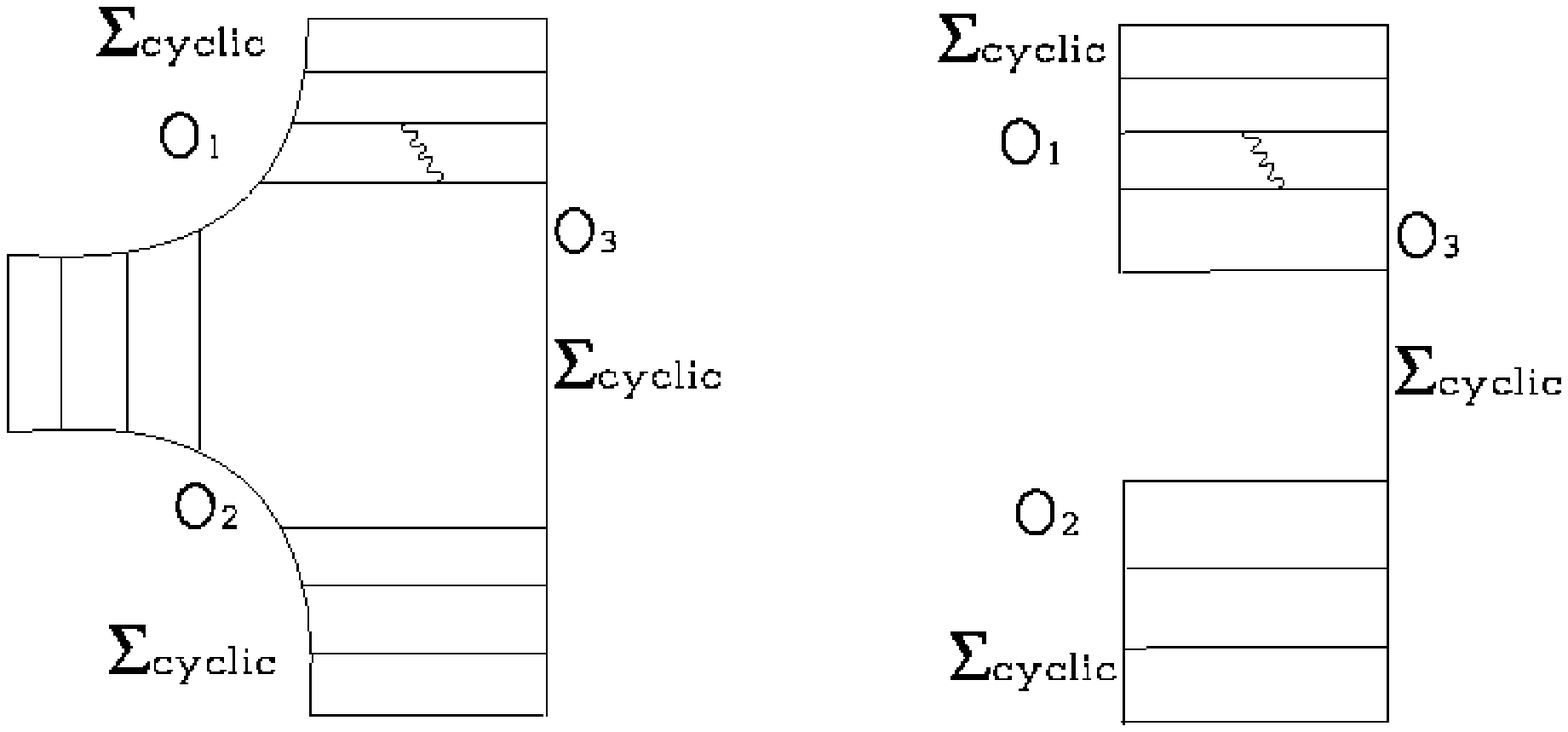}}
\caption{Two body corrections to 3-string overlap}
}
\FIGURE{
\label{3plogc}
\centerline{\epsfxsize=16.truecm \epsfbox{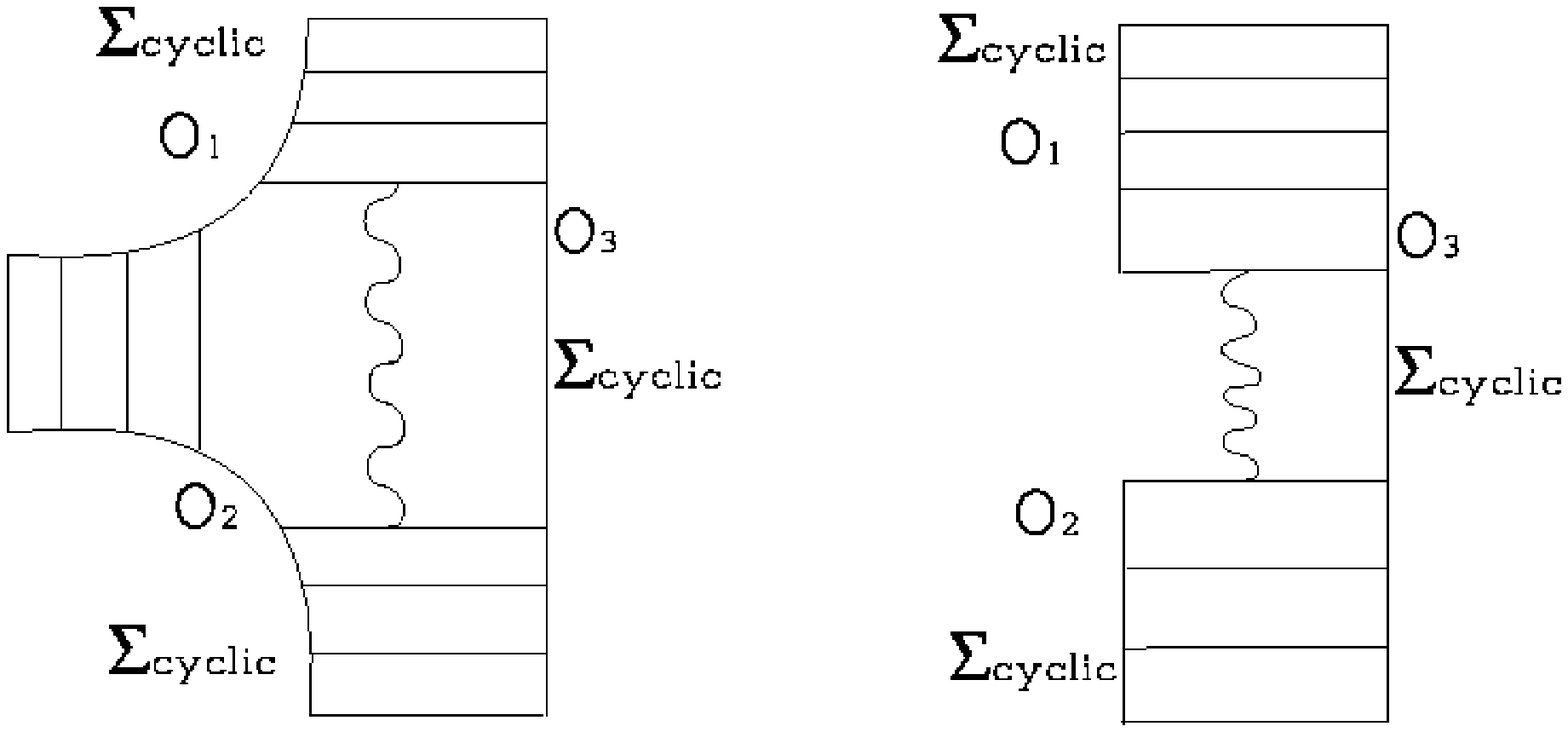}}
\caption{Genuine three body corrections to 3-string overlap}
}

The logarithmic corrections for the two-body interactions are
identical to the corrections encountered for the two-point functions.
The reason for this is as follows:  
let the two bit-strings involved in the two-body
interactions be at positions $x_1$ and $x_2$. 
Then the nearest neighbour bits in each of these strings 
are at these respective positions. The  corrections
to $U^\dagger U$ are then identical to that 
encountered for the two-point function. The discussion of the previous
section goes through and the logarithmic 
corrections are identical to those for the two-bit case, 
discussed in the previous section. 

As for the genuine three-body interactions, it is clear that one cannot
just use the corrections to $U^\dagger U$ to evaluate the 
logarithmic divergences to the overlap. This is because 
two of the bits belong to two different strings and therefore 
they are at two
different positions. Then  the corrections to the charges cannot be
just written as those given in \eq{corchr}, which assumes that the
two bits are the same position.  
The generic three-body interaction will involve two bits at two
different positions, which belong to two different strings, say string
$O_1$ and $O_2$, 
and two nearest neighbour bits which belong to the third string $O_3$, 
as indicated in the figure \eq{3plogc}.
The strategy we will employ to
obtain the corrections to the overlap is, first, to  obtain the two bits 
at different positions 
as an expansion of the same bits at a common  position and then 
perform the overlap onto the two nearest neighbour bits of the
third string $O_3$. When $\lambda =0$,   this expansion is just a
simple Taylor series expansion. But, at finite $\lambda$,
when the bits are at the same position, they carry anomalous dimensions.
The simple Taylor series expansion does not respect this scaling and
therefore we have find the appropriate expansion, which 
is the usual operator product expansion,
given, for example, in \cite{Ferrara}. 
In the language of the bits it is given by
\bea{opebi}
&\;& U_1\exp{(-ix_1 P_1)} |s_1\rangle U_2 \exp{(-ix_2 P_2)} |s_2\rangle
=  \\ \nonumber
&\;& x_{12}^{ \lambda H_{12}} 
F( 1 + \frac{1}{2}\lambda H_{12} ; 2 + \lambda H_{12};  -(x_{12}) \cdot
\partial_{2} )U\exp(- ix_2 (P_1 +P_2)  
| s_1\rangle  |s_2\rangle
+ \cdots
\eea
The subscripts on the operators $U$ and $P$ just refer to the 
site at which the operator acts. 
$F(a;c;z)$ refers to the confluent hypergerometric
function, which has the following integral representation 
\be{inhyp}
F(a; c;z ) = \frac{\Gamma{(c)}}{\Gamma{(a)}\Gamma{(c-a)}}
\int_0^1 du \cdot u^{a-1} ( 1- u)^{c-a-1} e^{uz} 
\ee
Note that on the left hand side of the equation \eq{opebi} 
the bits are at the same position $x_2$ and the operator
$U$ acts on both the sites. 
In performing the
operation  $(x_{12} \cdot \partial_2)^n$ only the symmetric 
and traceless components  are retained.
The dots in \eq{opebi} refer to the 
tensor primaries at higher levels, 
and we have also kept only the contribution from 
the scalar and its 
conformal descendents. This is sufficient if 
one is interested in only three point functions of operators belonging to the 
$SO(6)$ subsector. This is because, finally 
we have to take the
inner product of the expansion of the left-hand side of 
\eq{opebi} with two bits consisting of only scalars  at
position $x_3$. Therefore only the scalar component in the 
expansion survives. Since the 
operator on the left-hand side of \eq{opebi} is at the same position,
we can now evaluate 
the inner product with two nearest neighbour 
bits of the third string, obtaining
\bea{finc3pt}
& T(s_1', s_2', s_1, s_2) = 
\langle s_1'|  \langle s_2'| \exp(ix_3 P^\dagger)
U^\dagger U_1 \exp( -ix_1 P_1) | s_1\rangle  U_2 
\exp(-ix_2 P_2) |s_2\rangle, \cr
&=  \langle s_1'| \langle s_2'|
x_{12}^{ \lambda H_{12}} \int_0^1 du 
u^{\lambda \frac{H_{12} }{2}  } 
(1-u)^{\lambda \frac{H_{12}}{2}} \exp ( -u x_{12} \cdot 
\partial_{x_{2}} ) \frac{1}{x_{32}^{4 + 2 \lambda H_{12} } }
|s_1 \rangle |s_2\rangle.
\eea
Note that the overlap is identical to that  occurring in  
the two-point function. To obtain the second line, we have 
substituted the equation \eq{sol2pfir}. Performing the 
$u$ integral we obtain 
\be{3ptonef}
T(s_1', s_2', s_1, s_2 ) =
\frac{1}{x_{13}^2 x_{23}^2} ( 1 - \lambda \Delta 
( \ln |x_{13}| + \ln|x_{23}| - \ln|x_{12}|). 
\ee
where 
$ \Delta = \langle s_1'| \langle s_2' |H_{12} |s_1\rangle |s_2\rangle$.
The integral in \eq{finc3pt} is evaluated by first 
expanding all the terms 
one order in $\lambda$. The easiest way to 
argue the result of the integral 
is that given in \eq{3ptonef}, is to choose a holomorphic
direction and set $x_{12} = z_{12}$. Then the  
exponential 
with the derivative 
just acts as a translation with the parameter
$-u z_{12}$. This is because choosing a holomorphic direction
automatically ensures that there is no trace in the action 
of the derivative.
Now it is easy to act this on the function
$1/x_{34}^{4 + 2 \lambda \Delta}$ and perform the 
integrations. Then one can reinstate the value of $z_{12}$ as
$x_{12}$ to obtain the above result.
The result in \eq{3ptonef} is the required answer between 
for the genuine three body interactions to reproduce the
logarithms in the three-point functions. 
Thus we have shown that the overlap at one 
loop in $\lambda$ reproduces the 
logarithms for the three-point functions.

\subsection{Structure constants in the $SU(2)$ subsector}

As argued in the introduction to this section, to obtain the 
structure constants we need to deform the commutation relations
of the oscillators $\{a, b, \alpha,\beta \}$. 
The consistent way to do this is to obtain the deformed relations
from a world-sheet action. In this subsection we show that
in the $SU(2)$ subsector one can uniquely determine the 
nearest neighbour interactions in the world sheet that 
reproduce the structure constants to one loop.

The $SU(2)$ subsector consists of 
gauge invariant operators made of two complex scalars belonging 
to two different Cartans of $SO(6)$, say $z$ and $y$. 
In terms of 
the oscillators, only one of the $\alpha$'s and one of the $\beta$'s
are excited in this subsector. 
The generators for the $SU(2)$ subsector are given by
\be{gensu2}
J^i = \bar\varphi \sigma^i \varphi
\ee
where $\sigma^i$ refers to the Pauli matrices and
$\varphi$ is the $SU(2)$ spinor given by
\be{su2spin}
\varphi = \left(
\begin{array}{c}
\alpha \\
-\beta^\dagger
\end{array}
\right), 
\qquad \bar\varphi = \left(
\alpha^\dagger  , -\beta \right). 
\ee
The operators corresponding to the two states are 
$z\rightarrow |0\rangle$ and 
$y \rightarrow \alpha^\dagger\beta^\dagger |0\rangle$.
Note that the combination $\bar\varphi\varphi$ is 
$U(1)_Z$ invariant.

To obtain the corrections to the structure constants we start with the 
following bit world sheet Lagrangian in the $SU(2)$ sector. 
\bea{su2ws}
{\cal L} &=& \sum_s^l \left[ \frac{i}{2} \left( 
\bar\varphi_s
{\partial}_t \varphi_s
- 
\partial_t\bar\varphi_s
 \varphi_s \right)  \right. \\ \nonumber
&+& \lambda_1
\frac{i}{2} 
( 
\bar\varphi_s{\partial}_t \sigma^i \varphi_s 
-\partial_t\bar\varphi_s \sigma^i \varphi_s 
)
\left( 
\bar\varphi_{s+1} \sigma^i\varphi_{s+1} + 
\bar\varphi_{s-1} \sigma^i\varphi_{s-1}  \right)
\\ \nonumber
&+&
\lambda_2 \frac{i}{2} 
( 
\bar\varphi_s{\partial}_t  \varphi_s  -
\partial_t\bar\varphi_s  \varphi_s )
\left( 
\bar\varphi_{s+1} \varphi_{s+1} + 
\bar\varphi_{s-1} \varphi_{s-1}  \right)
\\ \nonumber
&+& \left.
 \lambda V(\varphi_s\varphi_{s+1})\right].
\eea
As argued in the introduction 
to this section, we need to  modify the kinetic energy
term in the action with nearest neighbour interactions 
so that we will obtain deformed commutation relations
for the oscillators, which will in turn change their norm. 
The terms proportional to $\lambda_1$ and $\lambda_2$ 
in \eq{su2ws} are the 
most general 
nearest neighbour interactions consistent with the following
symmetries:
\\
$\bullet$ Local $U(1)_Z$ symmetry.
All terms occur in combinations $\bar\varphi_s \varphi_s$ 
which respects local $U(1)_Z$ symmetry.
\\
$\bullet$ Orientation preserving
symmetry. That is, the interaction with 
the $s+1$ th site is the same as the interaction with 
the $s-1$ th site.
\\
$\bullet$ Global $SU(2)$ symmetry. In \eq{su2ws} we have parameterized 
the interaction into two components. $\lambda_1$ refers to the
strength of the spin 1 interaction  between the nearest
neighbours, and $\lambda_2$ refers to the strength of the
spin 0 interaction between the nearest neighbours.
Both $\lambda_1$ and $\lambda_2$ will in general be proportional
to the 't Hooft coupling.
\\
In \eq{su2ws}, the potential $V(\varphi_s, \varphi_{s+1})$, refers
to the anomalous dimension Hamiltonian. This is because  
we require the world sheet Hamiltonian to be the operator $E$, which
measures the conformal dimension. The zero point energy of the 
$a, b$ oscillators already provide the bare dimensions of the 
operators in the $SO(6)$ subsector.
For the first order action given in \eq{su2ws}, the Hamiltonian is 
just given by the potential $V(\varphi_s, \varphi_{s+1})$. 
Therefore, the potential is the anomalous dimension Hamiltonian.
In the next section we will provide a gauging principle
which determines this potential,
here we just focus on the 
structure constants. 

Our  strategy will be first to quantize the 
Lagrangian given in \eq{su2ws}, and determine the 
modified commutation relations. Then we construct the 
vertex which implements the delta function overlap consistent
with the modified commutation relations, and finally evaluate the
overlap of the states. We will  show that the structure constants
are identical, upto to an overall constant, 
to those obtained in the gauge theory calculation 
of \cite{Alday:2005nd}

The canonical conjugate momenta to $\varphi_x$ and 
$\bar\varphi_x$ are given by
\bea{conmom}
\Pi_{\varphi_s^a} &-& \left[ \frac{i}{2} \bar\varphi^a_s + \frac{i\lambda_1}{2}
( \bar\varphi_s \sigma^i)_a 
\left( 
\bar\varphi_{s+1} \sigma^i \varphi_{s+1}
+
\bar\varphi_{s-1} \sigma^i \varphi_{s-1} \right)
\right. \cr
&\;& \quad + \left. \frac{i\lambda_2}{2} 
 \bar\varphi^a_s  
\left( 
\bar\varphi_{s+1}  \varphi_{s+1}
+
\bar\varphi_{s-1}  \varphi_{s-1} \right)  \right] 
=0, \\ \nonumber
\Pi_{\bar\varphi_s^a} &-& \left[ \frac{i}{2} \varphi^a_s
+ \frac{i\lambda_1}{2}
( \sigma^i \varphi_s)_a
\left( 
\bar\varphi_{s+1} \sigma^i \varphi_{s+1}
+
\bar\varphi_{s-1} \sigma^i \varphi_{s-1} \right)
\right.
\cr 
&\;& \quad +\left.  \frac{i\lambda_2}{2}
\varphi^a_s
\left( 
\bar\varphi_{s+1}  \varphi_{s+1}
+
\bar\varphi_{s-1}  \varphi_{s-1} \right)  \right]
=0 ,
\eea
where $a, b, \ldots$ takes values $1, 2$.
We will denote the first and the second equations as 
$C_{\varphi_s^a}=0$ and
$C_{\bar\varphi_s^a} =0$ respectively. These
are second class constraints as their 
Poisson brackets are non-zero. To quantize a Lagrangian 
with second class constraints, we need to use the Dirac brackets. To
this end, let us compute the commutation relations among the constraints.
They are given by
\bea{comcons}
\{ C_{\varphi_s^a}, C_{\varphi_t^b} \} =0, \;\;\;&\;&\;\;
\{ C_{\bar\varphi_s^a}, C_{\bar\varphi_t^b} \} =0,
\\ \nonumber
\{ C_{\varphi_s^a}, C_{\bar\varphi_t^b} \} 
&=& \delta_{st}\delta_{ab}  \\ \nonumber
&+& \lambda_1 \delta_{st} \sigma_{ba}^i 
\left( 
\bar\varphi_{s+1} \sigma^i \varphi_{s+1}
+
\bar\varphi_{s-1}  \sigma^i \varphi_{s-1} \right) 
\cr
&+& \lambda_1 (\sigma^i\varphi_t)_b 
\left( 
\delta_{t+1, s} (\bar\varphi_{t+1}\sigma^i)_a 
+ 
\delta_{t-1, s} (\bar\varphi_{t-1}\sigma^i)_a  \right)
\\ \nonumber
&+& 
\lambda_2 \delta_{st} \delta_{ba} 
\left( 
\bar\varphi_{s+1}  \varphi_{s+1}
+
\bar\varphi_{s-1}  \varphi_{s-1} \right) 
\cr
&+& \lambda_2 \varphi_t^b
\left( 
\delta_{t+1, s} \bar\varphi_{t+1}^a 
+ 
\delta_{t-1, s} \bar\varphi_{t-1}^a  \right).
\eea
The Dirac bracket is then given by \cite{Dirac}
\be{dbdef}
\{ \bar\varphi_s^a, \varphi_t^b\}_{{\rm DB}} =
\{\bar\varphi_s^a, \varphi_t^b \} - \{\bar\varphi_s^a, C_{\bar\varphi_u^c}\}
\{ C_{\bar\varphi_u^c} , C_{\varphi_{u'}^d} \}^{-1} \{ \varphi_{u'}^d,
\phi_t^b \}.
\ee
Here summation over $u, u', c, d$ is implied, and we have also used the
fact that the matrix of the commutation relations of the constraints
is off diagonal \eq{comcons}.
Substituting the last equation of \eq{comcons} and evaluating the 
inverse in \eq{dbdef} upto first order in $\lambda$'s, we obtain
\bea{bascom}
\{ \bar\varphi_s^a, \varphi_t^b \}_{{\rm DB}} &=&
\delta_{s, t} \delta_{a, b}   \\ \nonumber 
&-& \lambda_1 \delta_{st} \sigma_{ba}^i 
\left( 
\bar\varphi_{s+1} \sigma^i \varphi_{s+1}
+
\bar\varphi_{s-1}  \sigma^i \varphi_{t-1} \right) 
\cr
&-& \lambda_1 (\sigma^i\varphi_t)_b 
\left( 
\delta_{t+1, s} (\bar\varphi_{t+1}\sigma^i)_a 
+ 
\delta_{t-1, s} (\bar\varphi_{t-1}\sigma^i)_a  \right)
\\ \nonumber
&-& \lambda_2 \delta_{st} \delta_{ba} 
\left( 
\bar\varphi_{s+1}  \varphi_{s+1}
+
\bar\varphi_{s-1}  \varphi_{s-1} \right) 
- \lambda_2 \varphi_t^b
\left( 
\delta_{t+1, s} \bar\varphi_{t+1}^a 
+ 
\delta_{t-1, s} \bar\varphi_{t-1}^a  \right). 
\eea
The rest of the commutators vanish
\be{restdb}
\{\varphi_s^a, \varphi_t^b\}_{\rm{DB}} =0, \;\;\;\;\; \;\;
\{\bar\varphi_s^a, \bar\varphi_t^b\}_{\rm{DB}} =0.  
\ee
From now on, (anti-)commutation relations involving the $\varphi$'s
will always be understood as Dirac brackets.
To prove that these relations are consistent, one can show that the 
following Jacobi
identities are satisfied to linear order in $\lambda$'s. 
\bea{jacob}
[\varphi^c_u, \{\bar\varphi^a_s, \varphi^b_t \} ]
+
[\bar\varphi^a_s, \{\varphi^b_t, \varphi^c_u \} ]
+
[\varphi^b_t, \{\varphi^c_u, \bar\varphi^a_s \} ] =0
\\ \nonumber
[\bar\varphi^c_u, \{\bar\varphi^a_s, \varphi^b_t \} ]
+
[\bar\varphi^a_s, \{\varphi^b_t, \bar\varphi^c_u \} ]
+
[\varphi^b_t, \{\bar\varphi^c_u, \bar\varphi^a_s \} ] =0.
\eea

To construct the delta function overlaps, it is convenient to work
with oscillators which are diagonal either in the position space
or in the momentum space. To this end 
we define the  oscillators $\chi$ which satisfy the usual
commutation rules.
\bea{defnewos}
\chi^a_s &=& \varphi^a_s + \frac{\lambda_1}{2} ( \sigma^i\varphi_s)_a
\left( \bar\varphi_{s+1}\sigma^i \varphi_{s+1} + 
\bar\varphi_{s-1} \sigma^i \varphi_{s-1} \right)
\cr
&+& 
\frac{\lambda_2}{2} \varphi_s^a 
\left( \bar\varphi_{s+1} \varphi_{s+1} +
\bar\varphi_{s-1} \varphi_{s-1}  \right),
\\ \nonumber
\bar\chi^a_s &=& \bar\varphi^a_s + \frac{\lambda_1}{2} (
(\bar\varphi_s \sigma^i )_a 
\left( \bar\varphi_{s+1}\sigma^i \varphi_{s+1} + 
\bar\varphi_{s-1} \sigma^i \varphi_{s-1} \right)
\cr
&+& 
\frac{\lambda_2}{2} \bar\varphi_s^a 
\left( \bar\varphi_{s+1} \varphi_{s+1} +
\bar\varphi_{s-1} \varphi_{s-1}  \right).
\eea
With these definitions, 
it is easy to verify that the commutation relations of the $\chi$'s
to first order in $\lambda$ are given by the usual relations
\be{uschi}
\{\bar\chi_s^a, \chi_t^b\} = \delta_{st}\delta_{ab}, \;\;\;
\{\chi_s^a, \chi_t^b\} = 0, \;\;\;
\{\bar\chi_s^a, \bar\chi_t^b\} = 0. 
\ee
From the definition of the $\chi$'s in \eq{defnewos}
it is clear that the vacuum state 
for the $\varphi$ oscillators remains a vacuum state for the 
$\chi$ oscillators.  
It is also useful to have the inverse relations
\bea{inuschi}
\varphi^a_s &=& \chi^a_s - \frac{\lambda_1}{2} ( \sigma^i\chi_s)_a
\left( \bar\chi_{s+1}\sigma^i \chi_{s+1} + 
\bar\chi_{s-1} \sigma^i \chi_{s-1} \right)
\cr
&-& 
\frac{\lambda_2}{2} \chi_s^a 
\left( \bar\chi_{s+1} \chi_{s+1} +
\bar\chi_{s-1} \chi_{s-1}  \right), \\ \nonumber
\bar\varphi^a_s &=& \bar\chi^a_s - \frac{\lambda_1}{2} (
(\bar\chi_s \sigma^i )_a 
\left( \bar\chi_{s+1}\sigma^i \chi_{s+1} + 
\bar\chi_{s-1} \sigma^i \chi_{s-1} \right)
\cr
&-& 
\frac{\lambda_2}{2} \bar\chi_s^a 
\left( \bar\chi_{s+1} \chi_{s+1} +
\bar\chi_{s-1} \chi_{s-1}  \right).
\eea
Note that, using the $\chi$ oscillators, it is easy to construct the 
global $SU(2)$ generators to the leading order in the $\lambda$'s.
They are given by $\sum_s^i \bar\chi(s) \sigma^i \chi(s) $. 

To obtain the corrected structure constants, following 
\cite{Kugo:1992md}, we
define the overlap using the diagonal oscillators $\chi$'s.
To be more explicit it is convenient to work with the 
$SU(2)$ invariant vacuum given by 
\bea{su2invv}
|0\rangle' &=& -\beta^\dagger |0\rangle,  \qquad {\rm then}  \quad
\varphi^a|0\rangle' = 0, \cr
\bar\varphi^a|0\rangle' &=&  \alpha^\dagger\beta^\dagger |0\rangle,
 \qquad {\rm for} \;\;a =1, 
\cr
&=& |0\rangle  \qquad\qquad {\rm for} \;\;a=2,
\eea
It is clear, using \eq{inuschi}, that the vacuum $|0\rangle'$ remains 
the vacuum for the $\chi$ oscillators. This is because the $\lambda$ 
corrections in \eq{inuschi} involve annihilation operators $\chi$, which
can be replaced by $\varphi$ to the leading order in $\lambda$. 
It is sufficient for us to deal with the two bit overlap
for the sites $s$ and $s+1$. This is given by
\be{newover}
{}_{s+1}\langle 0|' {}_{s}\langle 0|'
\chi_{s+1}^d \chi_s^c \bar\chi_s^a\bar\chi_{s+1}^b
|0\rangle_s' |0\rangle_{s+1}' = 
\delta^{ca}\delta^{db}.
\ee
The sites $s$ and $s+1$ can be on bits belonging to two different
strings. Such a situation will occur in a three-body term.
To obtain the $\lambda$ dependence of the two-point 
functions or three-point functions in string theory,  we need to 
evaluate the overlap constructed out of diagonal oscillators $\chi$,
as in \eq{newover}, on the old states constructed out of the $\varphi$
oscillators \cite{Kugo:1992md}.
We now give the results for the overlap of  two-bit states.
\bea{corstover}
\langle 0 | \varphi^d_{s+1} \varphi^c_s
\bar\varphi^a_s\bar\varphi^b_{s+1} |0\rangle &=& 
\delta^{ca}\delta^{db}
\\ \nonumber
&-& \lambda_1 (\sigma^i_{ca})_s (\sigma^i_{db})_{s+1} 
- \lambda_2 \delta_{ca}\delta_{db}, 
\eea
To evaluate the overlap, we write the $\varphi$ oscillators in terms of 
the $\chi$ oscillators using \eq{inuschi} and then apply the definition 
of the overlap given in \eq{newover}. 
We can now 
determine the relationship between the $\lambda$'s  using
the fact that on chiral primaries there should be no
corrections. This implies that, when all the indices in \eq{corstover}
are same the $\lambda$, the corrections have to vanish. This condition
gives $\lambda_2 =- \lambda_1 =\lambda$. We thus obtain
\be{finlcorov3}
\langle 0 | \varphi^d_{s+1} \varphi^c_s
\bar\varphi^a_s\bar\varphi^b_{s+1} |0\rangle = 
\delta^{ca}\delta^{db}
+ \lambda( 
1 - \sigma^i_s \sigma^i_{s+1} )_{ca, db}.
\ee
Note that the corrections to the norm are proportional to the 
anomalous dimension Hamiltonian of the $SU(2)$ sector which is given by
$\sum_s (1- \sigma_s\cdot \sigma_{s+1})$. 
To compute structure constants one needs to evaluate the
renormalization group invariant quantity given in \eq{schindep}.
As shown in \cite{Alday:2005nd} the corrections to the 
structure constants  from all 
the two body terms 
in the three point functions will vanish on subtracting the metric 
corrections. What remains are half the contribution of the 
genuine three body terms. The three body corrections 
can be read out from \eq{finlcorov3}. Therefore we find that
the structure constants in the $SU(2)$ sector is dictated by the 
anomalous dimension Hamiltonian, which was the result in the
gauge theory calculation of \cite{Alday:2005nd}. 
Here we have arrived at this conclusion entirely from the
symmetries in the bit picture and using the formalism of 
\cite{Kugo:1992md} to obtain the corrections to the overlap.

\section{Anomalous dimensions for the $SU(2)$
sector.}

Throughout this paper we have emphasized the role of 
the  oscillator variables for a string bit formulation 
of ${\cal N}=4$ Yang-Mills. 
One of the underlying symmetries in this formulation is the 
local $U(1)_Z = N_a+ N_\alpha - N_b -N_\beta$ .
In section 2.3 we saw that the spectrum
was obtained after gauging this symmetry. 
As we discussed in the previous section, this symmetry also played 
a role in determining the structure constants 
in the $SU(2)$ subsector, at one loop in $\lambda$. 
In this section  we present a linear $U(1)_Z$ gauged model
for the oscillators $\varphi^a$, which reproduces the anomalous
dimension Hamiltonian to two loops.
To be precise, we define the anomalous dimension
Hamiltonian to be given by
\be{defgaupro}
\exp{c\left( \sum_{n=1}^\infty \lambda^n H^{(n)}\right)  } 
= \int {\cal D} U \exp ( \sum_{ s}^l  g ( \varphi^{\dagger}_s  
U_{s, s+1}  \varphi_{s+1}  + {\rm {h. c} } ) ).
\ee
where $U_{s, s+1} $ are $U(1)$ link variables. 
Under local $U(1)_Z$ transformation the link variables 
transform as $U_{s, s+1} \rightarrow \exp{(i\theta_s)} U_{s,s+1} 
\exp{ (- i \theta_{s+1})}$, thus
the Hamiltonian on the left-hand side has local $U(1)_Z$  
symmetry. It  has global  $SU(2)$ symmetry and 
is also just a linear model in the spinor variables $\varphi_s$. 
Only nearest neighbour oscillators are coupled in the model.
The integration over the link variables is just the classical 
integration and ${\cal D} U = \prod_s^l \frac{1}{2\pi} 
(U^{-1}_sdU_s)$. 
The coupling $g$ depends on the 't Hooft parameter
$\lambda$. It will turn out to be proportional
to the square root of the 't Hooft coupling. $c$ is an overall constant.
In fact, the above model is a simplified version of the
one considered in 
\cite{Affleck:1988yi,Dagotto:1988rj} 
to obtain the Heisenberg spin chain.

At first sight it is not immediately 
obvious  that one would obtain
next nearest neighbour interaction of the kind required for the 
two-loop anomalous dimension Hamiltonian, since the model
involves only a nearest neighbour coupling.
Also, it is not
obvious one would obtain a local Hamiltonian after integrating
the link variables.
In this section we integrate out the link variables 
in \eq{defgaupro} to two loops and show
that we obtain the two-loop anomalous dimension
Hamiltonian for the $SU(2)$ subsector, as found in \cite{Beisert:2003tq}
from a gauge theory calculation.  
We also show 
that the gauged linear model obeys BMN scaling to all loops. 
In deriving both these facts we neglect the contributions of the 
modified commutation relations obtained in the previous section. 
These would in principle contribute from two loop and above.
A complete analysis would require to keep track of their contribution
at all loops. We leave this for later work, but independently of this,
one could ask the question if the model in \eq{defgaupro} reproduces 
the three-loop anomalous dimension Hamiltonian found in 
\cite{Beisert:2003ys,Eden:2004ua}
\footnote{We thank Jan Plefka who emphasized on testing the
model to three loops.}.
Unfortunately the answer turns out to be negative, but in the 
appendix E  we show  that the inclusion of non-nearest couplings
in \eq{defgaupro} is sufficient to reproduce the anomalous dimension
Hamiltonian to three loops.
It would be interesting to see if the simple model presented in 
\eq{defgaupro}, together with the modified commutation rules 
due to just the nearest neighbour interactions  introduced in 
\eq{su2ws}, would reproduce the anomalous dimension to three loops.

\subsection{Anomalous dimensions to two loops}

Consider the linear model in \eq{defgaupro} to two loops
\be{ontmod}
\exp( c( \lambda H^{(1)} + \lambda H^{(2)})) = 
\int {\cal D} U \exp (  g \sum_s^l ( A_s U_{s, s+1}  + 
 A_s^\dagger U_{s, s+1} ) )
\ee
where $A_s = \bar\varphi_s \varphi_{s+1} = 
\alpha_s^\dagger  \alpha_{s+1} + 
\beta_s \beta_{s+1}^\dagger$ and $c$ is an overall constant. 
As an illustration of the 
calculations involved  we first obtain the one-loop answer
in detail.
We start by  expanding the exponential and then perform the 
$U$ integrations. 
Due to the $U$ integrations only even powers of $g$ are retained, 
and for the one loop answer it is sufficient to 
retain only the second order term. This is given by
\bea{onexp2n}
\tilde H^{(1)}  &=& 
\int {\cal D}U \frac{g^2}{2} \sum_{s, t} 
\left( A_sA_t^\dagger +
A_s^\dagger A_t \right) U_{s,s+1} U_{t,t+1}^\dagger, \\ \nonumber
&=& \frac{g^2}{2} \sum_s ( A_s A_s^\dagger  + A_s^\dagger A_s ). 
\eea
On external states, which satisfy the $U(1)_Z$ constraint,
it is easy to see that $A_s^\dagger A_s = A_s A_s^\dagger$. This is 
due to the commutation relation
\be{comaar}
[A_s, A_t^\dagger] = \delta_{st} G_s =
\delta_{st} \left(  N_s^\alpha -N_s^\beta -
( N_{s+1}^\alpha -N_{s+1}^\beta) \right), 
\ee
where $N_s^\alpha, N_s^\beta$ refer to the number operators 
of the oscillators $\alpha, \beta$ respectively.
Note that 
the right hand side of this equation is the $U(1)_Z$
charge, which vanishes on physical external states. We 
will refer to this as the Gauss' law constraint
$G$. 
Therefore the 
last line  of equation \eq{onexp2n} reduces to 
\bea{onexp2n2}
\tilde H^{(1)} &=& \frac{g^2}{2} \sum_{s} A_s A_s^\dagger,
\\ \nonumber
&=& \frac{g^2 }{2}  \sum_s ( 1- \sigma_s \cdot \sigma_{s+1}). 
\eea
In the last line of the above expression we have 
expressed the operator $A_s A_s^\dagger$ in terms of
Pauli matrices. To do this one uses the identifications 
of $J^+, J^-, J^3$ in terms of 
oscillators  given in \eq{gensu2} and the $U(1)_Z$ constraint. 
Equating the leading order expansion on the left hand side of
\eq{ontmod}, we find
\be{frorte1}
c \lambda H^{(1)} = \frac{g^2}{2} \sum_s ( 1- \sigma_s\sigma_{s+1}).
\ee
We have thus obtained, up to a proportionality
constant, the one loop anomalous dimension
Hamiltonian found in \cite{Minahan:2002ve} for the $SU(2)$ subsector. 
See \cite{Plefka:2005bk} for 
a recent review on developments in this subject.

We now proceed to expand  \eq{ontmod} to order $g^4$ to obtain 
the two-loop Hamiltonian. At this order one encounters 
the following 
\bea{4thorexp}
\tilde H^{(2)} &=& \frac{g^4}{4!} 
\int {\cal D}U 
\left[\sum (A_s U_{s+1} + A_s^\dagger U_{s+1}^\dagger ) \right]^4
\\ \nonumber
&=& \frac{g^4 }{4!} \left[ \sum_{s,t} \left(
A_sA_tA_s^\dagger A_t^\dagger + A_sA_tA_t^\dagger
A_s^\dagger\right)  - \sum_s A_s A_s A_s^\dagger A_s^\dagger
 \right. \\ \nonumber
& \;& \left. +  ( 5 \;\;  \rm{ Permutations}) \right]
\eea
In the second line we have performed the $U$ integrations, 
which retains only terms with equal number of $A$'s and
$A^\dagger$'s. The permutations refer to all other possible
arrangements of 2 $A$'s and 2 $A^\dagger$'s. 
Using the commutation relation in \eq{comaar} we can 
bring all the other permutations to the 
standard form, in which all the $A$'s are to the left and 
$A^\dagger$'s are  to the right. In performing this 
operations there are additional terms which arise due to
right hand side of \eq{comaar}. It is easy to see that all
these terms cancel. We will illustrate this with the following
term which is the permutation of the standard term
\bea{permopr}
&\;&\sum_{s,t} \left( A_s A_s^\dagger A_t A_t^\dagger + 
A_s A_t^\dagger A_t A_s^\dagger \right) - \sum_s A_s
A_s^\dagger A_s A_s^\dagger \\ \nonumber
&=& \sum_{s,t} \left( A_s A_t A_s^\dagger A_t^\dagger
+ A_s A_t A_t^\dagger A_s^\dagger \right) 
- \sum_s A_s A_s A_s^\dagger A_s^\dagger
\\ \nonumber
&-& \sum_s A_s G_s A_s^\dagger - \sum_s A_s \sum_t G_t
A_s^\dagger + \sum_s A_s G_s A_s^\dagger.
\eea
Here $G_s$ refers to the Gauss' law constraint defined in
\eq{comaar}.The last line contains the terms obtained
after moving  the $A$'s to the left. 
Note that first and the third terms of the 
last line cancel each other and the middle term vanishes 
identically, since $\sum_y G_y =0$. 
Using this procedure, all the five permutations in 
\eq{4thorexp} can be brought to the standard form.

What remains now is to write the standard form in terms of
Pauli matrices. This is done by bringing together
the $A$'s and $A^\dagger$ which have the same position
index. Once they 
are together we can easily write the various terms as Pauli matrices.
Performing these manipulations we have
\be{manfourt}
\tilde H^{(2)} = 
\frac{6g^4 }{4!} \left( 
2 \sum_s A_s A_s^\dagger \sum_t A_tA_t^\dagger
+ 2\sum_s A_s G_s A_s^\dagger  + 
\frac{1}{2} \sum_{s,t} [A_s, A_t] [A_t^\dagger A_s^\dagger]
\right).
\ee
We have not included the term
$A_sA_sA_s^\dagger A_s^\dagger$,  as it can be shown to vanish
on any external state satisfying the $U(1)_Z$ constraint. 
To evaluate  the extra terms we need the following commutation 
relations 
\bea{2lcomrel}
[G_s, A_t^\dagger] = ( \delta_{s, t+1} + \delta_{t, s+1} -
 2\delta_{s,t} ) A_t^\dagger, \\ \nonumber
[A_s,A_t] = \delta_{s+1, t} A_s^{(2)} - 
\delta_{s, t+1} A_t^{(2)}, \\ \nonumber
[A_s^\dagger, A_t^\dagger]
= \delta_{t+1, s} A_t^{(2)\dagger}
 - \delta_{t, s+1} A_s^{(2)\dagger}.
\eea
where $A_s^{(2)} = \bar\varphi_s \varphi_{s+2} = 
\alpha_s^\dagger \alpha_{s+2} + \beta_s\beta^\dagger_{s+2}$. 
This involves oscillators of next to nearest neighbour
positions.
Substituting the above relations in \eq{manfourt}
and using the fact that when the Gauss' law $G$ 
is on the extreme right it vanishes on physical
external states,  we obtain
\bea{fin2anl}
\tilde H^{(2)} 
&=& \frac{g^4 }{2} 
 \sum_{s, t}  A_s A_s^\dagger  A_tA_t^\dagger +
\frac{g^4 }{4}\sum_s\left(  - 4  A_s A_s^\dagger 
+  A_s^{(2)} A_s^{(2)\dagger} \right),
\\ \nonumber
&=& \frac{g^4 }{2} \left[\sum_s \frac{1}{2}
( 1- \sigma_s\cdot \sigma_{s+1})\right] ^2 
+ 
 \frac{g^4}{4}   \sum_s 
\left(   -4 \frac{1}{2} ( 1- \sigma_s \cdot \sigma_{s+1}) 
 + \frac{1}{2} ( 1- \sigma_s \cdot \sigma_{s+2}) \right),
\\ \nonumber
&=& \frac{1}{2} c^2 \lambda^2( H^{(1)})^2  + c\lambda^2 H ^{(2)}.
\eea
Note that the non-local term is precisely the one 
due to the expansion of the one-loop term in the exponential.
The local term is proportional to the two-loop anomalous dimension 
Hamiltonian.  
The relative coefficient of the nearest
neighbour term and the next nearest neighbour term  ensures
BMN scaling.
To obtain the two-loop anomalous dimension Hamiltonian, one 
then chooses $c \lambda = g^2/2$ and $ c \lambda^2  = g^4/8$ 
which gives $\lambda = g^2/4$ and $c =2$
\footnote{Using a similar gauged linear model for 
the $SU(1|1)$ subsector we obtain the 
leading $1/J$ piece of the 2-loop anomalous dimension Hamiltonian
found in \cite{Callan:2003xr,Callan:2004uv,
Callan:2004ev,McLoughlin:2004dh,Alday:2005jm}. 
Restricting the  
model to the $SU(2|3)$ subsector we have verified that
we obtain the 1-loop anomalous dimension Hamiltonian for this
subsector.}. 

\vspace{.5cm}
\noindent
{\emph{ BMN scaling to all loops}}
\vspace{.5cm}

We now show that the 
nearest neighbour interaction in \eq{defgaupro} gives rise to 
anomalous dimensions satisfying BMN scaling to all loops
\footnote{We thank Jan Plefka for showing 
that the linear model in \eq{defgaupro} satisfies BMN scaling
to 5-loops numerically.}. Recall 
that the
nearest neighbour coupling 
$\bar\varphi_s U_{s,s+1} \varphi_{s+1} + {\rm h.c.}$ in 
\eq{defgaupro} can be written as $A+ A^{\dagger}$ where $A=\sum_s A_s 
U_{s,s+1}$ and $A^{\dagger}= \sum_s A^{\dagger}_s U^{\dagger}_{s,s+1}$, 
we
first note, using \eq{comaar} 
$[A, A^{\dagger}]=0$. 
This allows
one to write $e^{\sqrt{\lambda}V^{(1)}}= e^{\sqrt{\lambda}A^{\dagger}}.
e^{\sqrt{\lambda}A}$. The $n$-loop term then comes from the operator 
\begin{equation}
\frac{1}{n!^2} \lambda^n \int {\cal D}U {A^{\dagger}}^n A^n.
\end{equation} 
Our strategy will be to determine the representation of this operator on 
physical Hilbert space. In the BMN limit the relevant states are the ones
with large $J$ that have small number of impurities distributed sparsely.
Let us denote the two $SU(2)$ doublet states as $|0\rangle$, 
which we shall call 
ground state  and $\alpha^{\dagger}\beta^{\dagger}|0\rangle$ 
which will be called 
the impurity at a given site. In the BMN 
limit the distance between any two impurities is greater than $n$ where $n$ is 
the order of perturbation we are interested in. Since we will finally be 
taking the local terms in the interaction (the non-local terms canceling
from the lower order terms as explained above), it is sufficient to focus
on just one impurity say at site $s$. This state denoted by $|\psi_s\rangle$ 
is then  
$\alpha^{\dagger}_s \beta^{\dagger}_s|0
\rangle$ where $|0\rangle= \prod_t |0\rangle_t$. It is 
easy to see that the action of $A$ on $|\psi_s\rangle$ is
\begin{equation}
A |\psi_x\rangle = \alpha^{\dagger}_{s-1} U_{s-1,s} \beta^{\dagger}_s |0\rangle
- \alpha^{\dagger}_{s} U_{s,s+1} \beta^{\dagger}_{s+1} |0\rangle
\equiv |\psi_{s-1, s}\rangle - |\psi_{s,s+1}\rangle,
\end{equation}
where 
\begin{equation}
|\psi_{s,s+k} \rangle= \alpha^{\dagger}_{s} \prod_{j=0}^{k-1} U_{s+j,s+j+1} 
\beta^{\dagger}_{s+k} |0\rangle.
\end{equation}
By repeatedly applying $A$ we obtain
\begin{equation}
A^n |\psi_s\rangle = 
\sum_{k=0}^n (-1)^{n-k} ~^n C_k |\psi_{x-k, x+n-k}\rangle . 
\end{equation}
Using the fact that the inner products 
\begin{equation}
\langle\psi_{s,s+k} | \psi_{t,t+m}\rangle = \delta_{st} \delta_{km},
\end{equation}
we obtain for $0 \leq m \leq 2n$,
\begin{eqnarray}
\langle\psi_{s-n+m}|{A^{\dagger}}^n A^n |\psi_s\rangle 
&=& (-1)^{n+m} \sum_{k=|n-m|}^n ~^n 
C_k ~^n C_{k-|n-m|} ,\nonumber \\ &=& \frac{1}{2\pi i} \int\frac{dz}{z} 
z^{|n-m|}(1-z)^n
(1-\frac{1}{z})^n, \nonumber \\ &=& 
\frac{(-1)^n}{2\pi i} \int\frac{dz}{z} z^{|n-m|-n}
(1-z)^{2n} = (-1)^{m-n} ~^{2n} C_m,
\end{eqnarray}
where the $z$ integral is along a contour surrounding the origin.
For $m$ outside this range the inner product above is zero. Using the 
completeness relation we then find
\begin{equation}
{A^{\dagger}}^n A^n |\psi_s\rangle = \sum_{m=0}^{2n} (-1)^{m-n} ~^{2n}C_m
|\psi_{x-n+m}\rangle.
\label{bmnfinal}
\end{equation}
To see that this result satisfies BMN scaling we go to the states in the
momentum basis
\begin{equation}
|p\rangle = \sum_s e^{2\pi i p s/J} |\psi_s\rangle,
\end{equation}
where $J$ is the total length of the spin chain. 
Then (\ref{bmnfinal}) implies
\begin{eqnarray}
\int {\cal D}U e^{\sqrt{\lambda} V_1} |p\rangle, 
&=&\sum_n\frac{\lambda^n}{(n!)^2}
\int {\cal D}U {A^{\dagger}}^n A^n |p\rangle \nonumber \\&=& \sum_n
\frac{\lambda^n}{(n!)^2} \sum_{m=0}^{2n} (-1)^{m-n} ~^{2n}C_m 
e^{2\pi i p(m-n) /J} |p\rangle ,\\ \nonumber  
&=& \sum_n
\frac{\lambda^n}{(n!)^2}(2i \sin\frac{\pi p}{J})^{2n} |p\rangle
\rightarrow \sum_n \frac{1}{(n!)^2}(\frac{-4\pi^2 \lambda p^2}{J^2})^n 
|p\rangle,
\end{eqnarray}
where the last relation is in the large $J$ limit. Note that because of
the coefficients $(-1)^{m-n} ~^{2n}C_m $ in (\ref{bmnfinal}),
all the terms involving less powers of $1/J^2$ cancel and the leading term
goes like $1/J^{2n}$. This is the correct
$n$-loop BMN scaling behaviour that goes like $(\lambda/J^2)^n$. The effective 
Hamiltonian which is the logarithm of the above expression clearly also obeys
the BMN scaling. We can 
then ask whether the leading terms that survive, give rise to the BMN
anomalous dimension formula. The answer unfortunately turns out to be negative
for three and more loops.
In any case as, 
we will show  in Appendix E, the  model in \eq{defgaupro} with only the 
nearest interaction  does not 
give the right kind of terms that appear in the gauge theory 3-loop
anomalous dimension. We need to include also terms  that
are bilinear in $\varphi$ and connect $\bar\varphi$ and $\varphi$ 
at next nearest neighbour and next-next nearest sites. 
However as we mentioned in the introduction to this section, we have 
not taken into account the modified commutation rules 
given in \eq{bascom}. This would in principle contribute from 2-loop onwards.
It will be interesting to see if the model with just 
nearest neighbour interactions given in \eq{defgaupro} together with 
the modified commutation rules due to the nearest neighbour interactions
given in \eq{su2ws} is sufficient to reproduce the 3-loop anomalous 
dimensions.

\section{Conclusions}

In this paper we have shown that the ${\cal N}=4$ Yang-Mills
theory at weak coupling can be formulated as a theory of a
discretized string, whose world-sheet
$\sigma$ direction is a one-dimensional lattice of points, the
string bits. The degrees of freedom of the world-sheet
consists of 8 bosonic and 8 fermionic oscillators at each
lattice site. 
The spectrum of gauge invariant words of the Yang-Mills theory
at $\lambda =0$  is identical to the spectrum of states in the
Hilbert space of these oscillators,  together
with certain $U(1)$ gauge and cyclicity 
constraints. We showed how to 
obtain the two-point and three-points functions of
all gauge invariant words as the simple delta function overlap
of the discrete world-sheet.
At first order in $\lambda$, i.e. at one loop in
the gauge theory, we showed that 
logarithmic corrections and structure constants can be 
incorporated as nearest neighbour interactions between the
string bits. 
In fact, for the $SU(2)$ subsector, we could determine 
the structure constants from the symmetries of the bit picture, 
upto to an overall proportionality constant.
The symmetries of the bit picture also enabled us to 
obtain the two-loop anomalous dimension Hamiltonian from a 
$U(1)$ gauged linear model of the corresponding oscillators.
This model had only nearest neighbour interactions and it obeyed
BMN scaling to all loops. 

There are several open problems: at a more technical level,
it would be of some interest to see how the modified commutation
relations introduced in section 4 for the $SU(2)$ subsector alter
the calculations of section 5 for the two-loop anomalous dimensions
Hamiltonian.

More generally, throughout this paper we always worked in the planar limit
of the gauge theory, which corresponds to the tree-level
of the discretized string theory.
An interesting question would be to study the $1/N^2$ 
corrections to the (free) 
two-point and three-point functions, to see if
they  arise as a genus one contribution in the discrete world-sheet 
theory. In \cite{Gopakumar:2003ns,Gopakumar:2004qb,Gopakumar:2005fx},
the correlations functions in the free theory was rewritten
to appear as string like amplitudes in AdS. It would be interesting
to see if there is any relationship between this and the 
string bit formulation of this paper.

Perhaps the most pertinent  and obvious question 
that arises  is whether the
discretized string bit model 
discussed in this paper admits a continuum limit.

\acknowledgments

J.R.D. would like to thank Avinash Dhar and Rajesh Gopakumar for
useful discussions. 
L.F.A. thanks Gleb Arutyunov for reading the manuscript and suggesting
improvements.
We thank Jan Plefka for useful correspondences,
suggestions and interesting discussions.  
We  would like to thank the organizers for
the Third Regional Crete Meeting in String Theory for
the opportunity to present this work. 
The work of the authors is partially supported by 
the RTN European program: MRTN-CT-2004-503369.

\appendix

\section{Notations and Conventions}

We first summarize the conventions for the indices
\begin{center}
\begin{tabular}{ll}
$A, B, \ldots$ & $0,1,2, 3, 5, 6$: $SO(2,4)$ indices. \\
$\mu, \nu, \ldots$ & $0, 1, 2, 3$: 4d space time indices. \\
$m,n, \ldots$ & $1, 2, 3$ space indices. \\
$I, J, \ldots$ & $ 1, 2, 3, 4, 5, 6$: $SO(6)$ R-symmetry indices. \\
$i, j, \ldots$ & $1, 2, 3, 4$: $SO(4)\subset SO(6)$ indices. \\
$\gamma, \delta, \ldots$ & $SU(2)_R\subset SO(1,3)$ 
spinorial indices. \\
$\dot\gamma, \dot\delta, \ldots$ & $SU(2)_L\subset SO(1,3)$ 
spinorial indices.\\
$\tau, \upsilon, \ldots $ & 
$SU(2)_{L'} \subset SO(4)\subset SO(6)$ 
R-symmetry spinorial indices. \\
$\dot{\tau}, \dot\upsilon, \ldots$ & $SU(2)_{R'} 
\subset SO(4)\subset SO(6)$  R-symmetry 
spinorial indices.\\
$s, t, \ldots$ & Site labels on bit string. \\
\end{tabular}
\end{center}
We work in the signature diag$(-, +, +, +, + ,-)$ for $SO(2,4)$. 
Raising and lowering the $\gamma, \dot\gamma \ldots $ indices will be
done as in \cite{Wess:1992cp}.

\vspace{.5cm}
\noindent
{\emph{Four dimensional representation of $SO(2,4)$ generators}}
\vspace{.5cm}

Our conventions for the 
four dimensional Weyl representation of $SO(2,4)$ generators is
as follows.
\bea{weyl4d}
M^{\mu\nu} = -\frac{i}{4} [\gamma^\mu, \gamma^\nu], 
&\qquad&
M^{\mu 5}= -\frac{i}{2} \gamma^\mu \gamma^5,
\cr
M^{\mu 6} = \frac{1}{2} \gamma^\mu,
&\qquad&
M^{56} = \frac{1}{2} \gamma^5.
\eea
Here the $\gamma$ matrices are $4\times 4$, 
$SO(1,3)$  gamma matrices in the
the Dirac representation
they obey the algebra 
$\{\gamma^\mu, \gamma^\nu \} = - 2\eta^{\mu\nu}$, 
where $\eta^{\mu\nu} = {\rm diag} ( -1, 1, 1, 1)$. 
These matrices are given by
\bea{gammwb}
\gamma^0 = 
\left(
\begin{array}{cc}
1  & 0 \\
0 & -1
\end{array}
\right), \cr
\gamma^m = 
\left(
\begin{array}{cc}
0  & \sigma^m \\
-\sigma^m  & 0
\end{array}
\right), \cr
\gamma^5 = -i
\left(
\begin{array}{cc}
0  & 1 \\
1  & 0
\end{array}
\right), 
\eea
where $\sigma^m$ refers to the Pauli matrices. 
It is  easily  shown that the $4\times 4$ matrices in \eq{weyl4d}
satisfy the $SO(2,4)$ algebra given in \eq{fullageb}

\vspace{.5cm}
\noindent
{\emph{Four dimensional representation of $SO(6)$ generators}}
\vspace{.5cm}

The conventions for the four dimensional Weyl representation
of $SO(6)$ generators is as follows.
\bea{weyso6}
M^{ij} = \frac{i}{4} [\gamma^i, \gamma^j], &\qquad&
M^{i5} = \frac{i}{2} \gamma^i \gamma^5, \cr
M^{i6} = - \frac{1}{2} \gamma^i , &\qquad& 
M^{56} = -\frac{1}{2} \gamma^5.
\eea
Where $\gamma^i$ and $4\times 4$ $SO(4)$ gamma matrices in 
the Weyl representation, they are given by
\be{44so4}
\gamma^i = \left(
\begin{array}{cc}
0 & \sigma^i \\
\bar\sigma^i & 0 
\end{array}
\right).
\ee
$\sigma ^i = ( 1, i\vec \sigma)$ and 
$\bar\sigma^i = ( 1, -i \vec\sigma)$. These gamma matrices 
obey the algebra $\{ \gamma^i, \gamma^j \} = 2 \delta^{ij} $.
Finally $\gamma^5$ is given by 
\be{so6gamma5}
\gamma^5 = -\gamma^1\gamma^2\gamma^3\gamma^4
= \left(
\begin{array}{cc}
1 & 0 \\ 
0 & -1
\end{array}
\right)
\ee
One can easily verify that the $4\times 4$ matrices 
of \eq{weyso6} satisfy the
$SO(6)$ algebra given in \eq{compso}.

Though we use the symbol $\gamma$  and $\sigma$ to denote the
matrices that occur in both $SO(2,4)$ and $SO(6)$ generators
it will be clear from the context and the indices 
$\mu, \nu, \ldots $ or $i, j, \ldots $ which generator we will 
be dealing with.

\section{Properties of the transform $U$}

The easiest way to obtain the properties of 
$U = \exp(\frac{\pi}{2} M_{05})$ 
given in \eq{propu}
is to use the $4\times 4$ representation of $SO(2,4)$ generators.
We first verify the relation $U^- D U = i E$. 
Note that 
\be{def44gen}
M_{05} =  \frac{i}{2} \sigma^2 \otimes 1, \qquad
D = -M_{56} = -\frac{i}{2} \sigma^1 \otimes 1 , \qquad
E= M_{06} = \frac{1}{2}\sigma^3 \otimes 1. 
\ee
Therefore $U$ is a $\pi/2$ rotation about the $\sigma^2$-axis, given
by 
\be{simdefu}
U = \exp(\frac{\pi}{2} M_{05} ) = \frac{1}{\sqrt{2}} (1+i \sigma^2) \otimes
1,
\ee
using  the above expression for $U$ 
we obtain
\be{verpropu}
U^- D U = \exp{( -\frac{i\pi}{4} \sigma^2)}\,  \frac{i}{2} \sigma^1 \,
\exp{(  \frac{i\pi}{4} \sigma^2)} =  \frac{i}{2} \sigma^3 = i E.
\ee
The tensor product with the identity $2\times 2$ matrix is understood
in the above relation.
From the equation \eq{verpropu}  it   easy to see that the 
property $U^- P_\mu U = L^+$ and $U^-K_\mu U = L^-$ 
is also satisfied. 
$L^+$ is defined as operators which have $+1$ eigen value
with the generator $E$, since $E$ is conjugate to $D$ and 
$P_\mu$ has $+$ eigen value with the operator $D$, 
we obtain $L^+$ is conjugate to $P_\mu$.  A similar argument
can be applied to the operators $L^-$. 
To be more explicit, 
using $P_\mu = M_{\mu 5} + M_{\mu 6}$ and the $4\times 4$
representation of $SO(2,4)$ generators one can show that 
\be{transpw}
U^-P_\mu U = 
\left(
\begin{array}{cc}
0 & \bar\sigma_\mu \\
0 & 0 
\end{array}
\right),
\ee
where $\bar\sigma^\mu = ( -1, -\vec\sigma)$. 
The above relation implies that $\hat U^- \hat P_\mu \hat U 
= -a^\dagger \bar\sigma_\mu b^\dagger$ when one works with the 
Fock space generators.
Similarly it is easily 
shown that 
\be{transkw}
U^-K_\mu K =
\left(
\begin{array}{cc}
0 & 0 \\
\sigma_\mu &0 
\end{array}
\right),
\ee
where $\sigma^\mu = ( -1, \vec\sigma)$.  This implies
that for the Fock space generators we have $\hat U^- \hat K_\mu \hat U 
= b\sigma_\mu a$.

\section{The $PSU(2,2|4)$ algebra}

Let us define the following generators for the $PSU(2,2|4)$ algebra 
\bea{defraise}
A_{\dot\gamma\gamma} = a^\dagger_{\dot\gamma} b^\dagger_{\gamma},
&\qquad& A^{\gamma\dot\gamma} = b^\gamma a^{\dot\gamma}, \cr
L_{\dot\gamma}^{\dot\delta} = a^\dagger_{\dot\gamma} a^{\dot\delta}
-\frac{1}{2} \delta^{\dot\delta}_{\dot\gamma} N_a, &\qquad&
R_{\gamma}^{\delta} = b^\dagger_{\gamma} b^{\delta}
-\frac{1}{2} \delta^{\delta}_{\gamma} N_b, \cr
J^{\tau\dot\tau} = \alpha^{\tau\dagger}\beta^{\dot\tau\dagger},
&\qquad&
J_{\dot\tau\tau} = \beta_{\dot\tau} \alpha_{\tau}, \cr
L^{\prime \tau}_{\upsilon} &=& \alpha^{\tau\dagger}\alpha_{\upsilon}
-\frac{1}{2}\delta^\tau_\upsilon N_\alpha, \cr
R^{\prime \dot\tau}_{\dot\upsilon} &=& 
\beta^{\dot\tau\dagger}\beta_{\dot\upsilon}
-\frac{1}{2}\delta^{\dot\tau}_{\dot\upsilon} N_\beta.
\eea
Then the $PSU(2,2|4)$ algebra is given by the following commutation
relations.
\begin{eqnarray}
[A^{\delta\dot\delta},A_{\dot\gamma\gamma}]&=&
\delta_{\gamma}^{\delta} L_{\dot\gamma}^{\dot\delta} + 
\delta_{\dot\gamma}^{\dot\delta} R_{\gamma}^{\delta}+
\delta_{\gamma}^{\delta} \delta_{\dot\gamma}^{\dot\delta} E, \cr
~[ J^{\tau \dot\tau},J_{\upsilon \dot\upsilon} ] &=&
\delta_{\dot\upsilon}^{\dot\tau} L^{\prime\tau}_{\upsilon}+
\delta_{\upsilon}^{\tau} R^{\prime\dot\tau}_{\dot\upsilon}- 
\delta_{\upsilon}^{\tau}\delta_{\dot\upsilon}^{\dot\tau} J, \cr
\{Q^{- ~ \dot\tau}_{\dot\gamma},S^{+ ~ \dot\delta}_{\dot\upsilon}\}&=&
\delta^{\dot\tau}_{\dot\upsilon} L^{\dot\delta}_{\dot\gamma}-
\delta^{\dot\delta}_{\dot\gamma} 
R^{\prime\dot\tau}_{\dot\upsilon}
 +\frac{1}{2}\delta^{\dot\tau}_{\dot\upsilon} 
\delta^{\dot\delta}_{\dot\gamma}(E+J+Z),\cr
\{Q^{- ~ \tau}_{\gamma},S^{+ ~ \delta}_{\upsilon}\}&=& 
\delta^{\tau}_{\upsilon} R^{\delta}_{\gamma}-
\delta^{\delta}_{\gamma} 
L^{\prime\tau}_{\upsilon}
 +\frac{1}{2}\delta^{\tau}_{\upsilon} 
\delta^{\delta}_{\gamma}(E+J+Z),\cr
\{Q^{+}_{\dot\gamma\tau}, S^{- ~\dot\delta\upsilon} \}&=&
\delta_{\tau}^{\upsilon} L_{\dot\gamma}^{\dot\delta}+
\delta^{\dot\delta}_{\dot\gamma}L^{\prime\upsilon}_{\tau}  +
\frac{1}{2}\delta^{\dot\delta}_{\dot\gamma}\delta^{\upsilon}_{\tau}
(E-J+Z), \cr
\{Q^{+}_{\gamma\dot\tau}, S^{- ~\delta\dot\upsilon} \}&=&
\delta_{\dot\tau}^{\dot\upsilon} R_{\gamma}^{\delta}+
\delta^{\delta}_{\gamma}R^{\prime\dot\upsilon}_{\dot\tau}  +
\frac{1}{2}\delta^{\delta}_{\gamma}\delta^{\dot\upsilon}_{\dot\tau}
(E-J+Z). 
\end{eqnarray}
Here the $Q$'s are defined in Table 2. 
Note that the generator $B= N_\alpha -N_\beta$ does not 
appear on the left hand side of the (anti)-commutator relations,
therefore it acts as an external automorphism of the algebra.

\section{Bit string overlap and three-point functions}

\vspace{.5cm}
\noindent
\emph{Length conserving process}
\vspace{.5cm}

We will first consider examples of various length conserving
processes to test the vertex in \eq{3lcver}. Consider the following
Yang-Mill states
\bea{oper} 
O^{(1)} = \frac{1}{\sqrt{N^{l_1+1}} }{\rm Tr} ( \phi^i
z^{l_1}), 
\qquad O^{(2)} = \frac{1} { \sqrt{N^{l_2 +1}} }{\rm Tr}
( \phi^j z^{l_2} ), \\ 
\nonumber O^{(3)} =
\frac{1}{\sqrt{N^{l_1+l_2 +2}} } {\rm Tr} (\phi^i \bar{z}^{J}
\phi^j \bar{z} ^{l_1 +l_2 -l}), \qquad i \neq j, l_1<l_2. 
\eea 
The
normalizations are chosen such  that in the large $N$ limit the
leading term in their two point function is canonically
normalized. The two point functions of the
fields involved are given by
\be{field2ptfn}
\phi^i_{ab}(x_1)
\phi^j_{a'b'}(x_2) = \frac{\delta_{ij}\delta_{aa'} \delta_{bb'}}
{|x_1-x_2|^2}, \qquad 
z_{ab}(x_1) z_{a'b'}(x_2) 
= \frac{\delta_{aa'}\delta_{bb'}}{|x_1-x_2|^2},
\ee
where $a, b$ label the $U(N)$ indices.
The three point functions of these
operators in the planar limit is given by 
\bea{3ptoper} 
\langle
O^{(1)} (x_1) O^{(2)}(x_2) O^{(3)}( x_3) \rangle &=& 
\frac{l+1}{ N
|x_{23}|^{2(l_2+1)} |x_{31}|^{2(l_1+1)} }, \;\;\; l<l_1, \\
\nonumber &=& \frac{l_1+1}{ N |x_{23}|^{2(l_2+1)}
|x_{31}|^{2(l_1+1)} }, \;\;\; l_1\leq l \leq l_2, \\ \nonumber &=&
\frac{l_1+l_2 -l + 1}{ N |x_{23}|^{2(l_2+1)} |x_{31}|^{2(l_1+1)}
}, \;\;\; l>l_2, 
\eea
where $x_{13}= x_1 -x_2$, $x_{23} =x_2-x_3$.
The multiplicity occurring in the
structure constants are due to the cyclicity of the trace. To
evaluate this correlation function using the three
string vertex, we
first set up the dictionary of states. 
\bea{dict} 
  \langle O^{(1)}(x_1)| &=&
{}^{(1)} \langle
l_1+1| \sum_{s=0}^{l_1} \frac{1}{\sqrt{2}}
(\beta\bar\sigma^i \alpha)(s),
\cr 
\langle O^{(2)} (x_2) |
&=&
{}^{(2)} \langle l_2+1| \sum_{u=l_1+1}^{l_1+l_2 +1}
\frac{1}{\sqrt{2}}
(\beta \bar{\sigma}^i \alpha)(u), 
\\
\nonumber 
\langle O^{(3)} (x_3)|
&=&
{}^{(3)}\langle l_1 +l_2 +2 |
\sum_{t=0}^{l_1+l_2+1} \frac{1}{2} (\beta \bar\sigma^i
\alpha(t) (\beta \bar\sigma^j
\alpha)(t+l+1). 
\eea
Here the state ${}^{(1)}\langle l_1 +1|$ refers to the 
$l_1+1$ bit vacuum state at position $x_1$, explicitly it is 
given by
\be{defl1x1v}
 {}^{(1)} \langle l_1 +1| = \langle 0| 
\exp(\sum_{s=0}^{l_1+1} ix_1P^\dagger(s) ).
\ee
Similar definitions apply for the states
${}^{{2}} \langle l_2+ 1|$ and 
${}^{{3}} \langle l_1 +l_2+ 2|$.
Evaluating the three string
vertex on these states we obtain 
\bea{stringv3} 
\langle O^{(1)}|
\otimes \langle O^{(2)}| \otimes \langle O^{(3)}| \;\; | V_3
\rangle &=& \sum_{t=0}^{l_1+l_2+1} \sum_{u=l_1+1}^{l_1+l_2+1}
\sum_{s=0}^{l_1} \delta(s, t) \delta(u, t+l+1) 
\cr
&\;& \times 
\frac{1}{N|x_{23}|^{2(l_2 +1)} |x_{31}|^{2(l_1+1)}},
\\ \nonumber
&=& 
\frac{l+1}{N|x_{23}|^{2(l_2 +1)} |x_{31}|^{2(l_1+1)}},
\;\; l   < l_1, 
\\ \nonumber &=& 
\frac{l_1+1}{N|x_{23}|^{2(l_2 +1)} |x_{31}|^{2(l_1+1)}},
\;\;    l_1 \leq l \leq l_2, 
\\ \nonumber &=& 
\frac{l_1 +l_2 -l +1}{N|x_{23}|^{2(l_2 +1)} |x_{31}|^{2(l_1+1)}},
 \;\; l >l_2. 
\eea
Thus the overlap rules of the three string vertex agrees with the
gauge theory correlator given in \eq{3ptoper}. 

\vspace{.5cm}
\noindent
\emph{Length non-conserving processes}
\vspace{.5cm}

As an example for a length non-conserving process, consider the
correlation function of 
\bea{oper1} 
O^{(1)}(x_1) &=& \frac{1}{\sqrt{ N^{l_1
+2} }} {\rm Tr} ( \phi^j z^l \phi^i z^{l_1-l} ),
\\ \nonumber
O^{(2)} (x_2)
&=& \frac{1}{\sqrt{ N^{l_2 +2} }} {\rm Tr} ( \phi^i z^m \phi^k
z^{l_1-m} ),
\\ \nonumber
O^{(3)}(x_3) 
&=& \frac{1}{\sqrt{ N^{l_3 +2} }} {\rm Tr} ( \phi^j \bar{z}^n
\phi^k \bar{z}^{l_1-n} ), \;\;\; i \neq j\neq k, \;\; l_1 + l_2
=l_3 .
\eea 
The correlation function of the three operators  in the
large $N$ limit is given by 
\be{corop1} 
\langle O^{(1)}(x_1) O^{(2)}(x_2)
O^{(3)}(x_3) \rangle = \frac{\delta(l+m, n)}{N} 
\frac{1}{|x_{12} |^{2} |x_{23}|^{2(l_2+1)} |x_{13}|^{2(l_1+1)}  } 
\ee 
For
this process, the length violation is $2$ units. To evaluate the
correlation function in string variables we set up a dictionary
similar to the previous case. 
\bea{dict1} 
\langle O^{(1)} | &=&
\sum_{s=0}^{l_1+1} {}^{(1)} \langle l_1 +2| (\beta 
\bar{\sigma}^j\alpha)(s)
(\beta \bar{\sigma}^i\alpha)(s+l+1)
 , \\ \nonumber
\langle O^{(2)} | &=& \sum_{t=l_1+1 }^{l_3+2} {}^{(2)} 
\langle l_2 +2| (\beta  \bar{\sigma}^i\alpha)(t)
(\beta \bar{\sigma}^k\alpha) (t+m+1)
 , \\ \nonumber
\langle O^{(3)} | &=& \sum_{u=0}^{l_3+1} {}^{(3)} \langle l_3 +2|
(\beta  \bar{\sigma}^j\alpha)(u)
(\beta \bar{\sigma}^k\alpha)(u+n+1). 
\eea 
Evaluating the length non-conserving
vertex in \eq{ncv3} with $l=1$ on these states we obtain.
\bea{fincor} 
& &\langle O^{(1)}| \otimes \langle O^{(2)}| \otimes
\langle O^{(3)}| \;\; |V_3 \rangle =  \\ \nonumber
& &   \sum_{u=0}^{l_3+1}
\sum_{s=0}^{l_1+1} \left( \sum_{t=l_1 +1}^{l_3+2} \delta(s, u)
\delta(s+ l+1, l_1 +1) \delta(t, l_1 +1) \right. \\ \nonumber & &
\left. \delta(t+ m+1, u+s +n +2)\right) 
\times
\frac{1}{N |x_{12} |^{2} |x_{23}|^{2(l_2+1)} |x_{13}|^{2(l_1+1)}  } 
, \\ \nonumber &=&
\frac{\delta(m+l, n) 
}{N |x_{12} |^{2} |x_{23}|^{2(l_2+1)} |x_{13}|^{2(l_1+1)}  }. 
\eea 
Here the two delta functions with $l_1+1 $ in
the argument are due to the fact that the $1$ and $2$ strings
overlap only if the bits at the $l_1+1$ sites are equal. The
argument in the last delta function $u+s+n+2$ is due to the fact
the overlap of the $2$nd and $3$rd string are shifted by $l=1$
in \eq{ncv3}. 
Thus there is agreement with the field theory
calculation \eq{corop1}. The above calculation can easily be
extended to cases with $i=j$ etc.

\section{Anomalous dimensions at three loops}

As we mentioned in section 5, the model given in \eq{ontmod} is not 
sufficient to obtain the anomalous dimensions at three loops.
In this appendix we show that it is possible to reproduce the 
anomalous dimensions to three loops for the $SU(2)$ subsector
using a gauged linear oscillator model, but with 
bilinears of oscillators at positions
$(s, s+2)$, $(s, x+3)$ in addition to 
oscillators at $(s, s+1)$. 
Let us define the anomalous dimension Hamiltonian 
to be given by 
\be{defan3l}
\exp( c \sum_{n=1}^3 \lambda^n H^{(n)}) 
= \int {\cal D } U \exp( \sum_{n=1 2, 3, s}^3 g^n f_n
(  A_s^{(n)}U_{s,s+n}  + \rm{ h. c. } ) ), 
\ee
where $g= \sqrt{\lambda}$ and 
$f_n$ are functions of the 't Hooft coupling $\lambda$, with 
the expansions given by 
\bea{expgnt}
f_1 &=& f_{10} + \lambda f_{11} + \lambda ^2 f_{12}, \\ \nonumber
f_2 &=& f_{20} + \lambda f_{21}, \\ \nonumber
f_3 &=& f_{30}. 
\eea
In these expansions we have assumed that the couplings $f_n$ are 
themselves expansions of in the 't Hooft coupling $\lambda$. 
$U_{s,s+n}$ is a link variable which transforms as 
$U_{s,s+n}\rightarrow \exp(i\theta_s) U_{s,s+n} \exp(-i\theta_{s+n})$
under the local $U(1)_Z$ transformation. 
The anomalous dimensions to three loops is given by \cite{Beisert:2003ys}
\bea{ful3loandim}
H^{(1)} = a_1, \qquad H^{(2)} = a_2 - 4 a_1, \cr
H^{(3)} = 2 a_3 - 12 a_2 + 30 a_1 + t_2 -t_3,
\eea
where the $a_n$ and $t_n$ are defined by
\bea{defat3l}
a_n &=& ( 1- \sigma_s \cdot \sigma_{s+n} ),  \\ \nonumber
t_1 &=&  ( 1- \sigma_s \cdot \sigma_{s+1}) 
          ( 1-\sigma_{s+2}  \cdot \sigma_{s+3}),  \\ \nonumber
t_2 &=& ( 1- \sigma_{s} \cdot \sigma_{s+2}) 
( 1- \sigma_{s+1} \cdot \sigma_{s+3} ), \\ \nonumber
t_3 &=& ( 1-\sigma_{s} \cdot \sigma_{s+3}) 
( 1- \sigma_{s+1} \cdot \sigma_{s+2} ).
\eea
In the above equations summation over the sites $s$ is implied.
Now our strategy will be to perform the integrations over the 
link variables in \eq{defan3l} and determine the $f_{mn}$ and $c$
so that the anomalous dimensions to three loops, known
from gauge theory calculations, is reproduced.
At this point one might think that there is a  sufficient
number of unknowns in $f_{mn}$ and $c$ to allow to obtain
$H^{(n)}$. However we will see that the $f_{mn}$ are overdetermined
but they admit a unique consistent solution.
We now expand the exponential in \eq{defan3l} to the sixth order. 
Here we write the expressions obtained 
after the integrations over the link variables in \eq{defan3l} 
\bea{sigexp3}
\tilde  H  
&=&
\frac{1}{2} \left(
g^2 f_1^2   a_1 + g^4 f_2^2 a_2 + g^6 f_3^2a_3 \right)   
\\ \nonumber
&+&
\frac{3}{4!} f_1^4 g^4 
\left( - 4 a_1 + a_2 \right)
+ \left\{ \frac{g^4}{2}  \left( \frac{f_1^2 a_1}{2} \right)^2 \right\}
\\ \nonumber
&+&
\frac{1}{4!} 
f_3 f_1^3g^6 
\left[ 12 a_1 - 12 a_2 + 4 a_3 + 6( -t_1 + t_2 -t_3) \right]
\\ \nonumber
&+& \frac{6}{4!} f_1^2 f_2^2 g^6 
\left( -t_1 -t_2 + t_3 \right)
+
\\ \nonumber
&+& \frac{1}{6!}
g^3 f_1^6 
\left( 40( a_3 -6 a_2 + 15a_1) + 15 ( t_1 + t_2 -t_3) \right)
\\ \nonumber
&+&
\left\{ g^6 \frac{f_1^2 a_1}{2} \frac{3 f_1^4 }{4!} (-4a_1 + a_2) 
+ 
g^6 \left( \frac{ f_1^2 a_1}{2}\right) 
\left(\frac{f_2^2a_2}{2} \right) \right\}.
\eea
To obtain the above equation we have used various identities from 
Appendix F and performed similar manipulations as for the two loop
case discussed in section 5.
Note that in \eq{sigexp3}, the non-local terms are written in 
curly bracket. They correspond to the exponentiation of the 
lower order terms. For instance the first term in the curly 
brackets correspond to the exponentiation of the 
one-loop term $g^2 f_1^2 a_1/2$. Similar arguments 
hold for the other non-local terms. Therefore on exponentiating,
as in \eq{defan3l}, we will obtain a local Hamiltonian.

We will now deal only with the local terms and impose the 
condition that the local terms organize themselves to be
the anomalous dimensions Hamiltonian to 3-loops given in \eq{ful3loandim}. 
Note that the coefficient of $t_1$ for the anomalous dimension at 3-loop 
is zero, this implies the following equation for the couplings
\be{t10eq}
1 -12 \frac{f_{30}}{f_{10}^3} - 12 
\left( \frac{f_{20}}{f_{10}^2}\right)^2 =0.
\ee
We also have the condition that the coefficient of $a_3$ is 
twice the coefficient of $t_2$, this leads to the following
equation
\be{a32t2}
12 \left( \frac{f_{30}}{f_{10}^3}\right)^2 + 
4 \frac{f_{30}}{f_{10}^3} + \frac{4}{3} 
= 12 \frac{f_{30}}{f_{10}^3} -12 
\left( \frac{f_{20}}{f_{10}^2} \right) ^2 + 1.
\ee
Combining \eq{t10eq} and \eq{a32t2} we obtain a quadratic 
equation for the ratio $f_{30}/f_{10}^3$.
\bea{quadreq}
12 \left( \frac{f_{30}}{f_{10}^3}\right)^2 - 
20 \frac{f_{30}}{f_{10}^3} + \frac{4}{3} 
  = 0 , \\ \nonumber
\frac{f_{30}}{f_{10}^3} = \frac{5\pm \sqrt{21}}{6}.
\eea
Among the two roots we choose the latter one as that ensures the
ratio $f_{20}/{f_{10}^2}$ to be real. Substituting this in 
\eq{t10eq} we obtain
\be{solf20}
\left( \frac{f_{20}}{f_{10}^2} \right)^2  = \frac{2\sqrt{21} -9}{12}.
\ee
There are two further conditions to be met so that 
the result of the integrations over the link variables result
in the anomalous dimension to 3-loops: (i) The ratio 
of the coefficient of the $g^2$ term to that of 
$a_2$ in the $g^4$ term
must be one; (ii)  the ratio of the 
coefficient of the $t_2$ to that of the $g^2$ term also should be 
one. These two conditions ensure that three loop anomalous dimension
is obtained upto an overall scaling constant $c = f_{10}^2/2$ in
the exponential of \eq{defan3l}
These conditions give two independent equations for the 
coupling $f_{10}$ which admit a common solution. The first 
condition gives 
\be{1stcondt}
\frac{1}{f_{10}^2} = 
\left( \frac{f_{20}}{f_{10}} \right)^2 + \frac{1}{4} = 
\frac{\sqrt{21} -3}{6},
\ee
while the second condition gives  
\bea{2ndcondt}
\frac{1}{4!} \left( 6 \frac{f_{30}}{f_{10}^3} 
- 6 \left( \frac{f_{10} }{f_{20} }\right)^2 + \frac{1}{2}\right) 
&=& \frac{1}{2 f_{10}^2 } \left(
\left( \frac{f_{20} }{f_{10}^2} \right)^2 + \frac{1}{4}  \right),
\\ \nonumber
&=& \frac{1}{2} \frac{1}{f_{10}^4}. 
\eea
Substituting  the equation \eq{t10eq} in the left hand side of
the first line we obtain the consistency condition
\be{constcondt}
\frac{f_{30}}{f_{10}^3} = \left( \frac{1}{f_{10}^2}\right)^2.
\ee
This condition is clearly satisfied by the solution in 
latter solution in \eq{quadreq}
and \eq{2ndcondt}. We now can easily fix the remaining
constants $f_{11}, f_{12}, f_{21}$ by requiring BMN scaling.
Though we have obtained the anomalous dimensions to three loop, 
the model in 
\eq{defan3l} is not entirely satisfactory. 
This is because we do not have a principle to determine 
the constants $f_n$, further more we have not taken the
modified commutation relations of section 4. into account. 
Taking this into account might change the values of
the $f_n$'s found in this appendix.

\section{Properties of bilinears in the $SU(2)$ sector}

We define a oscillator  bilinear of length $l$ to be
given by
\be{defbilsl}
A^{(l)}_s = \bar\varphi_s \varphi_{s+l} =
\alpha^\dagger_s \alpha_{s+l} + \beta_s\beta_{s+1}^\dagger.
\ee
Then the various 
commutation relations which are required 
to obtain the anomalous dimension Hamiltonian upto three loops
are given by
\bea{lotscomm}
[A_s^{(1)},  A_t^{(1)\dagger} ]
&=& \delta_{st}( N_s^\alpha - N_s^\beta - 
(N_{s+1}^\alpha -N_{s+1}^\beta ) ) 
= \delta_{st} G_s, 
\\  \nonumber
[G_s ,  A_t^{(1)\dagger} ] 
&=& ( \delta_{s, t+1} + \delta_{t,s+1} - 
2\delta_{st} ) A_t^{(1)\dagger}, \\
\nonumber
[A_s^{(1)}, A_t^{(1)} ] &=& \delta_{s+1, t} A_s^{(2)} -
\delta_{s, t+1} A_t^{(2)}, \\
\nonumber
[A_s^{(1)\dagger}, A_t^{(1)\dagger}]
&=& 
\delta_{t+1, s} A_t^{(2)\dagger} - \delta_{t, s+1} A_s^{(2)\dagger},
\\
\nonumber
[A_s^{(2)}, A_t^\dagger]
&=& \delta_{s+1, t} A_s^{(1)} -\delta_{st} A_{s+1}^{(1)}, 
\\
\nonumber
[A_s, A_t^{(2)}] &=& 
\delta_{s+1, t} A_s^{(3)} -\delta_{s, t+2} 
A_s^{(3)}, \\
\nonumber
[A_s^{(1)} , A_t^{(2)\dagger} ]
&=& 
\delta_{s, t+1} A_t^{(1)\dagger} -\delta_{st} A_{t+1}^{(1)\dagger}
,\\
\nonumber
[A_s^{(3)}, A_t^{(1)\dagger} ] 
&=& \delta_{s+2, t} A_x^{(2)} - \delta_{st} A_{s+1}^{(2)}.
\eea
One also can obtain other commutation relations by 
taking the Hermitian conjugates of the above.

The identities below relate the bilinears to Pauli matrices, these 
relations are always valid on physical states which satisfy the 
$U(1)_Z$ constraint.
Let us define 
then we have
\begin{eqnarray}
\label{paulirel}
A_s^{(l}  A_s^{(l)\dagger} &=& \frac{1}{2}
(1-\sigma_s\cdot \sigma_{s+l}),
\cr
A_s^{(l)\dagger} A_s^{(l)} &=& \frac{1}{2}
( 1- \sigma_s \cdot \sigma_{s+1}),  
\cr 
A_s^{(2)} A_{s}^{(1)\dagger}
A_{s+1}^{(1)\dagger} &=& - \frac{1}{4}
(1-\sigma_s\cdot \sigma_{s+2})(1-\sigma_{s+1} \cdot\sigma_{s+2}), 
\cr
A_{s+1}^{(2)} A_{s}^{(1)}
A_{s}^{(3)\dagger} &=& - \frac{1}{4}
(1-\sigma_{s+1}\cdot \sigma_{s+3})(1-\sigma_{s} \cdot\sigma_{s+3}), 
\cr
A_{s}^{(2)} A_{s+2}^{(1)}
A_{s}^{(3)\dagger} &=&  \frac{1}{4}
(1-\sigma_{s+1}\cdot \sigma_{s+2})(1-\sigma_{s} \cdot\sigma_{s+3}), 
\cr
A_s^{(2)\dagger} A_{s}^{(1)}
A_{s+1}^{(1)} &=&  \frac{1}{4}
(1-\sigma_s\cdot \sigma_{s+2})(1-\sigma_{s+1} \cdot\sigma_{s+2}), 
\cr
A_{s+1}^{(2)\dagger} A_{s}^{(1)\dagger}
A_{s}^{(3)} &=&  \frac{1}{4}
(1-\sigma_{s+1}\cdot \sigma_{s+3})(1-\sigma_{s} \cdot\sigma_{s+3}), 
\cr
A_{s}^{(2)\dagger} A_{s+2}^{(1)\dagger}
A_{s}^{(3)} &=&  - \frac{1}{4}
(1-\sigma_s \cdot \sigma_{s+2}) ( 1-\sigma_s \cdot
\sigma_{s+3}). 
\eea
In all the above equations 
and the remaining equations summation over the site label $s$ is
implied.
The identities below relate 4 bilinears to quartic terms in 
Pauli matrices which are are different sites. 
We first define the various quartic terms in 
Pauli matrices
\bea{defqaarpa}
t_1 &=& 
 (1- \sigma_{s}\cdot\sigma_{s+1})
 (1- \sigma_{s+2}\cdot\sigma_{s+3}),
\\ \nonumber
t_2 &=&
 (1- \sigma_{s}\cdot\sigma_{s+2})
 (1- \sigma_{s+1}\cdot\sigma_{s+3}),
\\ \nonumber
t_3 &=& 
 (1- \sigma_{s}\cdot\sigma_{s+3})
 (1- \sigma_{s+1}\cdot\sigma_{s+2}).
\eea
We  have the following identities
\bea{qarbipu}
A_{s+1} ^{(2)} A_s^{(2)\dagger} A_s^{(1)}
A_{s+1}^{(1)\dagger} + {\rm h.c. }
&=& \frac{1}{4}( t_3 - t_2 -t_1),
\cr 
A_{s}^{(3)}  A_{s+1}^{(1)\dagger} 
A_s^{(1)\dagger} A_{s+2}^{(1) \dagger} + {\rm h.c }
&=& \frac{1}{4} ( -t_3 + t_2 -t_1). 
\eea
We also have the following identity
\be{finidenu}
A_s^{(2)} A_{s+1}^{(1)} A_{s+1}^{(1)\dagger} A_s^{(2)\dagger}
+ A_s ^{(2)} A_s A_s^\dagger A_s^{(2)\dagger} =
0.
\ee
Note  that in all manipulation whenever the Gauss law $G_s$ 
occurs at either the extreme left or right of a string
of bilinears one can set that term to zero as it vanishes 
on any physical state. Also whenever higher $A_sA_s$ or higher
powers and $A_s^\dagger A_s^\dagger$ or higher powers
occur at either the extreme left or right 
that term can be set to zero
as one can easily verify it vanishes on any physical state. 

\bibliographystyle{utphys}
\bibliography{bits}

\end{document}